\documentclass{article}

    \PassOptionsToPackage{numbers, sort&compress}{natbib}

    \usepackage[final]{neurips_2025}

\usepackage[utf8]{inputenc}
\usepackage[T1]{fontenc}
\usepackage{shortex}
\usepackage{xcolor}

\begin{document}

\title{Asymptotically exact variational flows\\ via involutive MCMC kernels}

\author{
  Zuheng Xu \qquad Trevor Campbell\\
  Department of Statistics\\
  University of British Columbia\\
  \texttt{[zuheng.xu | trevor]@stat.ubc.ca} 
}

\maketitle

\begin{abstract}
Most expressive variational families---such as normalizing flows---lack practical convergence guarantees, 
as their theoretical assurances typically hold only at the intractable global optimum.
In this work, we present a general recipe for constructing tuning-free, asymptotically exact
variational flows on \emph{arbitrary} state spaces from involutive MCMC kernels.
The core methodological component is a novel representation of general involutive MCMC kernels 
as invertible, measure-preserving \emph{iterated random function} systems, 
which act as the flow maps of our variational flows.
This leads to three new variational families with provable total variation convergence.
Our framework resolves key practical limitations of existing variational families with similar guarantees (e.g., MixFlows), 
while requiring substantially weaker theoretical assumptions.
Finally, we demonstrate the competitive performance of our flows 
across tasks including posterior approximation, Monte Carlo estimates, and normalization constant estimation,
outperforming or matching No-U-Turn sampler (NUTS) and black-box normalizing flows.
\end{abstract}

\vspace{-1em}
\section{Introduction}
\vspace{-0.5em}

Variational inference (VI) \citep{Jordan99,Wainwright08,Blei17} is a general
methodology for approximate probabilistic inference, where the goal is to
approximate a target distribution (e.g., a Bayesian posterior) within a
specified variational family. This variational family is typically chosen to be
a parametric family that enables tractable inference---allowing for i.i.d. sampling and density evaluation
\citep{Wainwright08,Blei17,Rezende15,Ranganath16,Papamakarios21,margossian2024variational}.
This tractability offers key benefits: it enables the evaluation and
optimization of approximation quality via unbiased estimates of the evidence
lower bound (ELBO) \citep{Blei17}, which corresponds to the Kullback-Leibler
(KL) divergence \citep{Kullback51} to the target distribution up to a constant. 
Moreover, it facilitates downstream tasks such as importance sampling
\citep{kahn1953methods,kahn1955use} and normalization constant estimation.

The quality of a variational approximation is fundamentally determined by the expressiveness of its variational family. 
Significant progress has been made in constructing flexible families,
including boosted mixtures
\citep{Guo16,Miller17,Wang16,Locatello18,Locatello18b,Campbell19b} and
normalizing flows
\citep{Rezende15,Papamakarios21,Kobyzev21,Agrawal20,Dinh17,vansylvester}. 
These families often exhibit \emph{universal approximation guarantees} 
\citep{Campbell19b,koehler2021representational,kong2020expressive}:
as the number of mixture components or flow layers grows, the family can
approximate any distribution arbitrarily well under mild assumptions.
However, 
a major limitation remains---theoretical guarantees pertain only to the
\emph{optimal} variational approximation, which is rarely obtained in practice
due to non-convex optimization.
In contrast, Markov chain Monte Carlo (MCMC) [\citealp{Robert04,Robert11}; \citealp[Ch. 11, 12]{Gelman13}]
is \emph{asymptotically exact}, meaning it is guaranteed to produce arbitrarily accurate results given sufficient computation
for any valid choice of tuning parameters (though some values may yield higher efficiency than others).

This has motivated extensive work aimed at bridging VI and MCMC. 
In particular, many recent approaches fall under the categories of differentiable
annealed importance sampling (DAIS) or differentiable sequential Monte Carlo (DSMC)
\citep{Salimans15,Wolf16,Geffner21,Zhang21,Thin21,Jankowiak21,geffner2023langevin,Arbel21,matthews2022continual,naesseth2018variational,maddison2017filtering}.
These methods can be interpreted as gradient-based tuning of AIS/SMC
exploration or backward kernels to improve approximation quality. 
However, their theoretical guarantees in the limit of flow length
\citep{Jankowiak21,Zhang21,zenn2024differentiable} often rely on idealized 
assumptions---such as perfect transitions or diminishing stepsizes---that
rarely hold in practice or only apply under optimal tuning.
Moreover, AIS/SMC methods are known to be highly sensitive to tuning, and
DAIS/DSMC methods inherit substantial tuning cost to ensure robust performance \citep{Kim25}.

\citet{Xu22mixflow} introduced MixFlow, 
an asymptotically exact variational family that does not require optimal tuning. 
A MixFlow is constructed by averaging pushforwards of a reference distribution under
repeated application of an invertible map. 
When this map is both \emph{ergodic} and \emph{measure-preserving} (e.m.p) with respect to
the target distribution $\pi$, MixFlows converge to $\pi$ in total variation as
the number of steps increases, 
while retaining the tractability of standard variational inference. 
However, its practical applicability is limited by the
challenge of designing an invertible $\pi$-e.m.p. map for general continuous targets
(several solutions exist for discrete spaces \citep{Neal12,Diluvi2024MADMix}).
The main obstacles are: 
(1) continuous e.m.p.~maps often involve simulation of ODEs, 
which requires discretized numerical methods that destroy the e.m.p.~property;
(2) exactness often requires discrete Metropolis--Hastings (MH) corrections that are not invertible;
and (3) proving ergodicity of such maps is very challenging.
For example, \citet{Xu22mixflow} proposed a map based on the \emph{uncorrected} Hamiltonian Monte Carlo (HMC), 
which is neither exactly measure-preserving nor provably ergodic.
Other existing Hamiltonian-based methods \citep{ver2021hamiltonian} also suffer from discretization error and are non-ergodic \citep{robnik2023microcanonical}.
Attempts via deterministic Gibbs samplers based on measure-preserving ODEs 
\citep{neklyudov2021deterministic} or CDF/inverse-CDF transformations \citep{Neal12} 
are also limited by the intractability of computing the exact transformations.
MH corrections used to restore exactness \citep{Neal12, Murray12}
result in non-invertible transformations due to the accept-reject mechanism;
recall that invertibility is required by variational flows to enable tractable density evaluation.
To date, there is no framework for constructing variational
families whose \emph{practical implementation} achieves an MCMC-like asymptotic exactness.

In this work, we address the challenges mentioned above 
and propose a new framework for developing practical, asymptotically exact variational flows.
Rather than relying on e.m.p dynamics as in MixFlow \citep{Xu22mixflow},
our framework leverages \emph{iterated random functions} (IRF)\footnote{IRFs are also referred 
to as iterated function systems (IFS) in some literature, e.g., \citep{morita1988deterministic}.} \citep{diaconis1999iterated}---a type of 
\emph{random} dynamical system.
The main contributions of this work are as follows:
\benum
\item We develop a method for deriving exact measure-preserving transformations
    from general \emph{involutive MCMC} kernels \citep{Tierney94,Neklyudov20},
    while preserving invertibility of the transformation.

\item We introduce a more general framework for constructing asymptotically
    exact flows, leading to three novel variational families beyond
    the original MixFlow for general state spaces.

\item We establish total variation convergence guarantees for these new
    families under significantly weaker assumptions than those required in
    MixFlow theory \citep{Xu22mixflow}, notably relaxing the ergodicity conditions of the flow maps. 
\eenum

\vspace{-1em}
\section{Background}
\vspace{-0.5em}

Throughout, let $\pi$ be a target distribution on a measurable space $(\scX, \scB)$ 
equipped with a $\sigma$-finite base measure $m$.
All distributions are assumed to have densities with respect to the base
measure on their corresponding spaces, and we use the same symbol to denote
both a distribution and its density. 
Given a transformation $f$ and a
distribution $p$, we write $f(p(x))$ for the function $f$ evaluated at $p(x)$,
and $fp(x)$ for the density of the pushforward distribution $fp$ evaluated at $x$.

\vspace{-0.5em}
\subsection{Homogeneous MixFlows} \label{sec:deterministicmixflow}

A mixed variational flow (MixFlow) \citep{Xu22mixflow,Diluvi2024MADMix} is built from a \emph{deterministic}, 
$\pi$-ergodic (\cref{def:ergodic}) and measure-preserving (e.m.p.) 
diffeomorphism $f$\footnote{$f$ is a diffeomorphism if it is continuously differentiable and has a continuously differentiable inverse.}. 
Given such a map $f$ and a reference distribution $q_0$ on $\scX$ that enables \iid sampling and density evaluation, 
the MixFlow density is given by
\vspace{-1em}
\[ \label{eq:deterministicmixflow}
    \forall x \in \scX,    
    \quad
    \sbq_T(x) 
    = \frac{1}{T}\sum_{t=1}^{T} f^t q_0(x) 
    = \frac{1}{T}\sum_{t=1}^{T} \frac{q_0\left(f^{-t}x\right)}{\prod_{i=1}^t J\left(f^{-i}x\right)}, 
    \quad
    J(x) = |\det \nabla f(x)|,
\]
where $f^t x$ and $f^t q_0$ denote mapping $x$ or pushing $q_0$ through $t$ ($t > 0$) iterations of $f$.
We use the convention that $\sbq_0 = q_0$ (MixFlow of length $0$ is just the reference distribution $q_0$).
\cref{eq:deterministicmixflow} is tractable if $f^{-1}$ and the Jacobian $J$ can be evaluated.
To generate $X \sim \sbq_T$, we first draw $X_0 \sim q_0$ and a flow length $K \sim \Unif\{1, 2, \dots, T\}$, 
and then map $X_0$ through $K$ iterations of $f$, i.e., $X = f^K(X_0)$.
Since $\sbq_T$ is built from a time-homogeneous e.m.p dynamical system, 
we label it a \emph{homogeneous MixFlow},
to distinguish it from our proposed \emph{random} dynamical system flows (see \cref{sec:flow}).
The asymptotic exactness of homogeneous MixFlows comes from the fact that 
$\lim_{T\to \infty} \TV(\sbq_T, \pi) = 0$ regardless of the tuning of the flow map $f$ \citep[Theorem 4.2]{Xu22mixflow}.

In practice, the map $f$ is typically designed to mimic familiar MCMC kernels
\citep{Xu22mixflow,Diluvi2024MADMix}, so that its trajectories have similar
statistical behavior to the corresponding Markov chain. 
Despite this, general constructions of \emph{exact} e.m.p. MixFlow maps for
continuous target distributions remain unavailable. As discussed in the
introduction, achieving both exact measure preservation and ergodicity is
highly non-trivial in practice. Consequently, practitioners often rely on
approximate maps, leading to a gap between theoretical guarantees and practical
implementations. These approximations can introduce numerical instability and
degrade performance as $T$ increases \citep{Xu22mixflow,Xu23chaos}.
In \cref{sec:improveddetermixflow}, we show how to design homogeneous
MixFlows that are exact in practice. Additionally, we present a refined
characterization of the density $\sbq_T$ by leveraging the measure-preserving
property of $f$, which simplifies implementation, improves robustness, and
provides a more intuitive convergence analysis.

\vspace{-0.5em}
\subsection{Involutive MCMC}

An involutive MCMC kernel \citep{Tierney94,tierney1998note,Neklyudov20} is a Metroplis-type
Markov kernel with a deterministic proposal defined by an \emph{involution} 
$g$, i.e., a self-inverse function satisfying $g = g^{-1}$.
This framework encompasses a broad class of MCMC algorithms, 
with many popular algorithms appearing as special cases
\citep{Neklyudov20,Liu24,BironLattes24,bou2024gist,bou2024incorporating} (see \cref{apdx:example}).
The detailed transition procedure of involutive MCMC is described in \cref{alg:involutivemcmc} of \cref{apdx:codeimcmc}.
Consider an auxiliary variable $v$ defined on a space $\scV$, with conditional density 
$\rho(v \mid x)$ given $x \in \scX$ with respect to a base measure $m_v$ on $\scV$, 
and the augmented target density $\sbpi(x, v) := \pi(x) \rho(v|x)$. 
Let $\sbm := m\times m_v$ be the joint base measure on $\scX\times \scV$.
For an involution  $g\!: \!\scX\! \times\! \scV\! \to\! \scX\! \times\! \scV$,
each transition from state $x$ proceeds in three steps:
\benum
  \item Sample an auxiliary variable $v \sim \rho(\d v\mid x)$;
  \item Propose a new state $(x', v') = g(x, v)$;
  \item Accept $x'$ with probability 
  $\min\left(1, \frac{\sbpi(x', v')}{\sbpi(x, v)} J_g(x, v)\right)$ where 
  $J_g(x, v) := \frac{\d g\sbm}{\d\sbm}(x, v)$\footnote{
    For differentiable $g$ on continuous state spaces (e.g., $\reals^d$),
$J_g(x, v) = \left|\det \nabla g(x, v)\right|$ is its Jacobian determinant.
We adopt the measure-theoretic formulation of \citet{tierney1998note} to handle arbitrary state spaces.
  }.
\eenum

An involutive Markov kernel $K$ defined this way is \emph{reversible} with respect to
both the augmented target $\sbpi(x, v)$ and its marginal $\pi(x)$ \citep[Theorem 2]{tierney1998note}.

\bprop \label{prop:involutivemcmc}
The involutive MCMC kernel $K(x', v' | x, v)$ (defined in \cref{alg:involutivemcmc}) satisfies that
\[
K(x', v' | x, v) \sbpi(x, v) = K(x, v | x', v') \sbpi(x', v'), 
\quad 
\shK(x'|x) \pi(x) = \shK(x|x') \pi(x'), 
\]
where $\shK$ is the marginalized kernel defined as: $\shK(x'|x) := \int K(x', v'\mid x, v) \rho(\d v | x) \d v'$.
\eprop

\vspace{-0.5em}
\subsection{Iterated random functions}

An \emph{iterated random function} (IRF) system \citep{diaconis1999iterated} on $\scX$ consists of a
sequence of \emph{random} maps:
\vspace{-0.5em}
\[ \label{eq:ifs}
  \forall t\in\nats, \,\, X_{t+1}  = f_{\theta_{t+1}}(X_t), \quad X_0\in\scX, \quad (\theta_t)_{t\in \nats} \distiid \mu,
\] 
where $\{f_\theta: \scX \to \scX: \theta \in \Theta\}$ is a set of parametrized functions, 
with each $\theta$ drawn \emph{randomly} from a distribution $\mu$ on the parameter space $\Theta$.
The above IRF induces a Markov kernel given by:
\vspace{-0.5em}
\[\label{eq:markovkernel}
\forall x \in \scX, \quad \forall B \in \scB, \quad 
P(x, B) := \int_\Theta \1_{B}(f_\theta(x)) \mu(\d \theta).
\]
This yields a simple characterization of the action of the Markov process $P$ on a distribution $q$:
\vspace{-0.5em}
\[ \label{eq:irfpushforward}
    Pq(y) := \int_{\scX} P(x, y) q(\d x) = \E\left[ f_\theta q(y) \right], 
    \quad
    \theta \sim \mu, 
    \quad  
    \text{$f_\theta q$: pushforward of $q$ under $f_\theta$}.
\]
Throughout this work, we focus on IRFs where the family $\{f_\theta:\theta\in\Theta\}$ satisfies \cref{assump:niceirf}.
\bassum \label{assump:niceirf}
For $\mu$-a.e.~$\theta \in \Theta$, $f_\theta(\cdot)$ is bijective and $\pi$-measure-preserving ($\pi$-m.p.).
Furthermore, the Markov kernel $P$ induced by the IRF is $\pi$-irreducible.
\eassum

\cref{assump:niceirf} implies that the sequence of iterates $X_t$ produced by the IRF behave like a
$\pi$-invariant, irreducible Markov chain. Therefore,
long-run averages of
IRF iterates converge to expectations under $\pi$, 
following the standard law of large numbers (LLN) for MCMC \citep{Roberts04}, 
also known as the \emph{random Birkhoff ergodic theorem} in the IRF literature [\citealp{kakutani1950random}; \citealp[Cor. 2.2.]{kifer2012ergodic}]. 
\cref{thm:randomergodic} synthesizes these results under \cref{assump:niceirf}, providing a unified statement for convenient use in our framework; 
proof can be found in \cref{apdx:irf}.

\bthm \label{thm:randomergodic}
Suppose that IRF $f_\theta$ satisfies \cref{assump:niceirf}.
Then, given $\phi \in L_1(\pi)$, we have that 
\vspace{-0.5em}
\benum
\item for $\pi$-\aev $x \in \scX$ and $\mu$-almost all $(\theta_t)_{t\in \nats}$, as $T \to \infty$:
\vspace{-0.5em}
    \[\label{eq:LLN}
    \frac{1}{T} \sum_{t=0}^{T-1} \phi\left(f_{\theta_t}\circ \cdots\circ f_{\theta_1}(x)\right) 
    & \longrightarrow \E[\phi(X)], 
    \quad X\sim \pi;
    \]
\vspace{-1.2em}
\item for $\mu$-almost all $(\theta_t)_{t\in \nats}$, as $T \to \infty$:
\vspace{-0.5em}
    \[\label{eq:randomergodic}
    \frac{1}{T} \sum_{t=0}^{T-1} \phi\left(f_{\theta_t}\circ \cdots\circ f_{\theta_1}(x)\right) 
    & \stackrel{L^1(\pi)}{\longrightarrow} \E[\phi(X)], 
    \quad X\sim \pi.
    \]
\eenum
\vspace{-1em}
Moreover, the same results hold for the inverse IRF $\{f^{-1}_{\theta}: \theta \in \Theta\}$.
\ethm

\vspace{-0.5em}
\section{Invertible measure-preserving IRF from involutive MCMC} \label{sec:constructirf}

\vspace{-0.5em}
In this section, we provide a concrete, general construction of 
invertible and exactly measure-preserving IRFs 
based on involutive MCMC kernels.
The key idea, originally developed for the MH sampler \citep{Neal12,Murray12}, is 
to further augment the space with two additional variables $u_v \in [0, 1]^d, u_a\in [0, 1]$. 
The variable $u_v$ pairs with the auxiliary variable $v$ of dimension $d$, 
and $u_a$ encodes the randomness in the accept/reject decision.
Let the augmented target $\sbpi$ and space $\scS$ be defined as:
\[
    \sbpi(s) = \pi(x)\rho(v \mid x)\1_{[0, 1]^d}(u_v)\1_{[0,1]}(u_a),
    \quad s = (x, v, u_v, u_a) \in \scS 
    := \scX \times \scV \times [0, 1]^d \times [0, 1].
\]
The two uniform auxiliary variables $u_v$ and $u_a$ will be refreshed with two random parameters
$(\theta_v,\theta_a) \sim \mu = \Unif[0,1]^d\times\Unif[0,1]$.
Without loss of generality, we describe the IRF construction assuming a one-dimensional target $\pi(x)$.
The IRF $f_\theta(s) := f_\theta(x, v, u_v, u_a)$ is defined by the following steps (\cref{alg:IFSinvlution}):
\benum
\item Uniform auxiliary refreshment: 
        $u_v \gets \left(u_v + \theta_v \right) \mod 1, 
        \quad
        u_a \gets \left(u_a + \theta_a \right) \mod 1$

\item Update $(v, u_v)$ pair via CDF/inverse-CDF of $\rho(\cdot|x)$
    \footnote{Typically, $\rho(v|x)$ lies in a simple family; for instance, in HMC with a diagonal mass matrix, $\rho$ is a diagonal Gaussian, whose CDF and inverse-CDF can be computed stably.
For multidimensional $v$, a Gibbs-style update on the conditionals of $\rho(\cdot | x)$ can be used. }:
    $
        u_v' \gets F_{\rho(\cdot|x)}(v), 
        \quad 
        \stv \gets F_{\rho(\cdot|x)}^{-1}(u_v)
        $
    
\item Propose and compute the MH-ratio: 
    $
        (x', v') \leftarrow g(x, \stv), 
        \quad 
        r \gets \frac{\pi(x') \rho(v'\mid x')}{\pi(x) \rho(\stv\mid x)}J_g(x, \stv)
    $

\item Accept or reject: 
    If $u_a > r$, reject and stay at the \emph{pre-involution} state $s' = (x, \stv, u_v', u_a)$. 
    Otherwise, set $u_a' \gets \frac{u_a}{r}$ and accept the \emph{post-involution} state $s' = (x', v', u_v', u_a')$.
\eenum

The correspondence with involutive MCMC (\cref{alg:involutivemcmc}) is:
Step 2 simulates $v\sim \rho(\d v|x)$ via inverse CDF sampling, Step 3 mirrors the involution and MH ratio computation, 
and Step 4 performs the accept/reject step while explicitly tracking the randomness $u_a$ involved in the decision. 
Furthermore, as mentioned in \cref{sec:improveddetermixflow}, 
one can use this map in a homogeneous MixFlow by simply fixing $\theta_v, \theta_a$ to some pre-specified constant
values (rather than sampling from $\mu$).
And finally, using the same Jacobian computation as in \citep[Eq. (25)]{Murray12}, 
one can show that $\forall \theta = (\theta_v, \theta_a) \in \Theta$, 
the IRF $f_\theta$ (\cref{alg:IFSinvlution} of \cref{apdx:codeirf}) is $\sbpi$-measure-preserving.
\bprop\label{prop:measurepreserving}
The map given by \cref{alg:IFSinvlution} satisfies \cref{assump:niceirf} for $\sbpi$ if its induced Markov kernel $P$ is irreducible.
\eprop

\vspace{-0.5em}
One must be able to compute $f_\theta^{-1}(\cdot)$ if $f_\theta$ is to be used as a flow layer in a MixFlow. 
Steps 1--3 are straightforward to invert. 
The main challenge lies in inverting the accept/reject Step 4---we 
need to recover the accept/reject decision based solely on the output state $s'$.
Depending on different decisions, $s'$ could either  
be the pre-involution state $(x, \tilde{v}, u_v', u_a)$ or the post-involution state $(x', v', u_v', u_a')$.
Since the transformation $u_a' \gets u_a / r$ only present in the acceptance branch, 
inferring the branch incorrectly would lead to the failure of recovering $u_a$ (hence the entire state $s$).

We address this challenge (pseudocode in \cref{alg:invIFSinvlution} of \cref{apdx:codeirf})
by exploiting the self-inverse property of the involution $g$. 
First note that $g(x, \stv) = (x', v')$ and $g(x', v') = (x, \stv)$.
Suppose that $s' = (x^\#, v^\#, u_v^\#, u_a^\#)$. 
$\{g(x^\#, v^\#), (x^\#, v^\#)\}$ is exactly the unordered pair $\{ (x, \stv),  (x', v')\}$.
Then from the property of the Jacobian of $g$ (i.e., $J_g(x, \stv) = J_g^{-1}(x', v')$ and vice versa), we observe
\[
\frac{\pi(x', v')}{\pi(x,\stv)}J_g(x, \stv)
= 
\left(\frac{\pi(x,\stv)}{\pi(x', v')}J_g(x', v') \right)^{-1}.
\]
Hence, recomputing the MH-ratio as in Step 3 yields
\[
    \str := \frac{\pi(x^\#, v^\#)}{\pi\left(g(x^\#, v^\#)\right)}\cdot J_g\left(g(x^\#, v^\#)\right) \in \{r, r^{-1}\},
\]
where $r$ corresponds to the true MH-ratio as computed in the forward pass.
The key observation to infer the accept/reject decision then follows:
If $u_a^{\#} \cdot \str < 1$, then the acceptance branch was taken, so $u_a = u_a^{\#} \cdot \str$;
otherwise the move was rejected as $u_a$ cannot be larger than $1$. 

\begin{figure*}[t!]
    \centering 
\begin{subfigure}[b]{.24\textwidth} 
    \scalebox{1}{\includegraphics[width=\textwidth, trim=50 250 200 0, clip]{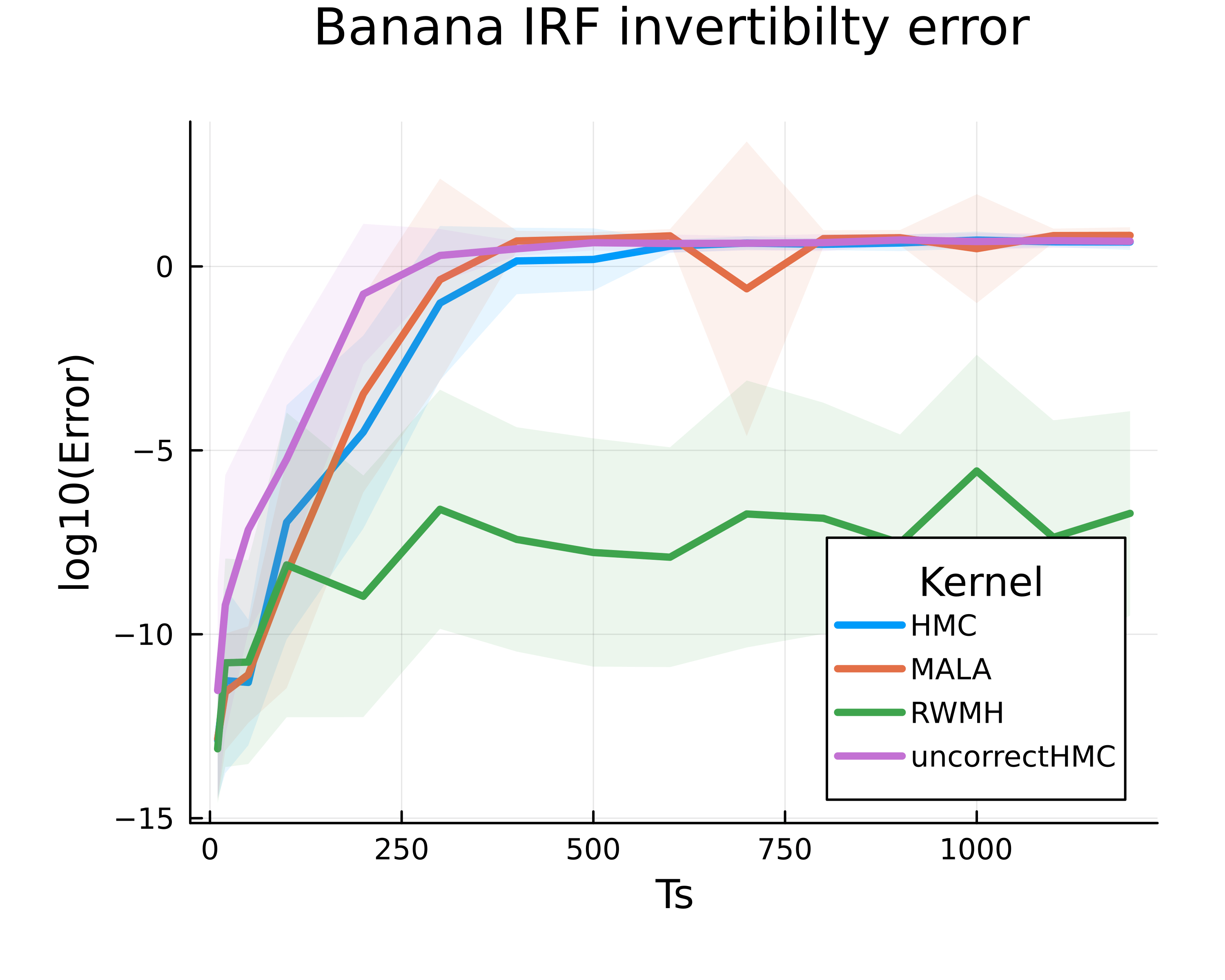}}
\end{subfigure}
\hfill
\centering 
\begin{subfigure}[b]{.24\textwidth} 
    \scalebox{1}{\includegraphics[width=\textwidth, trim=50 250 200 0, clip]{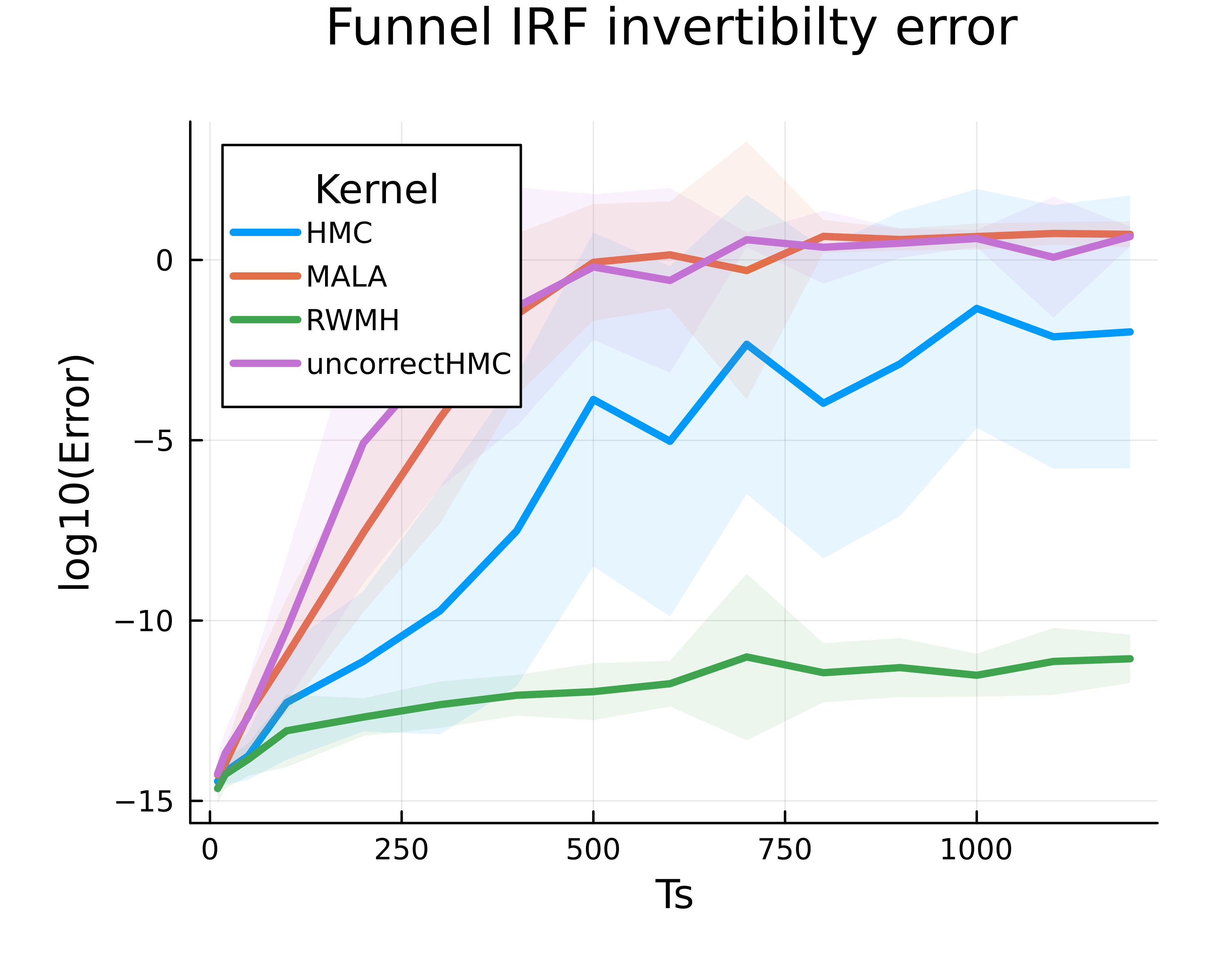}}
\end{subfigure}
\hfill
\begin{subfigure}[b]{.24\textwidth} 
    \scalebox{1}{\includegraphics[width=\textwidth, trim=50 250 200 0, clip]{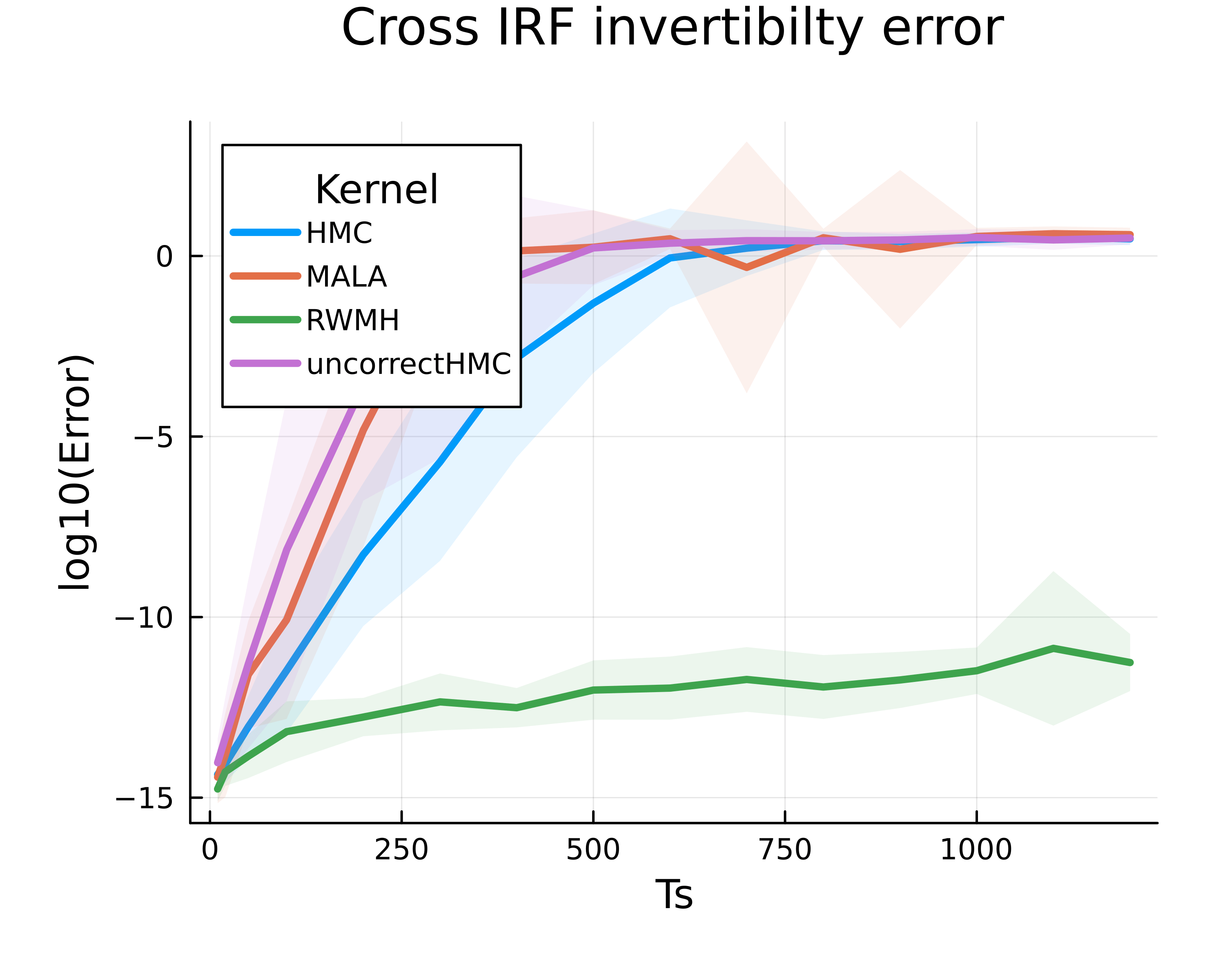}}
\end{subfigure}
\hfill
\centering 
\begin{subfigure}[b]{.24\textwidth} 
    \scalebox{1}{\includegraphics[width=\textwidth, trim=50 250 200 0, clip]{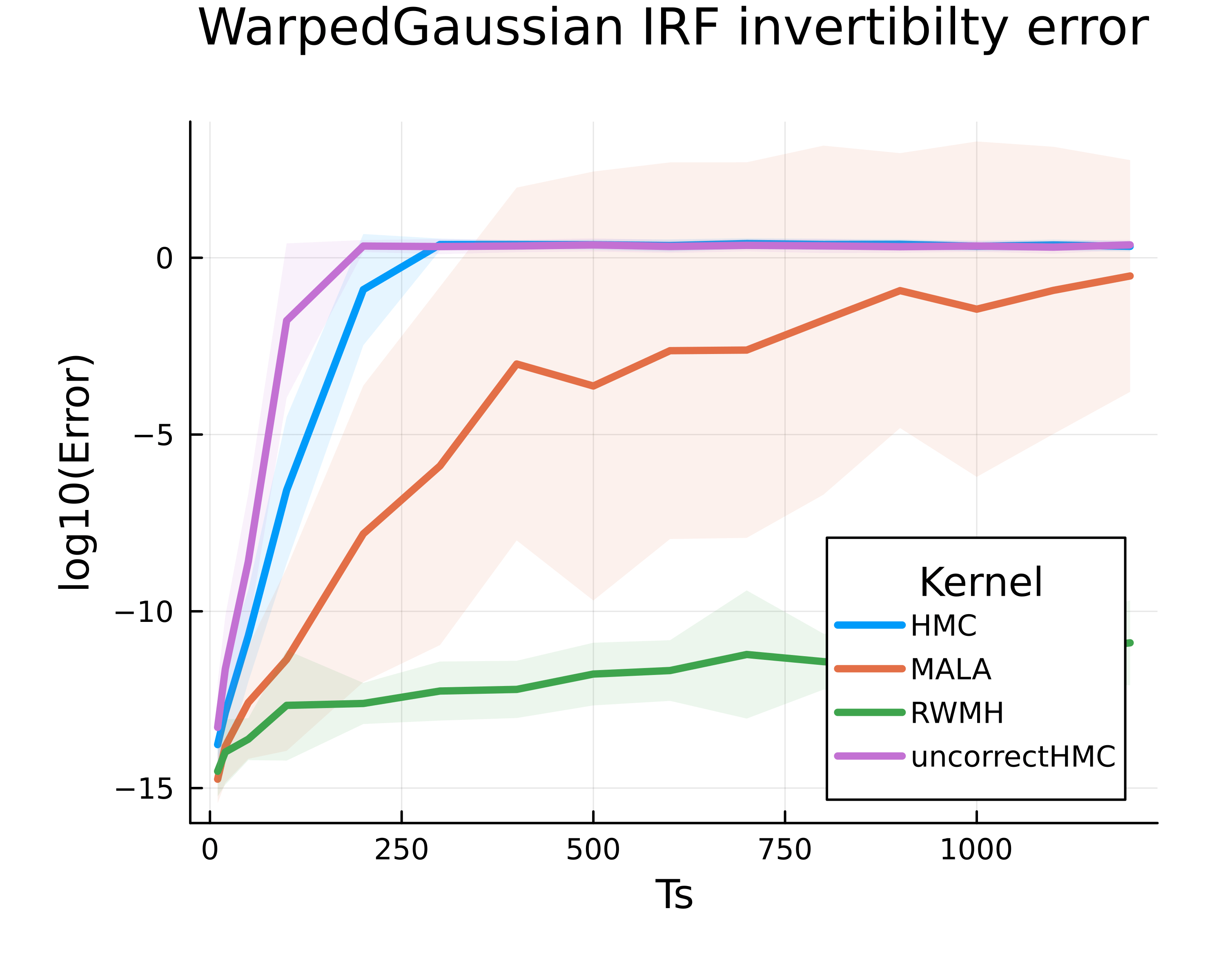}}
\end{subfigure}
\caption{
    Inversion error of IRFs (based on HMC, uncorrected HMC, MALA, and RWMH) over increasing flow length $T$ . 
    Verticle axis shows the 2-norm error of reconstructing $s = (x, v, u_v, u_a)$  ($s = (x, v)$ for the uncorrected HMC IRF)
    sampled from $q_0$ by the composing the forward simulation $f_{\theta_T}\circ \cdots \circ f_{\theta_1}(s)$ and its inverse.
    The lines indicate the mean, and error regions indicate the standard deviation over 32 independent initializations from $q_0$.
}
\label{fig:stability}
\end{figure*}

\cref{fig:stability} empirically verifies that 
one can successfully invert the proposed IRF map
for four MCMC-based IRFs---HMC\citep{Neal11,duane1987hybrid}, uncorrected
HMC \citep{Xu22}, MALA\citep{rossky1978brownian}, and RWMH \citep{livingstone2021geometric}---on four synthetic targets defined in \cref{apdx:expt}.
The same hyperparameters are used for every example: each (uncorrected) HMC
transition consists of 50 leapfrog steps with step size $0.02$; MALA uses step
size $0.25$; RWMH uses step size $0.3$. 
We evaluate the 2-norm error of reconstructing $s = (x, v, u_v, u_a)$ 
sampled from a mean-field Gaussian variational approximation $q_0$ 
by the composing the forward simulation $f_{\theta_T}\circ \cdots \circ f_{\theta_1}(s)$ and its inverse.
Both HMC variants and MALA remain reliably invertible up to $T \approx 200$ iterations, 
while the RWMH IRF remain invertible up to $T\approx 1000$ iterations.
Notably, the corrected HMC IRF is consistently more stable than its uncorrected counterpart used in past MixFlows work;
the additional MH step discards trajectories with large numerical
error that would otherwise cause the dynamics to diverge.
Although \cref{alg:IFSinvlution} and \cref{alg:invIFSinvlution} are exact
inverses in theory, floating-point round-off accumulates with $T$ and 
exact reconstruction can fail \citep{Xu22mixflow,Xu23chaos}. 
In practice, however, the resulting statistical error in downstream variational
inference is often negligible, thanks to the \emph{shadowing} property of
chaotic dynamical systems \citep{Xu23chaos}.

\captionsetup[subfigure]{labelformat=empty}

\vspace{-0.5em}
\section{Variational flows based on IRFs} \label{sec:flow}
\vspace{-0.5em}

In this section, we present a methodology that transforms any IRF system
satisfying \cref{assump:niceirf} into an \emph{asymptotically exact}
variational family. Alongside the exact homogeneous MixFlows derived from IRFs and their refined analysis,
we introduce three additional families---each constructed from the same IRF but
combined differently---and show that all converge to the target in total
variation. Proofs are deferred to \cref{apdx:proof}.
For simplicity, we present the methodology and theory using IRFs defined directly on the
original space $\scX$ rather than the augmented space $\scS$. 
We also assume access to a reference distribution $q_0$ supporting \iid sampling and tractable density evaluation.

\vspace{-0.5em}
\subsection{Improved homogeneous MixFlows} \label{sec:improveddetermixflow}

As reviewed in \cref{sec:deterministicmixflow}, the homogeneous MixFlow $\sbq_T$ is defined as
$\sbq_T = \frac{1}{T}\sum_{t=1}^{T} f^t q_0$ with the convention $\sbq_0 = q_0$.
Given an IRF $f_\theta$ satisfying \cref{assump:niceirf},
one can construct a homogeneous flow map $f$ 
by fixing the parameter $\theta$ to a constant value $\theta^\star$ (e.g., $\pi/16$), 
rather than sampling from the distribution $\mu$.
This provides a generic way of building exact $\pi$-m.p. flow maps. 
A key property not noted in prior MixFlow work
\citep{Xu22mixflow,Diluvi2024MADMix} is a simplified expression for the density
of $\sbq_T$ on arbitrary state spaces, enabled by a measure-theoretic formulation of the 
pushforward density under a measure-preserving map.
Specifically, for any $\pi$-m.p. bijection $f$, $fq_0(x) = \pi(x)\frac{q_0}{\pi}(f^{-1}x)$\footnote{
An implication of this result in continuous state space is that for any $\pi$-m.p. diffeomorphism $f$, 
the Jacobian determinant must satisfy $\abs{\det \nabla f^{-1}(x)}= \frac{\pi(f^{-1}x)}{\pi(x)}$, as established in \cref{prop:mpjacobian}. 
}, 
as introduced in \cref{apdx:changeofvar}.
This yields a simplified form for the density of $\sbq_T$ (in contrast to \cref{eq:deterministicmixflow}):
\[\label{eq:densitymixflow}
    \sbq_T(x) = \frac{1}{T}\sum_{t=1}^{T} f^t q_0(x) 
    = 
    \pi(x)\cdot \frac{1}{T}\sum_{t=1}^{T} \frac{q_0}{\pi}(f^{-t}(x)), 
    \quad \forall x\in \scX.
\]
In practice, this expression is particulary useful for evaluating the flow density; 
practitioners can evaluate $\sbq_T(x)$ without tracking the Jacobians of $f$ explicitly,
which simplifies implementation and avoids numerical instability from accumulating Jacobians over long trajectories.

Moreover, the explicit expression \cref{eq:densitymixflow} offers an intuitive
understanding of why $\sbq_T$ converges. While the original convergence result
in \citep[Theorem 4.2]{Xu22mixflow} relied on general operator theory for
e.m.p. systems \citep{Eisner15}, the density-based perspective is more transparent. 
If $f$ is $\pi$-e.m.p, the Birkhoff ergodic theorem [\citealp{Birkhoff31}; \citealp[p.~212]{Eisner15}] implies that 
$\frac{1}{T}\sum_{t=1}^{T} \frac{q_0}{\pi}(f^{-t}(x)) \to 1$. 
Consequently, for $\pi$-\aev $x \in \scX$, $\sbq_T(x) \to \pi(x)$ as $T\to \infty$.
This enables a substantially simplified proof of the convergence of homogeneous MixFlow.
The proof of \cref{thm:convergenceofmixflow} can be found in \cref{apdx:homomixflowconv}.

\bthm\label{thm:convergenceofmixflow}
Suppose that $f$ is a $\pi$-e.m.p diffeomorphism, and $q_0 \ll \pi$. Then, as $T \to \infty$,
\[
    \sbq_T(x) \to \pi(x), \quad \pi\text{-\aev} \,  x \in \scX, 
    \quad 
    \text{and} 
    \quad
    \TV(\sbq_T, \pi) \to 0.
\]
\ethm
It is worth noting that \cref{assump:niceirf} does not guarantee the ergodicity of a specific $f_{\theta^\star}$, 
leaving a gap between theory and the practical implementation of homogeneous MixFlows. 
In the remainder of this section, we introduce three new MixFlow families designed to address this limitation.

\vspace{-0.5em}
\subsection{IRF MixFlows}
An IRF MixFlow is a mixture of pushforwards of a reference $q_0$ through an IRF sequence:
\vspace{-0.5em}
\[
    \overrightarrow{q_T} := \frac{1}{T}\sum_{t=1}^{T}f_{\theta_t}\circ \cdots\circ f_{\theta_1} q_0,
    \quad
    \text{with the convention that } \overrightarrow{q_0} = q_0,
\]
\vspace{-0.2em}
where $\theta_1, \dots, \theta_T$ is a \emph{cached} \iid sequence drawn from $\mu$.
When constructing the flow, we first sample and freeze the random stream
$\theta_1, \dots, \theta_T$, yielding an \emph{inhomogeneous} sequence of $T$
parameterized bijections. 
Then to draw $X \sim \overrightarrow{q_T}$, we treat $\overrightarrow{q_T}$ as a mixture of $T$ distributions:
\[
    K & \sim \Unif\{1, 2, \dots, T\} 
    & 
    X_0 &\sim q_0 
    & 
    X & = f_{\theta_K}\circ \cdots\circ f_{\theta_1} (X_0)
\]
Note crucially that each sample $X$ is generated using the \emph{same frozen} sequence $\theta_1,\dots,\theta_T$.
For density evaluation, we compute the inverse IRF $f^{-1}_{\theta_T}, \cdots, f^{-1}_{\theta_1}$. 
Because each $f_\theta$ is $\pi$-m.p., by \cref{prop:mpjacobian},
the density takes a similar form as in a homogeneous MixFlow
(\cref{eq:densitymixflow}):
\[ \label{eq:densityirfmixflow}
    \overrightarrow{q_T}(x) 
    = \pi(x)\cdot \frac{1}{T}\sum_{t=1}^{T} \frac{q_0}{\pi}\left(f_{\theta_1}^{-1} \circ \cdots \circ f_{\theta_t}^{-1}(x)\right), 
    \quad \forall x\in \scX.
\]
However, note that this density requires simulating the \emph{backward process} of the inverse IRF (\citep{diaconis1999iterated})
\[
    \overleftarrow{X_t}(x) := f_{\theta_1}^{-1} \circ \cdots \circ f_{\theta_t}^{-1}(x) \quad \text{for } t \in [T],
\]
which cannot be computed sequentially. As a result, IRF MixFlows incur a
quadratic density evaluation cost $O(T^2)$. Fortunately, this backward
process can be computed in a parallel fashion, as the computation 
of each $\overleftarrow{X_t}(x)$,  $t\in [T]$ is independent. 
We recommend deploying IRF MixFlows on modern parallel hardware (e.g., GPUs) for efficient density evaluation.

IRF MixFlows share the total variation convergence guarantee (\cref{thm:tvconvergence}) of
homogeneous MixFlows. 
The key of the proof (\cref{apdx:irfflowconv}) is to show
$\int d_{\mathrm{TV}}\left(\frac{1}{m} \sum_{k=1}^m P^k(x, \cdot), \pi\right) \pi(\mathrm{d} x) \to 0$ as $m \to \infty$.
Note that the requirement that $P$ is $\pi$-irreducible is 
significantly weaker than the ergodicity assumption of \cref{thm:convergenceofmixflow}.
See \cref{sec:weakerassumption} for a detailed discussion.

\bthm \label{thm:tvconvergence}
Let $\mathbb{P}$ denote the joint distribution over the \iid sequence $(\theta_t)_{t\in\nats} \distiid \mu$.
If \cref{assump:niceirf} holds and $q_0 \ll \pi$, then
\[\label{eq:condconv}
    \TV\left(  \overrightarrow{q_T}, \pi\right) \stackrel{\mathbb{P}}{\longrightarrow} 0 
\quad
\text{ as } \,T \to \infty.
\]
\ethm

\vspace{-0.5em}
\subsection{Backward IRF MixFlows}

To address the $O(T^2)$ density cost of IRF MixFlows, we propose a simple modification: 
constructing the flow from the \emph{backward process}. 
Specifically, we define the \emph{backward IRF MixFlow} as:
\[ \label{eq:bwdirfmixflow}
    \overleftarrow{q_T} := \frac{1}{T}\sum_{t=1}^{T}f_{\theta_1}\circ \cdots\circ f_{\theta_t} q_0, 
    \quad
    \text{with the same convention that } \overleftarrow{q_0} = q_0.
\]
This construction retains $O(T)$ complexity of sampling $X \sim \overleftarrow{q_T}$ via:
\[
    K & \sim \Unif\{1, 2, \dots, T\} 
    & 
    X_0 &\sim q_0 
    & 
    X & = f_{\theta_1}\circ \cdots\circ f_{\theta_K} (X_0),
\]
while reducing the density computation cost to $O(T)$. The density of $\overleftarrow{q_T}$ is given by:
\[\label{eq:densitybwdirf}
    \overleftarrow{q_T}(x) 
    = \pi(x)\cdot \frac{1}{T}\sum_{t=1}^{T} \frac{q_0}{\pi}\left(f_{\theta_t}^{-1} \circ \cdots \circ f_{\theta_1}^{-1}(x)\right), 
    \quad 
    \forall x\in \scX.
\]
This mirrors the density formula of homogeneous MixFlows
(\cref{eq:densitymixflow}), enabling the use of the random ergodic theorem
(\cref{thm:randomergodic}) to establish the same pointwise and total variation convergence.

\bthm \label{thm:bwdirfconvergence}
If \cref{assump:niceirf} holds and $q_0 \ll \pi$, 
then for $\pi$-\aev $x\in \scX$ and $\mu$-almost all $(\theta_t)_{t\in \nats}$:
\[
    \overleftarrow{q_T}(x) \longrightarrow \pi(x)
    \quad
    \text{and}
    \quad
    \TV(\overleftarrow{q_T}, \pi) \longrightarrow 0
    \quad
    \text{as } T\to \infty.
\]
\ethm

\subsection{Ensemble IRF MixFlows}

All MixFlow variants discussed above---including homogeneous MixFlows---are
based on \emph{ergodic averaging} along the flow. This inherently limits their convergence rate
to $O(1/T)$, as the first component always retains a $1/T$ mixing
weight. In contrast, MCMC methods often exhibit geometric convergence in their
marginal distributions under suitable conditions [\citealp{Roberts04};
\citealp[Ch.~15]{douc2018markov}]. 
Motivated by this, we propose the \emph{ensemble IRF MixFlows}, which instead uses an \emph{ensemble average}
of the endpoint of multiple IRF trajectories in an attempt to match $T$-step MCMC marginal distribution:
\[
    \stq^{(M)}_T := \frac{1}{M} \sum_{m=1}^M q_T^{(m)} = \frac{1}{M} \sum_{m=1}^M f_{\theta_T^{(m)}} \circ \cdots \circ f_{\theta_1^{(m)}} q_0,
\]
where each ${\theta_1^{(m)}, \dots, \theta_T^{(m)}}$ corresponds to an independent IRF realization.
As in the case of the previous MixFlows, the $M$ streams of randomness $\theta_t^{(m)}$ are cached (i.e., \emph{frozen}) when
sampling and computing densities.
The resulting density of the ensemble IRF MixFlow is given by: 
\[\label{eq:densityensemble}
    \stq^{(M)}_T(x) 
    = \pi(x) \cdot \frac{1}{M} \sum_{m=1}^M \frac{q_0}{\pi}\left(f_{\theta_1^{(m)}}^{-1} \circ \cdots \circ f_{\theta_T^{(m)}}^{-1}(x)\right),
\]
whose computation costs $O(TM)$ (or $O(T)$ when parallelized across the $M$ streams). Drawing $X \sim \stq_T^{(M)}$ takes $O(T+M)$ operations: 
\[
    K & \sim \Unif\{1, 2, \dots, M\} 
    & 
    X_0 &\sim q_0 
    & 
    X & = f_{\theta_T^{(K)}} \circ \cdots \circ f_{\theta_1^{(K)}} q_0.
\]
Intuitively, the flow length $T$ controls the bias of the IRF system, while the
ensemble size $M$ controls the variance of the Monte Carlo average. 
This tradeoff is formalized in the following result.

\bthm \label{thm:tvensemble}
Suppose that \cref{assump:niceirf} holds, and that  $\forall x \in \scX$, $\frac{q_0}{\pi}(x) \leq B < \infty$.
Then, 
\[
    \E_\theta\left[\TV\left(\stq^{(M)}_T, \pi\right)\right] 
    \leq
    &
    \frac{1}{\sqrt{M}} 
    \E\left[
        \sqrt{ 
            \Var_{\theta_{1:T}}
           \left[
               \frac{q_0}{\pi}\left(f_{\theta_1}^{-1} \circ \cdots \circ f_{\theta_T}^{-1}(X)\right) \mid X
           \right]
        }
    \right] \\
    &
    + 
    B\cdot \E\left[\TV(R^T\delta_X, \pi)\right], \quad X \sim \pi,
\]
where $R$ is the Markov kernel induced by the inverse IRF $f^{-1}_\theta$, and $\delta_X$ is the Dirac measure at $X$.
\ethm
In the setting where $\TV(R^T\delta_X, \pi) = O(\rho^T)$ for some $\rho \in (0,1)$, 
the convergence rate of $\stq_T^{(M)}$ can be heuristically characterized as
$\TV\left(\stq^{(M)}_T, \pi\right) = O\left(\rho^T \vee \frac{1}{\sqrt{M}}\right)$,
capturing the tradeoff between the bias (via $T$) and variance (via $M$).
Given a fixed computational budget, choosing the balance between
flow length $T$ and ensemble size $M$ is critical. 
In the extreme case of $M = 1$, convergence will fail entirely---any $\pi$-measure-preserving $f$ satisfies
$\TV(f q, \pi) = \TV(q, \pi)$ \citep[Theorem 1]{qiao2010study}. 
On the other hand, small $T$ leads to high bias due to insufficient mixing. 
This tradeoff closely relates to recent studies on parallel MCMC algorithms
\citep{margossian2024nested, sountsov2024running}.

\vspace{-0.5em}
\subsection{Discussion} \label{sec:weakerassumption}

\paragraph{Relaxing ergodicity.}
A major advantage of IRF-based MixFlows over homogeneous MixFlows is that IRF-based MixFlows 
require only that the kernel $P$ is $\pi$-irreducible (\cref{assump:niceirf}),
a significantly weaker condition than the ergodicity assumed by homogeneous MixFlows.
In fact, whenever the set $\Theta^\star := \{\theta: f_{\theta} \text{ is $\pi$-ergodic}\}$ has positive $\mu$-measure, 
\cref{assump:niceirf} automatically holds \citep[Corollary 3.3]{morita1988deterministic}.
The IRFs we construct in \cref{sec:constructirf} correspond to involutive MCMC
kernels that are known to be irreducible, whereas establishing
ergodicity in MixFlows is typically so difficult that it is assumed without proof \citep{Tupper05,ver2021hamiltonian,Xu22mixflow}.

\vspace{-0.5em}
\paragraph{Which flow to choose?}
All four flows are asymptotically exact, yet their density formulae reveal
different bias-variance and cost-accuracy trade-offs. In every case the density
ratio takes the form $\frac{\text { flow density }}{\pi}(x)=\frac{1}{N} \sum_{n=1}^N \frac{q_0}{\pi}\left(T_n(x)\right)$, 
where $T_n$ is a composition of inverse IRF/ergodic maps, and $N$ can be the flow length or ensemble size. 
Hence practical convergence of each flow is dictated by how quickly $\frac{1}{N}
\sum_{n=1}^N \frac{q_0}{\pi}\left(T_n(x)\right)$ converges to a constant.
Empirically (see \cref{apdx:comparemixflows}) we find that IRF MixFlows often reach a given accuracy at shorter flow
lengths than homogeneous or backward IRF MixFlows,
but a full theoretical comparison study is deferred to future work.

\vspace{-0.5em}
\section{Experiments} \label{sec:expt}
\vspace{-0.5em}
This section presents an empirical evaluation of the four proposed flows---three IRF variants and homogeneous MixFlows 
(collectively referred to as ``IRF flows'' since homogeneous MixFlows can be viewed as a special case).
We compare them against two normalizing flows, RealNVP \citep{Dinh17} and Neural Spline Flow (NSF) \citep{Durkan19}, 
and against the No-U-Turn Sampler (NUTS) \citep{Hoffman14}.
Variational methods are assessed by their 
(i) ELBO and 
(ii) accuracy of the importance sampling estimate of the normalization constant $\log Z$ for the
unnormalized density $\gamma$:
\[
Z \approx \frac{1}{N} \sum_{n=1}^N \frac{\gamma}{q_T}\left(X_n\right), 
\quad
\left(X_n\right)_{n=1}^N \stackrel{\text { iid }}{\sim} q_T, 
\quad \text { where } 
q_T \in\left\{\bar{q}_T, \overrightarrow{q}_T, \overleftarrow{q}_T, \widetilde{q}_T^{(M)}\right\},
\pi = \frac{\gamma}{Z}
\]
and (iii) importance sampling effective sample size (ESS)
\citep{kong1992note,kong1994sequential,Liu96}. 
Sampling methods are evaluated via their Monte Carlo estimation error.
In all cases, all flows start
from the same reference distribution $q_0$: a mean-field Gaussian trained for
$10$K Adam steps with batch size $10$ and learning rate $10^{-3}$.  All IRF
flows are evaluated with $64$ \iid draws, while normalizing flows use $1024$.
Full experimental details appear in \cref{apdx:expt}.

\vspace{-0.5em}
\subsection{Synthetic examples}
\begin{figure}[t!]
    \begin{subfigure}{0.24\columnwidth}
    \includegraphics[width=\columnwidth, trim=250 250 250 250, clip]{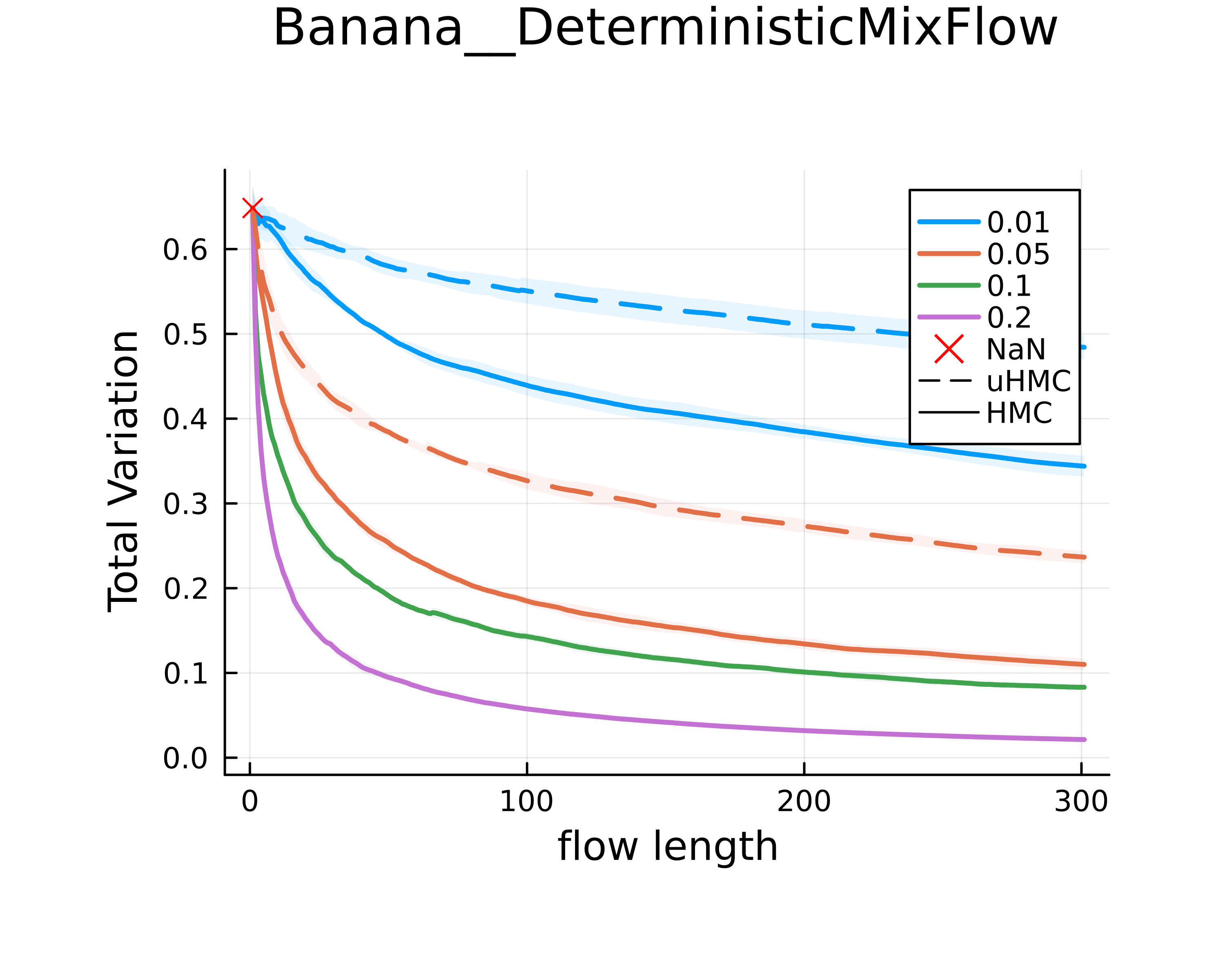}
		\caption{(a) Banana }
	\end{subfigure}
    \begin{subfigure}{0.24\columnwidth}
    \includegraphics[width=\columnwidth, trim=250 250 250 250, clip]{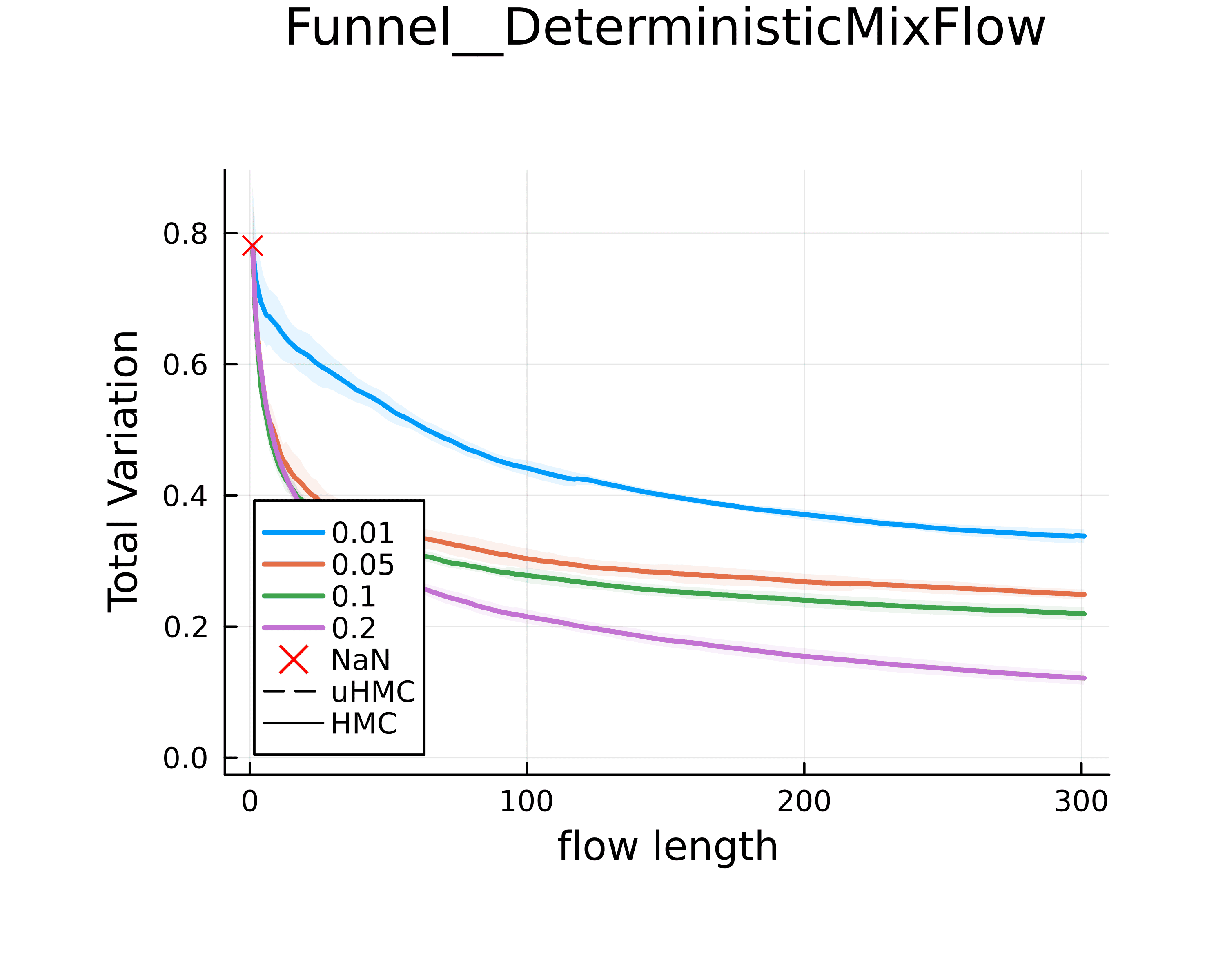}
		\caption{(b) Funnel }
	\end{subfigure}
    \begin{subfigure}{0.24\columnwidth}
    \includegraphics[width=\columnwidth, trim=250 250 250 250, clip]{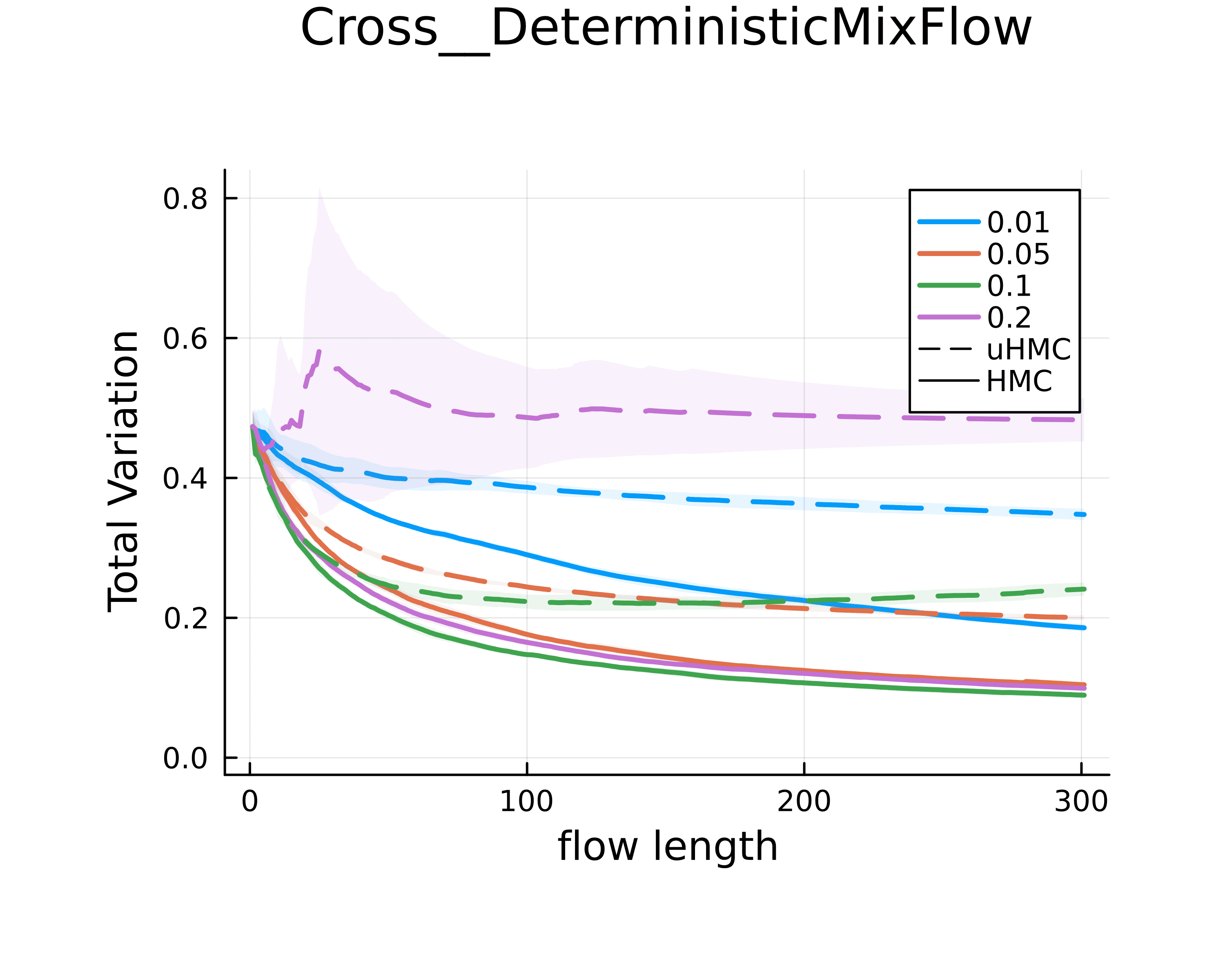}
		\caption{(c) Cross }
	\end{subfigure}
    \begin{subfigure}{0.24\columnwidth}
    \includegraphics[width=\columnwidth, trim=250 250 250 250, clip]{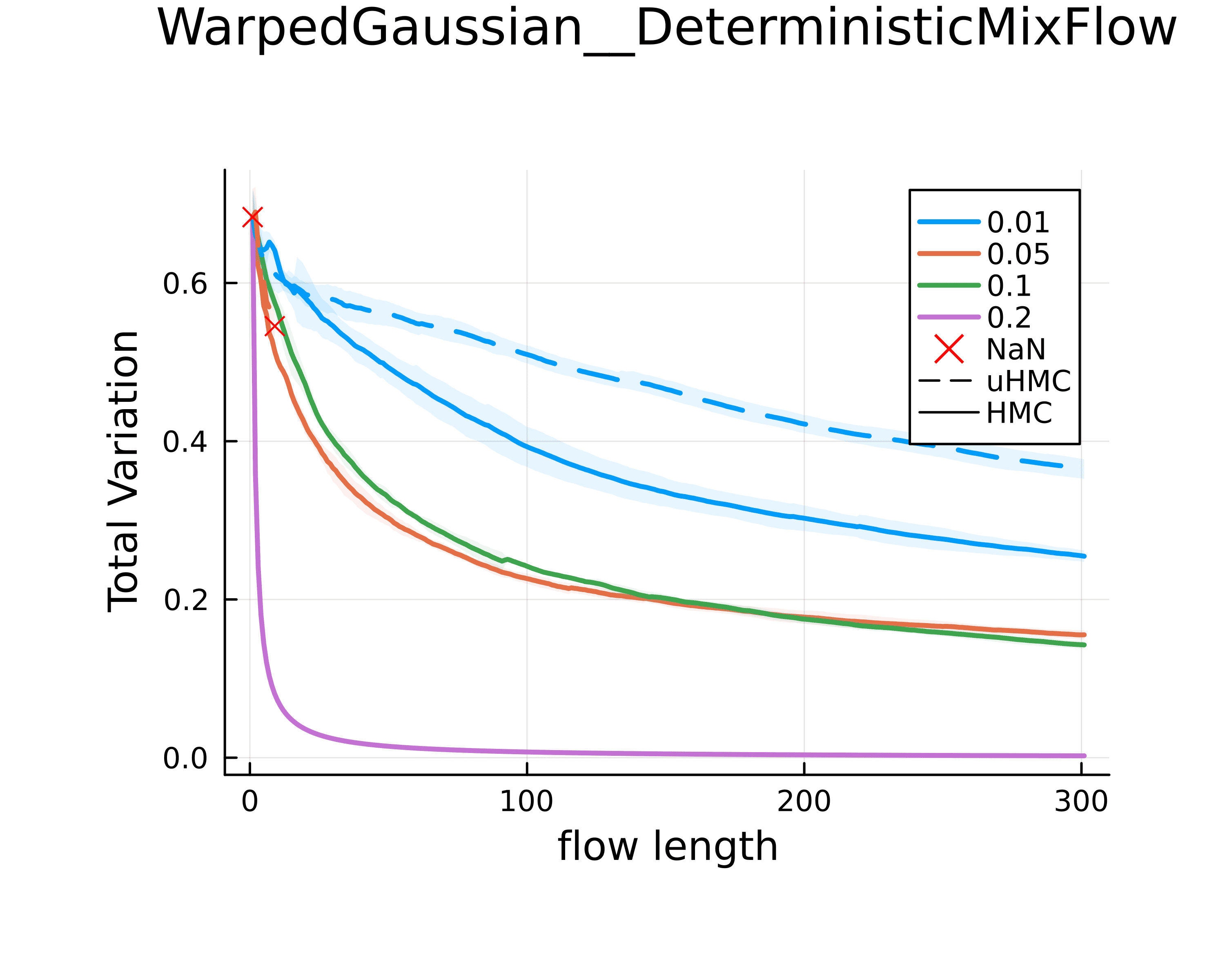}
		\caption{(d) Warped Gaussian }
	\end{subfigure}
    \caption{
        Total-variation error for homogeneous MixFlow built on
        \emph{corrected} (solid) versus \emph{uncorrected} (dashed) HMC
        kernels, plotted against flow length $T$ for several step sizes. 
        Each curve is the mean over 32 independent runs; shaded bands ($\pm 1$ SD) show run-to-run variability. 
        A cross marks any setting where at least one run returned a \texttt{NaN} (instability), at which point the trace is terminated.
}\label{fig:uhmc_hmc}
\end{figure}

Our synthetic experiments consist of four 2-dimensional targets used by \citet{Xu22mixflow}:
the Banana \citep{haario2001adaptive},
Neal's funnel \citep{neal2003slice}, a cross-shaped Gaussian mixture, and a warped Gaussian distribution.
\cref{fig:uhmc_hmc} shows
a comparison of the original Hamiltonian-MixFlow---built on an \emph{uncorrected} HMC kernel---with our \emph{corrected} version
including the MH step. For each target we run both
flows with identical hyper-parameters (50 leapfrog steps per transition,
several step-sizes) and estimate the total-variation (TV) distance to the
ground truth using 512 \iid samples.
Across all targets and step-sizes, the corrected HMC-based MixFlow consistently
achieves lower TV error and remains robust as the step-size grows. In contrast,
the uncorrected variant often deteriorates with longer flows because the inexact map
error accumulates (e.g., the green dashed curve in the third panel). At larger
step sizes the uncorrected flow frequently diverges, producing NaNs (marked by
crosses), whereas the corrected flow remains stable---echoing the inversion stability results in \cref{fig:stability}.

We next compare the four IRF flows with \texttt{RealNVP} and \texttt{NSF}.
Two IRF variants are examined: HMC-based (50 leapfrog steps per transition; $T = 200$) and 
RWMH-based ($T = 4000$). Each normalizing flow consists $6$ flow layers, and is trained via $50{,}000$ Adam steps with batch size $32$; 
we tune the learning rates in the grid $\{10^{-4}, 10^{-3}, 10^{-2}\}$, 
and report the results of the setting with smallest median TV distance over $5$ runs.
Additional implementation details can be found in \cref{apdx:synthetic_expt}.

\begin{figure}[t!]
    \begin{subfigure}{0.33\columnwidth}
        \includegraphics[width=\columnwidth, trim=0 0 50 0, clip]{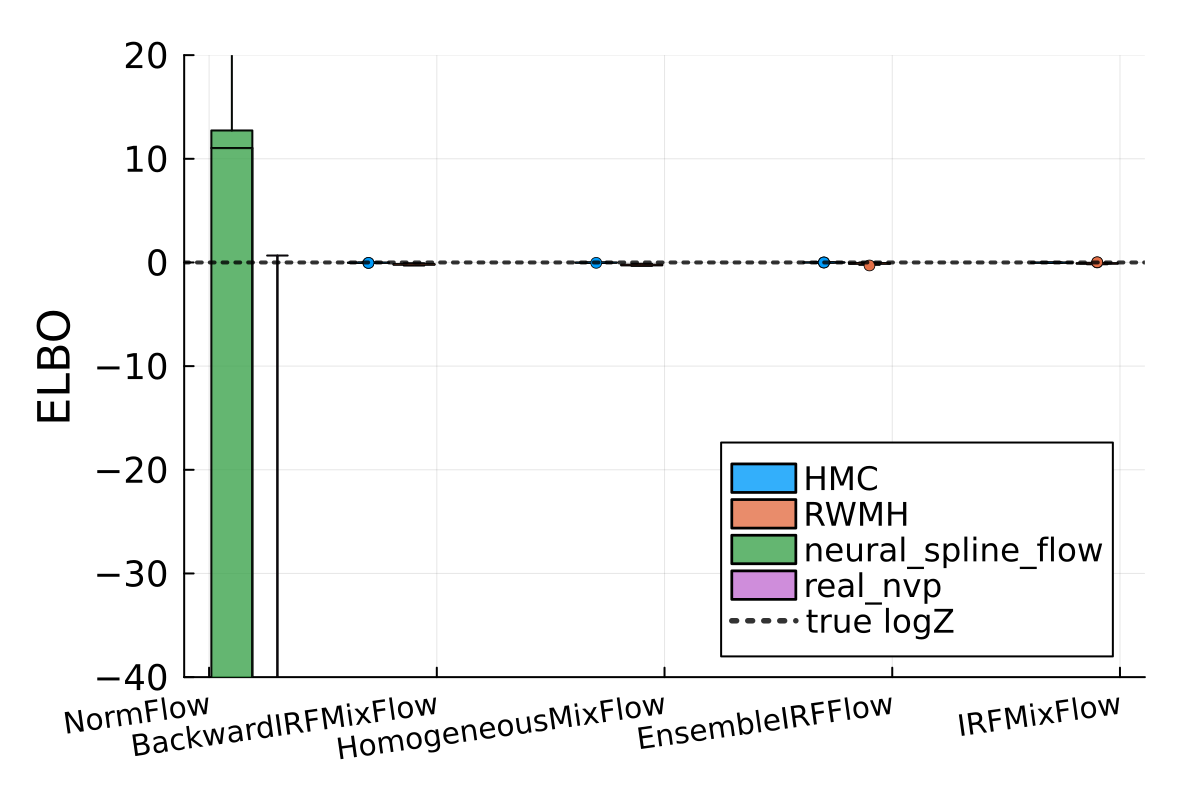}
		\caption{(a) ELBO }\label{fig:banana_elbo}
	\end{subfigure}
    \begin{subfigure}{0.33\columnwidth}
        \includegraphics[width=\columnwidth, trim=0 0 50 0, clip]{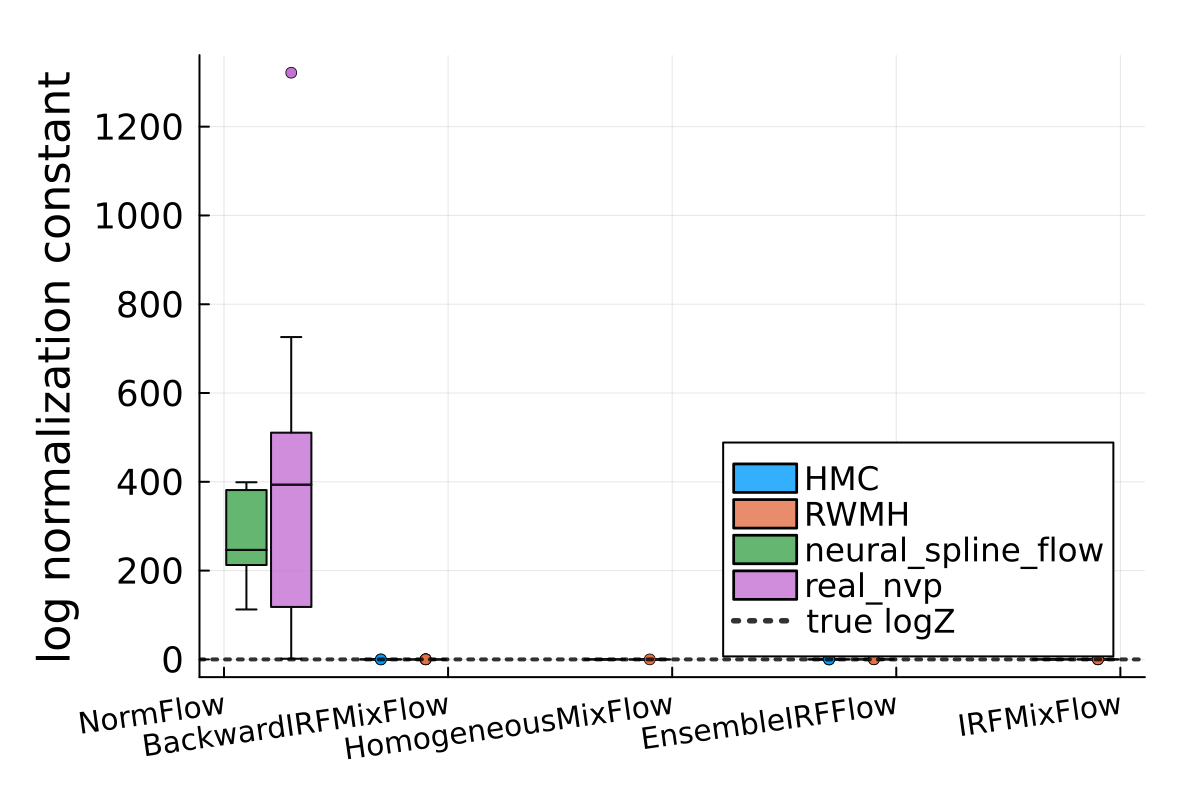}
		\caption{(b) $\log Z$ estimates.}\label{fig:banana_logz}
	\end{subfigure}
    \begin{subfigure}{0.33\columnwidth}
        \includegraphics[width=\columnwidth, trim=0 00 50 0, clip]{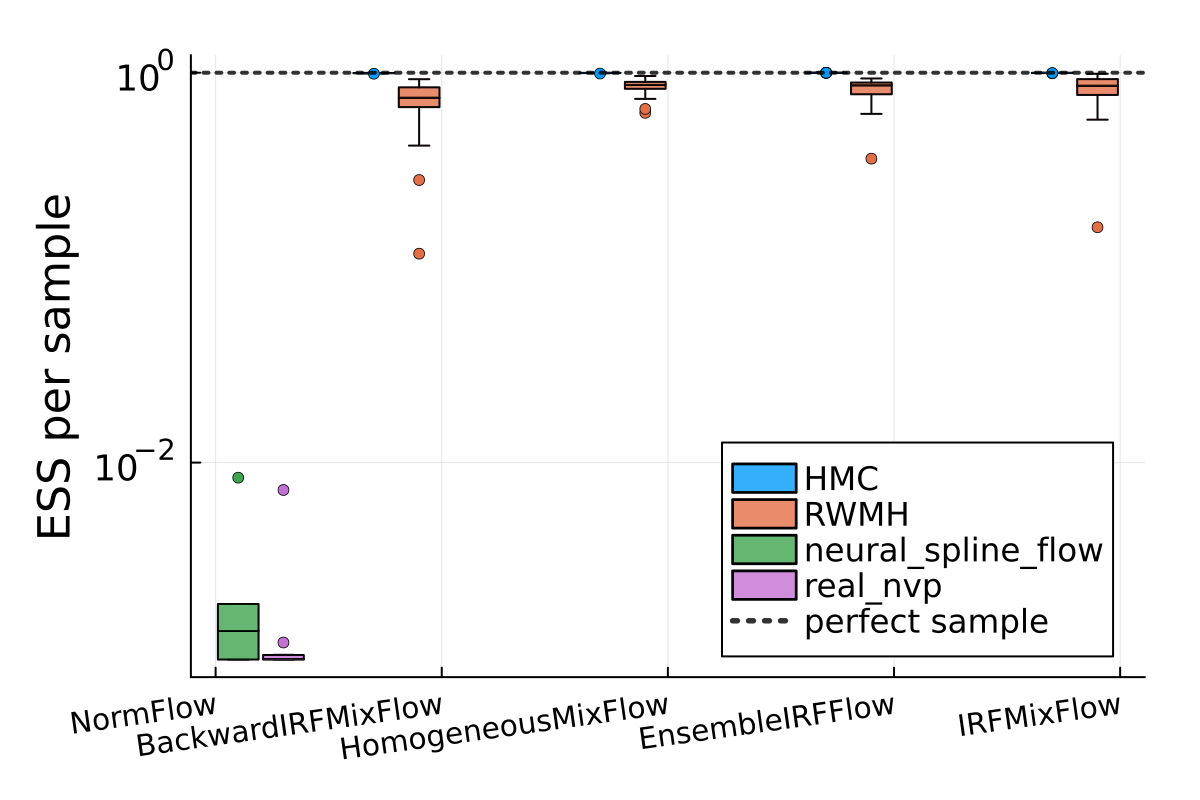}
		\caption{(c) Per-sample ESS.}\label{fig:banana_ess}
	\end{subfigure}
    \caption{
        Variational approximation quality of IRF Flows versus \texttt{RealNVP} and \texttt{NSF}. 
        Box plots for IRF flows are based on $32$ independent runs, and $10$ runs for the normalizing flows.
        The black dashed line in (c) indicates the optimal ESS of perfect \iid samples.
    } \label{fig:banana}
\end{figure}

\cref{fig:banana_elbo,fig:banana_logz} display the ELBO and $\log Z$ estimates (via importance sampling) 
for the Banana target; the remaining synthetic cases show the same pattern (\cref{fig:syn_no_banana} in \cref{apdx:additional_synthetic_results}).
As synthetic targets are normalized, a perfect variational approximation has both metrics near $0$.
The IRF flows meet this mark consistently across runs, 
whereas \texttt{RealNVP} and \texttt{NSF} exhibit high variability and 
often produce extreme ELBO or $\log Z$ values.
We restrict the vertical range of the ELBO plot for better visualization; full-range plots are in \cref{fig:syn_elbo_raw}.
We also note that training instability is common for the normalizing flows: on the Funnel example, 10 of 15 \texttt{RealNVP} runs and all \texttt{NSF} runs diverged.

\cref{fig:banana_ess} further examine the per-sample importance sampling ESS
(see \cref{fig:syn_ess_no_banana} on similar results for other examples), 
which reflects the $\chi^2$ divergence from the variational distribution to the target \citep{agapiou2017importance}.
The ESS is orders of magnitude higher for IRF flows than for the normalizing flows.
Additionally, we provide comparisons among the three ergodic averaging MixFlow variants in
\cref{apdx:comparemixflows}, and ensemble-size/length trade-offs for
ensemble IRF MixFlows are explored in \cref{apdx:ensembles}.

\vspace{-0.7em}
\subsection{Real-data experiments}

The real-data experiments include the Student-t-regression (\texttt{TReg};
4-dimensional), and the Sparse linear regression (\texttt{SparseReg}; 83-dimensional)
from \citep{Xu22mixflow}, and a latent Brownian motion model
(\texttt{Brownian}; 32-dimensional) and the Log-Gaussian Cox process model
(\texttt{LGCP}; 1600-dimensional) from the Inference Gym library \citep{inferencegym2020}.
Each normalizing flow is trained via $50{,}000$ Adam steps of batch size $32$; 
we grid-search both the learning rates $\{10^{-4}, 10^{-3}, 10^{-2}\}$ and flow layers $\{6, 10\}$, 
and report the configuration with the highest median ELBO over $5$ runs.
An additional mean-field Gaussian baseline is optimized for the same number of steps and batch size with learning rate $10^{-3}$.

All IRF variants use RWMH kernel, with the step size tuned
to achieve a $0.8$ acceptance rate using bisection search between $0.001$ and $10$.
In each search step, we estimate acceptance rate with $5{,}000$ RWMH-IRF iterations. 
We set $T = 5000$ for the backward IRF and homogeneous MixFlow and ensemble IRF MixFlow, and set $T = 4000$ for the IRF MixFlow.
Normalizing flow results are omitted for \texttt{LGCP}, which did not finish training within 48 hours on the same computation cluster.
Ground truth values are estimated using AIS with a dense temperature grid; see the details in \cref{apdx:real_expt}.

\begin{figure}[t!]
    \begin{subfigure}{\columnwidth}
        \includegraphics[width=0.24\columnwidth, trim=0 100 100 0, clip]{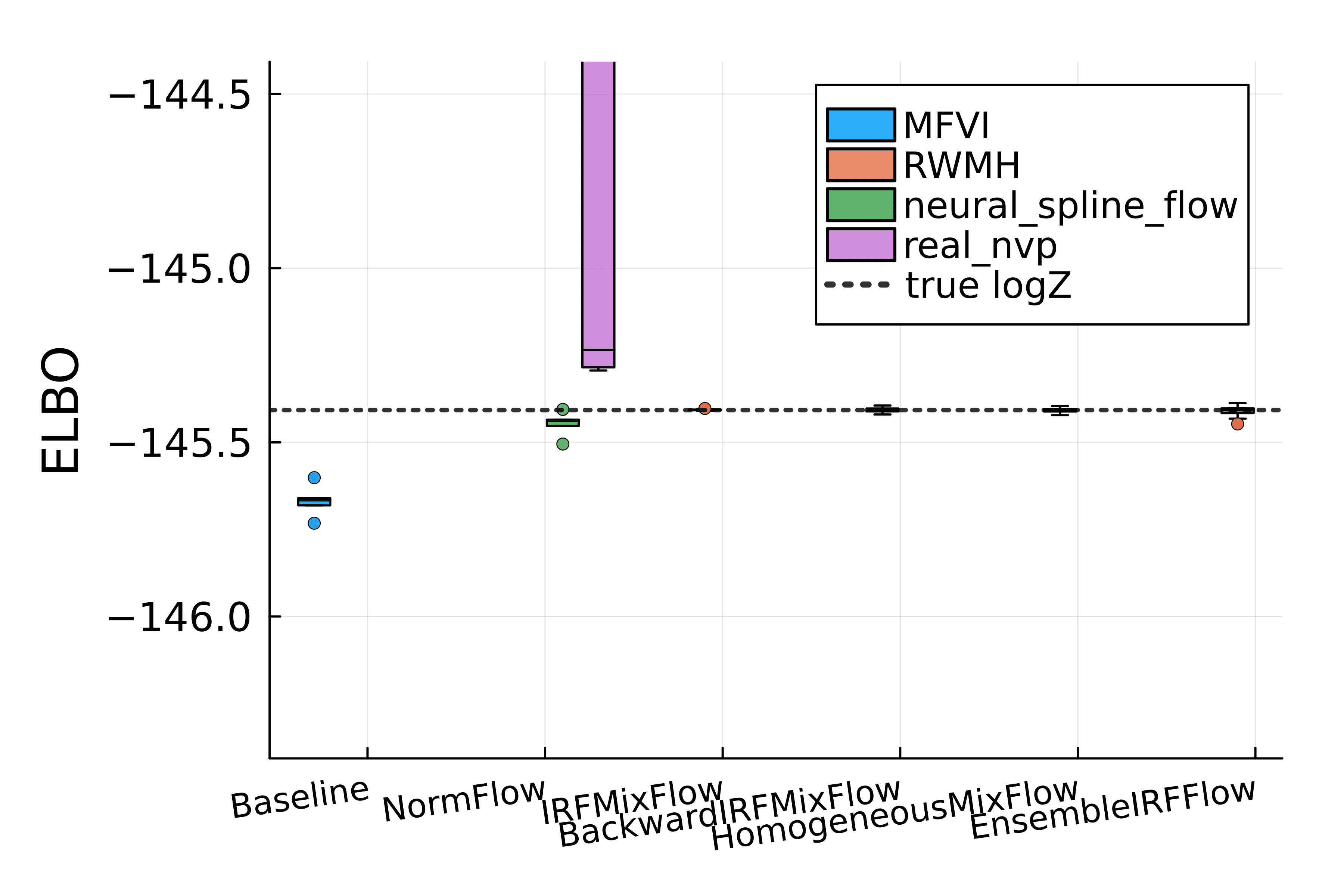}
        \includegraphics[width=0.24\columnwidth, trim=0 100 100 0, clip]{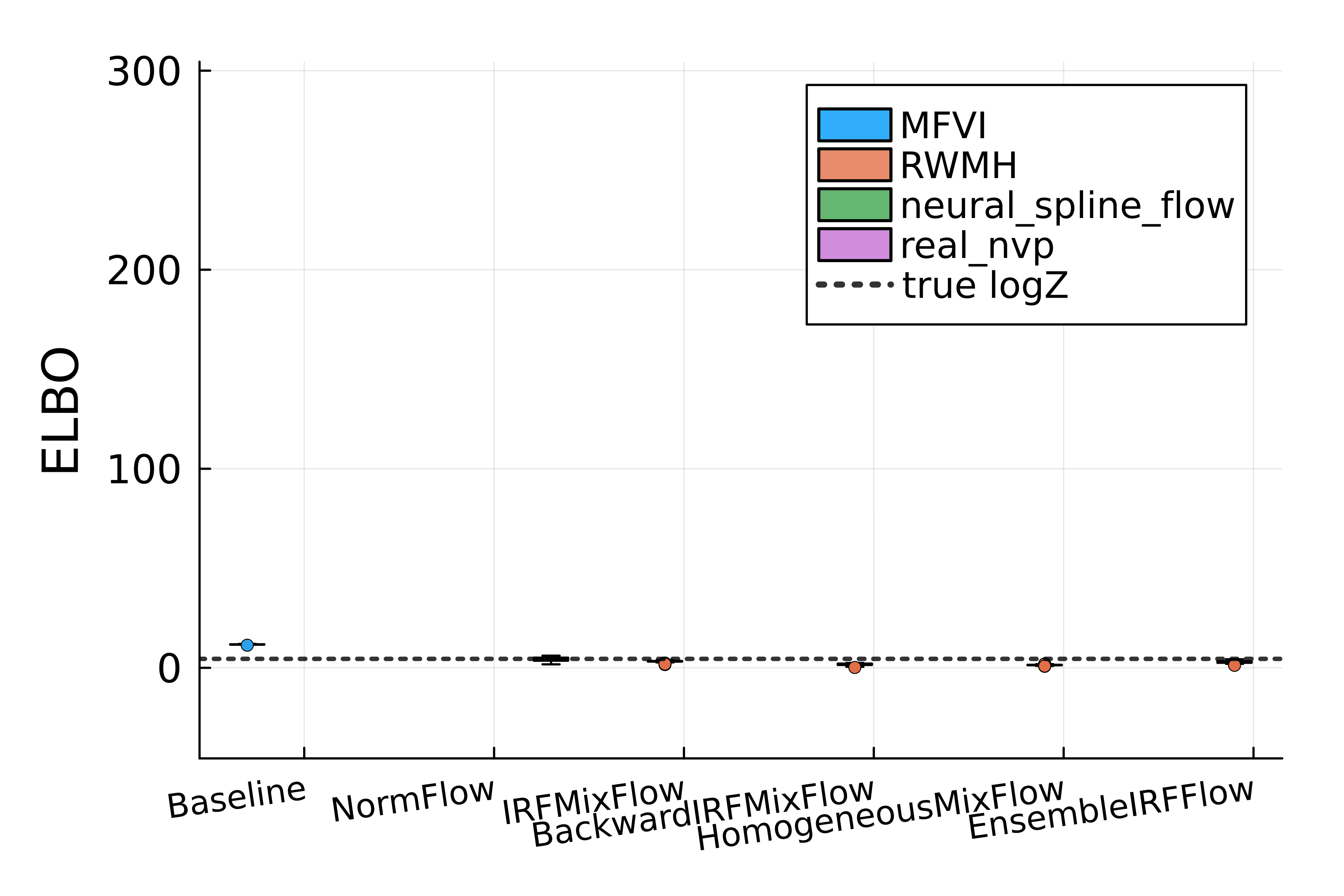}
        \includegraphics[width=0.24\columnwidth, trim=0 100 100 0, clip]{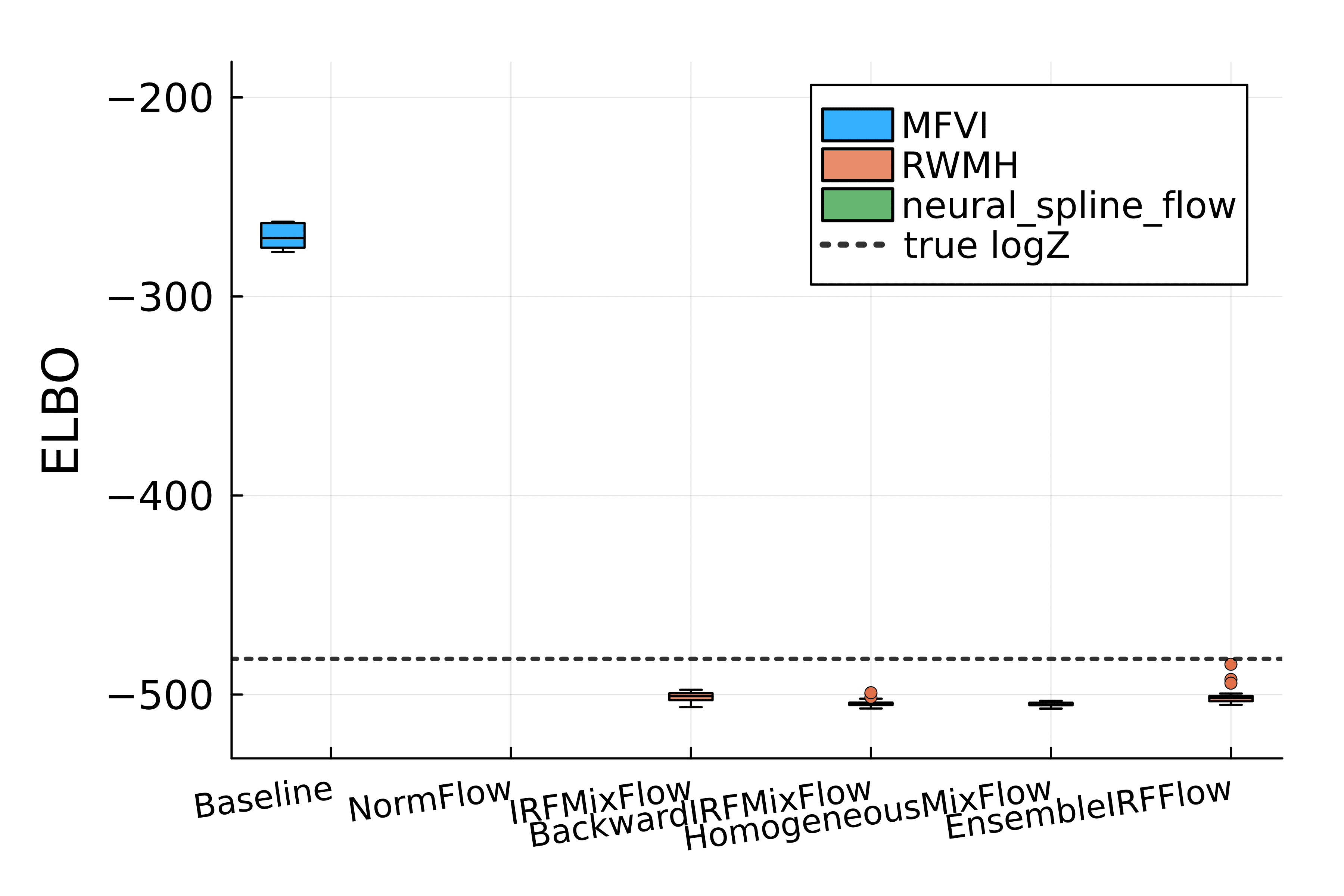}
        \includegraphics[width=0.24\columnwidth, trim=0 100 100 0, clip]{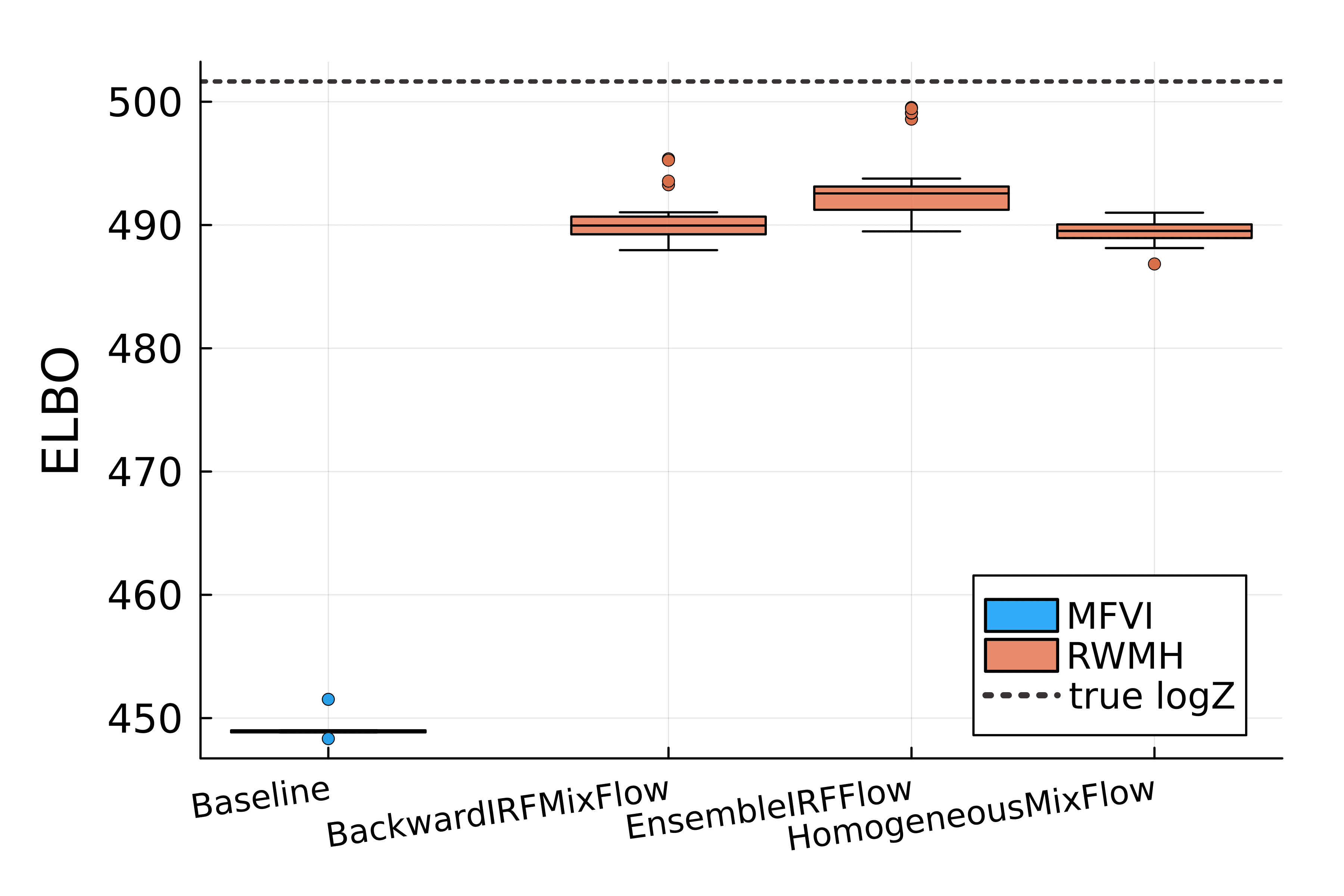}
		\caption{(a) ELBO}\label{fig:real_elbo}
	\end{subfigure}
    \begin{subfigure}{\columnwidth}
        \includegraphics[width=0.24\columnwidth, trim=0 100 100 0, clip]{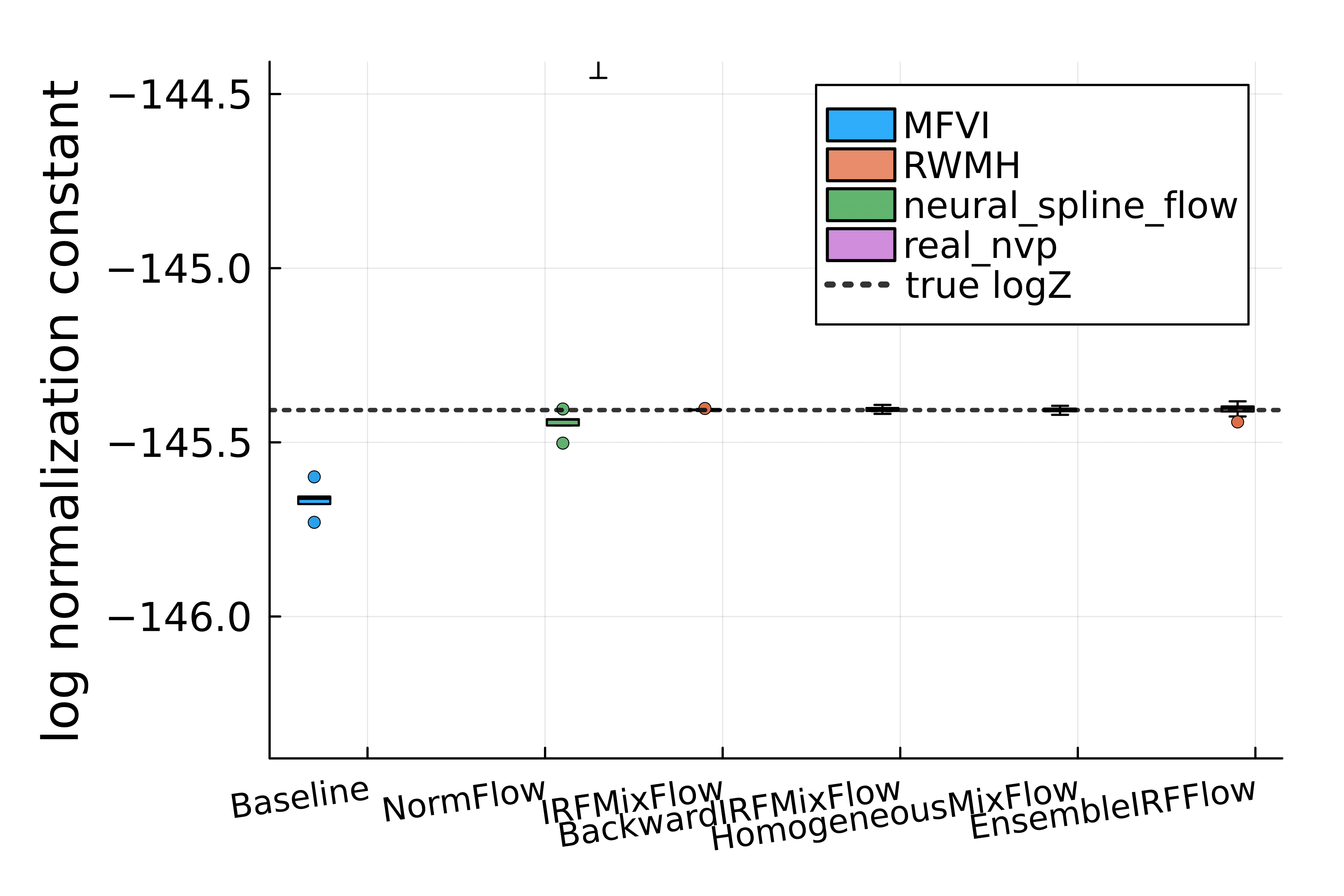}
        \includegraphics[width=0.24\columnwidth, trim=0 100 100 0, clip]{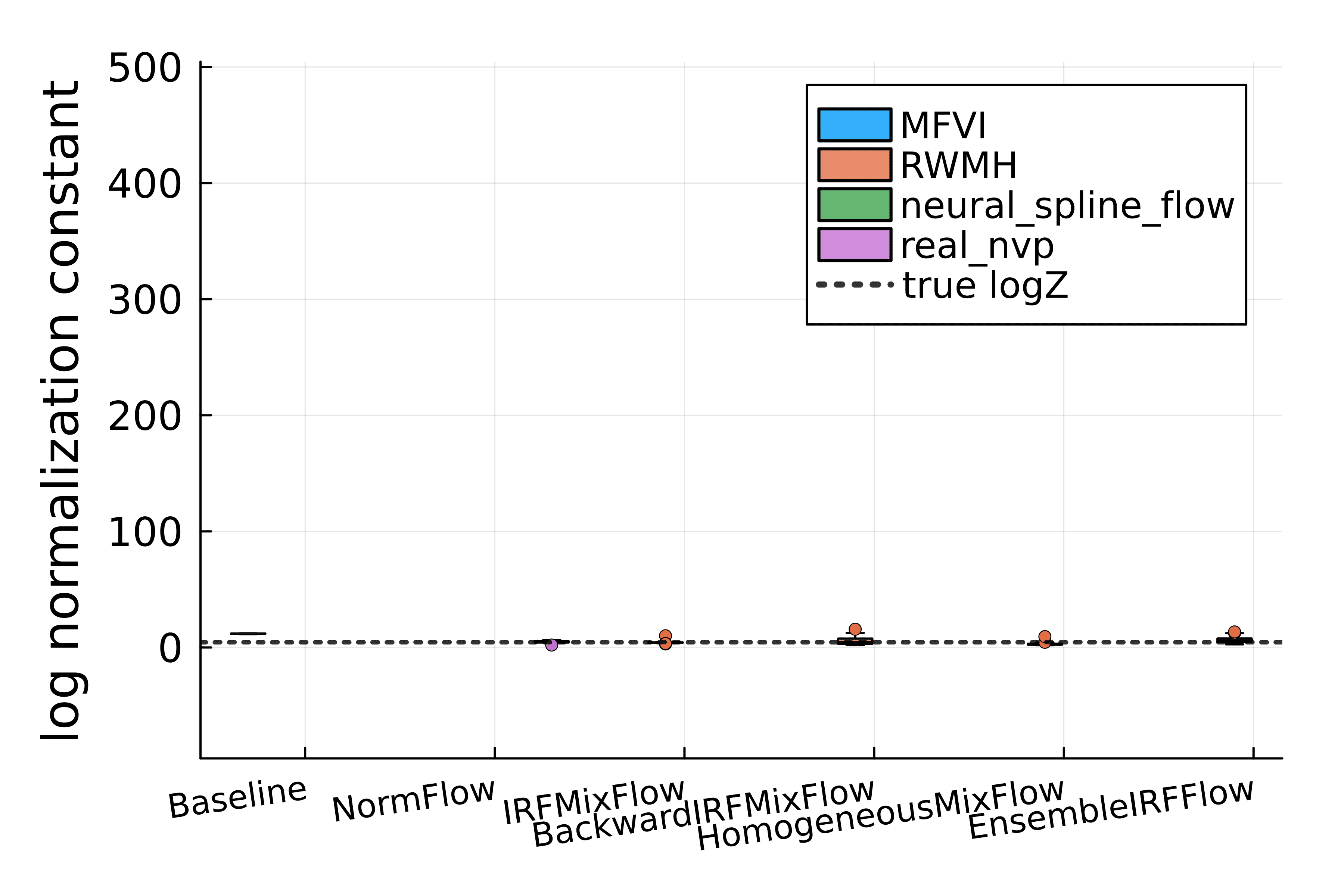}
        \includegraphics[width=0.24\columnwidth, trim=0 100 100 0, clip]{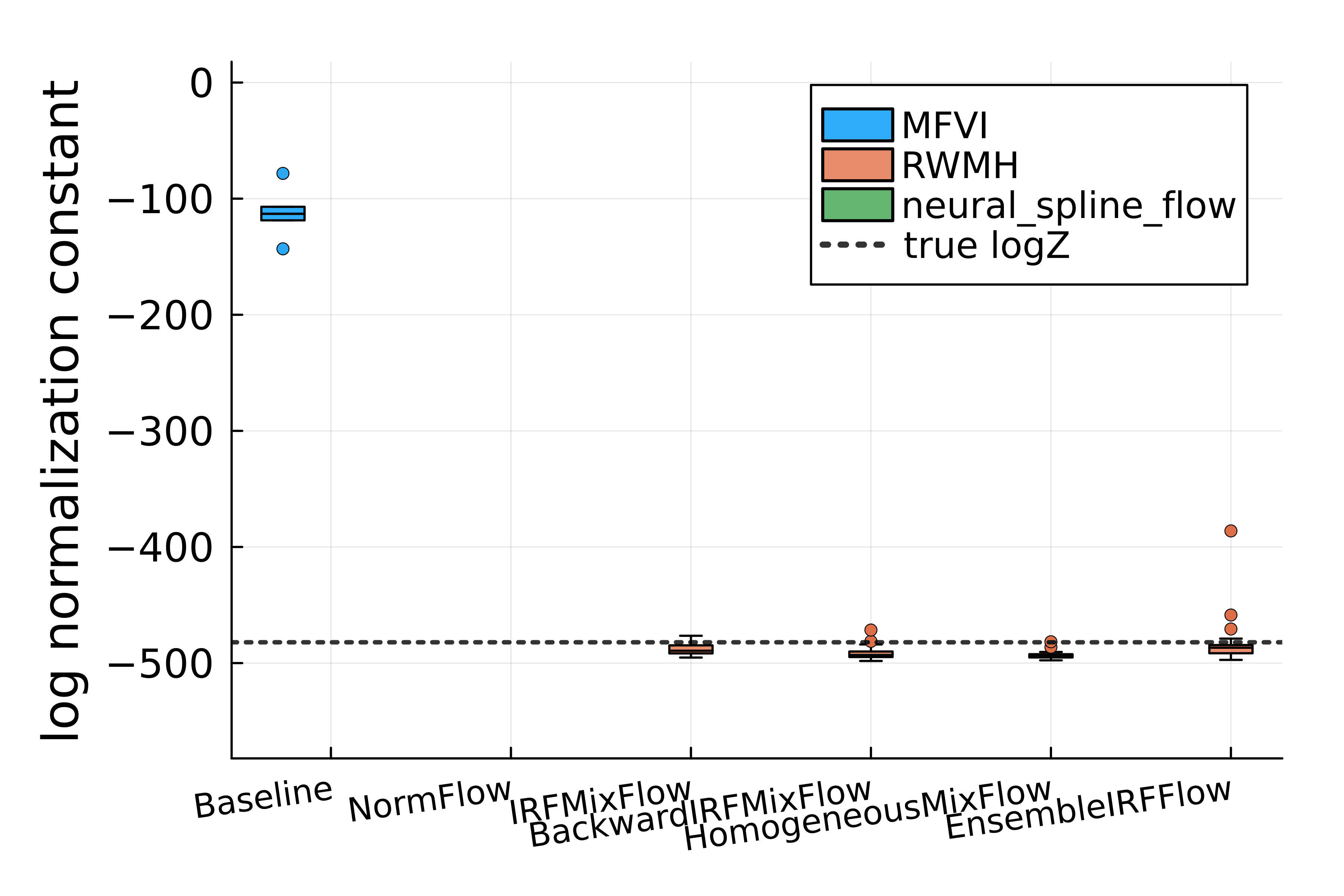}
        \includegraphics[width=0.24\columnwidth, trim=0 100 100 0, clip]{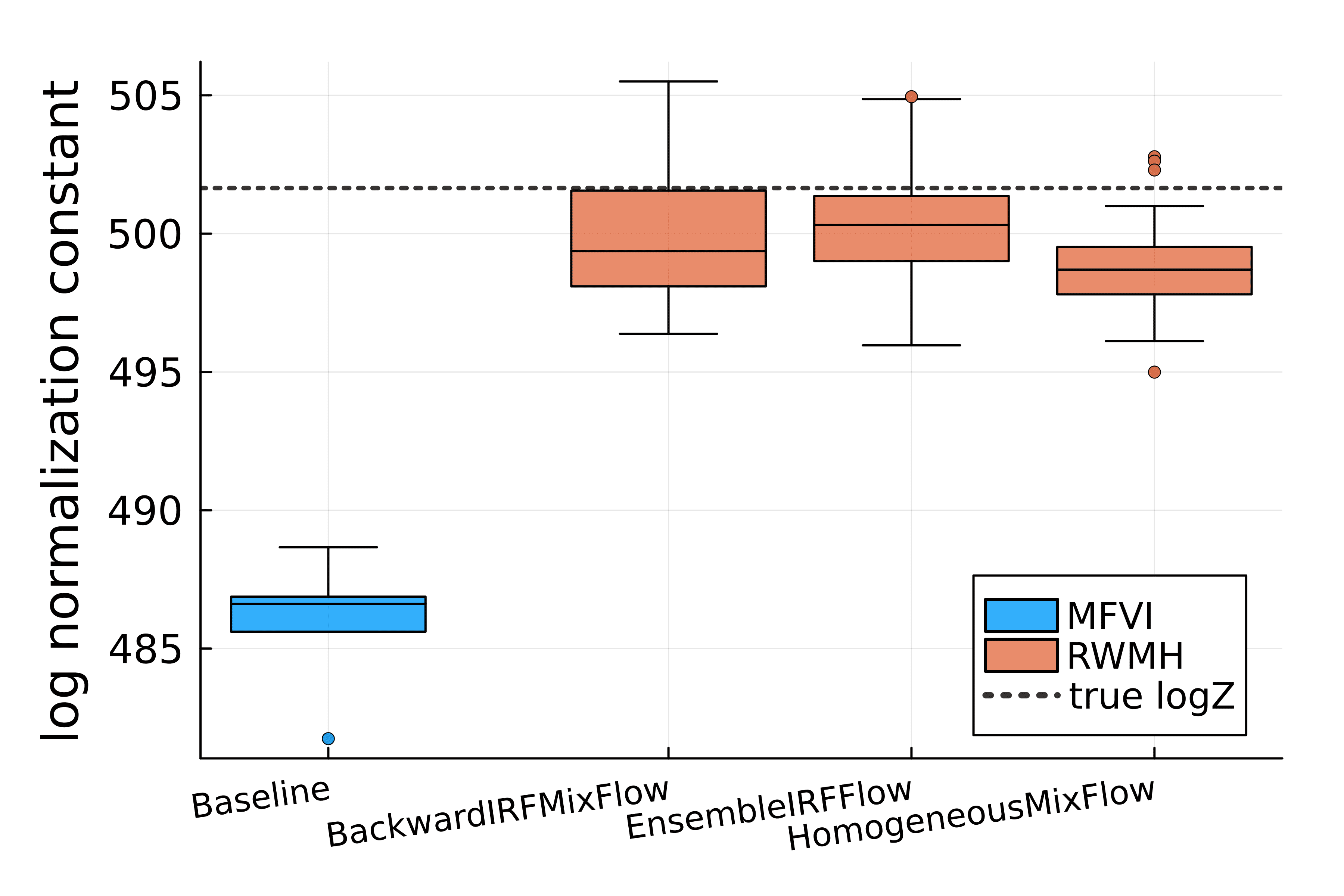}
		\caption{(b) $\log Z$ estimates}\label{fig:real_logz}
	\end{subfigure}
    \begin{subfigure}{\columnwidth}
        \includegraphics[width=0.24\columnwidth, trim=0 0 0 0, clip]{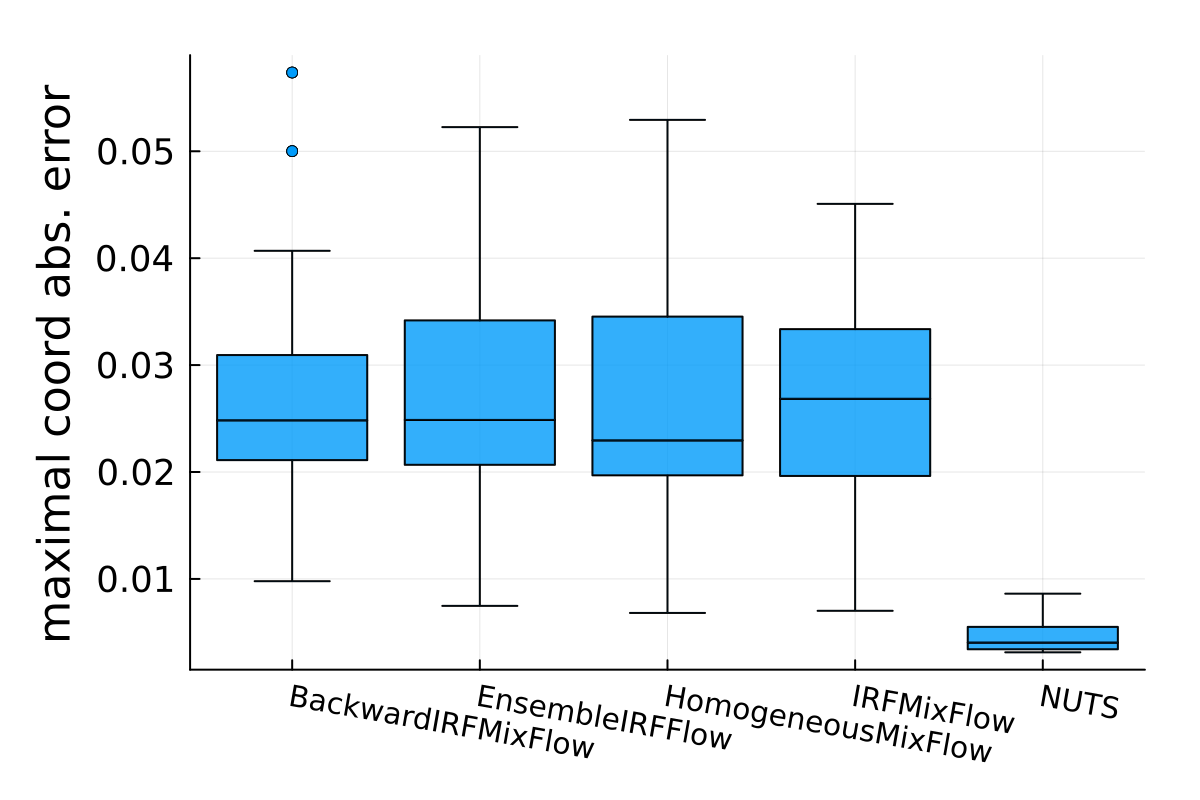}
        \includegraphics[width=0.24\columnwidth, trim=0 0 0 0, clip]{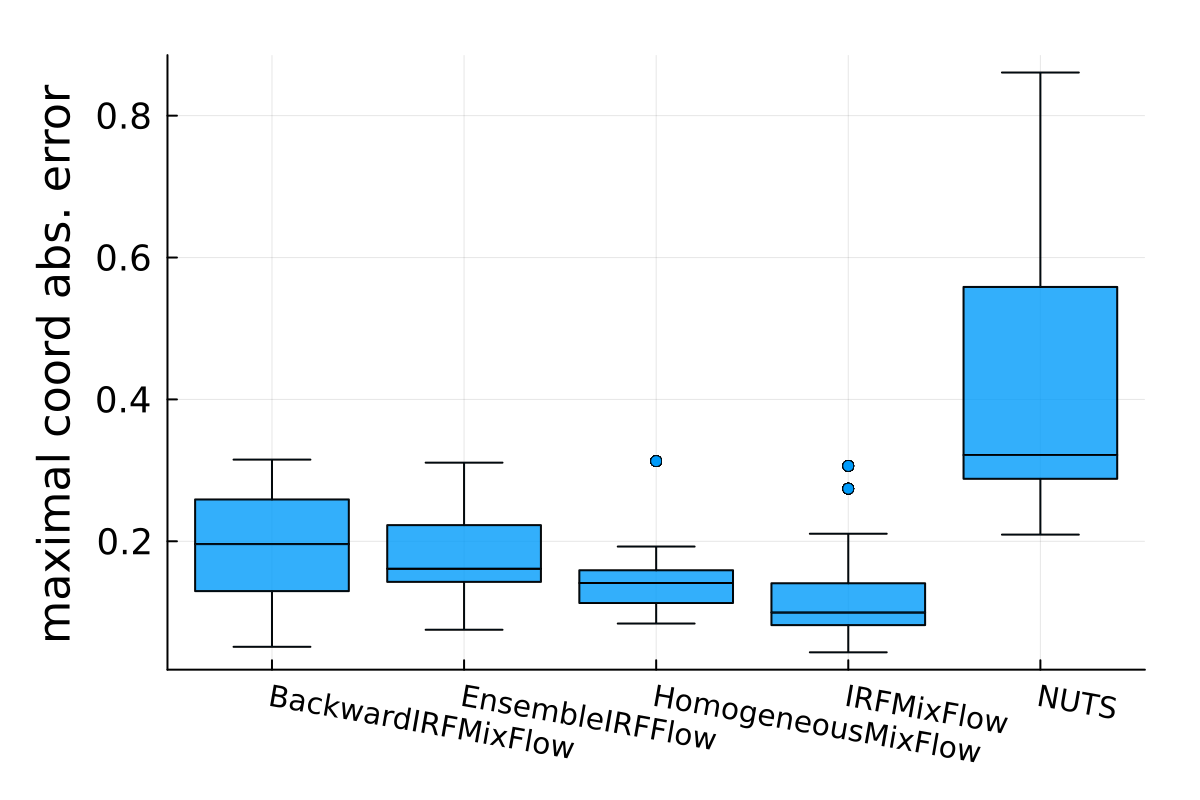}
        \includegraphics[width=0.24\columnwidth, trim=0 0 0 0, clip]{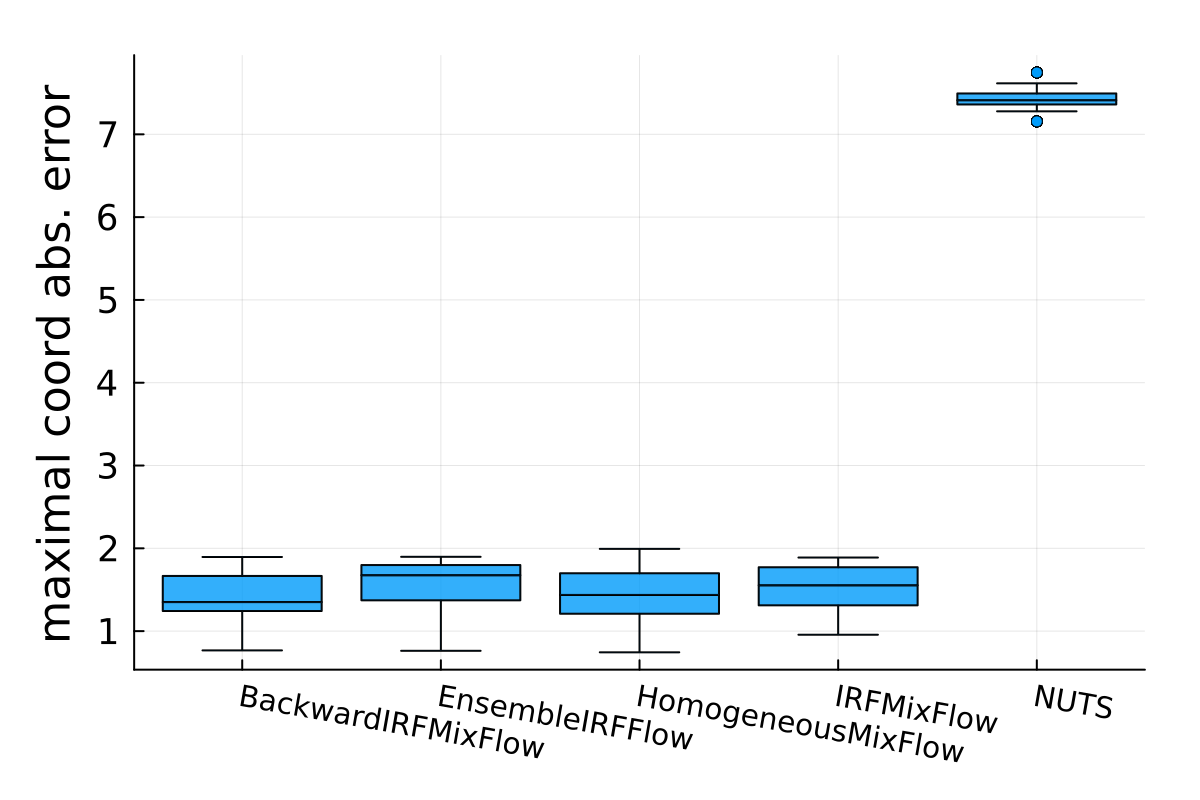}
        \includegraphics[width=0.24\columnwidth, trim=0 0 0 0, clip]{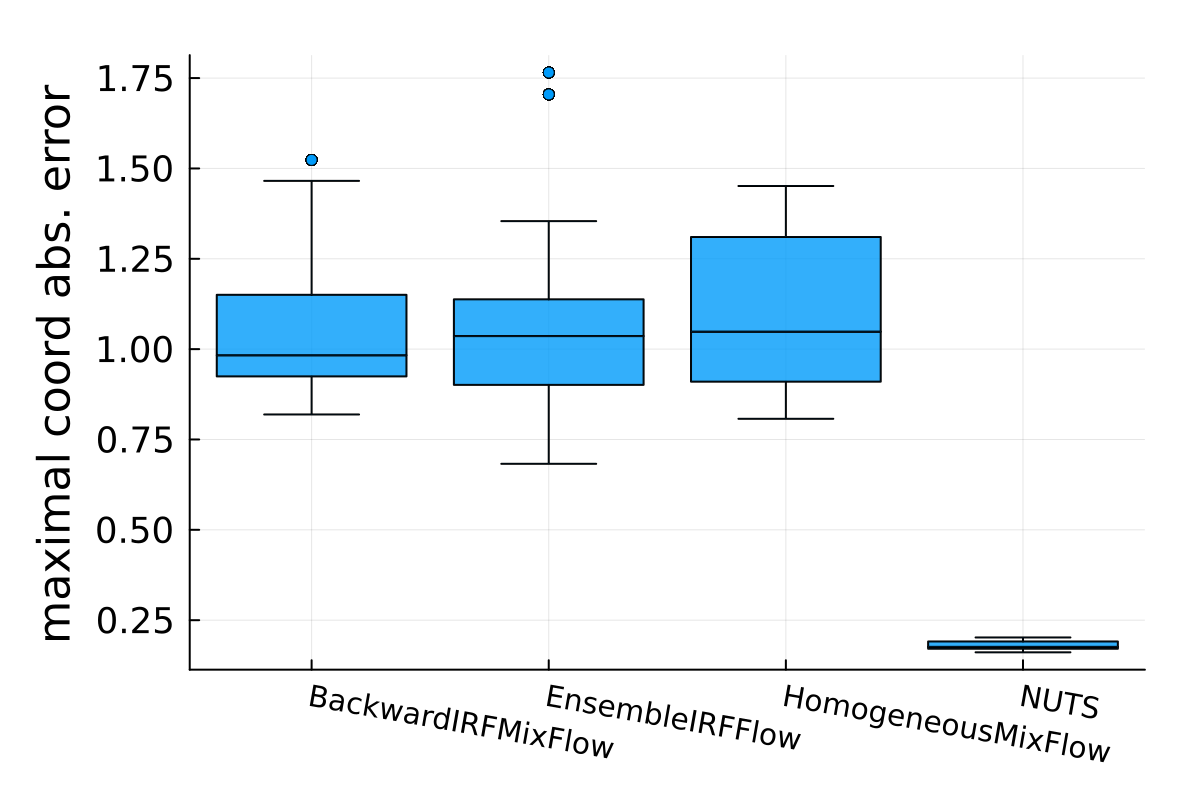}
		\caption{(c) Maximum absolute error of coordinate-wise posterior standard-deviation estimates relative to NUTS}\label{fig:real_std}
	\end{subfigure}
    \begin{subfigure}{\columnwidth}
    \includegraphics[width=0.24\columnwidth, trim=0 0 100 0, clip]{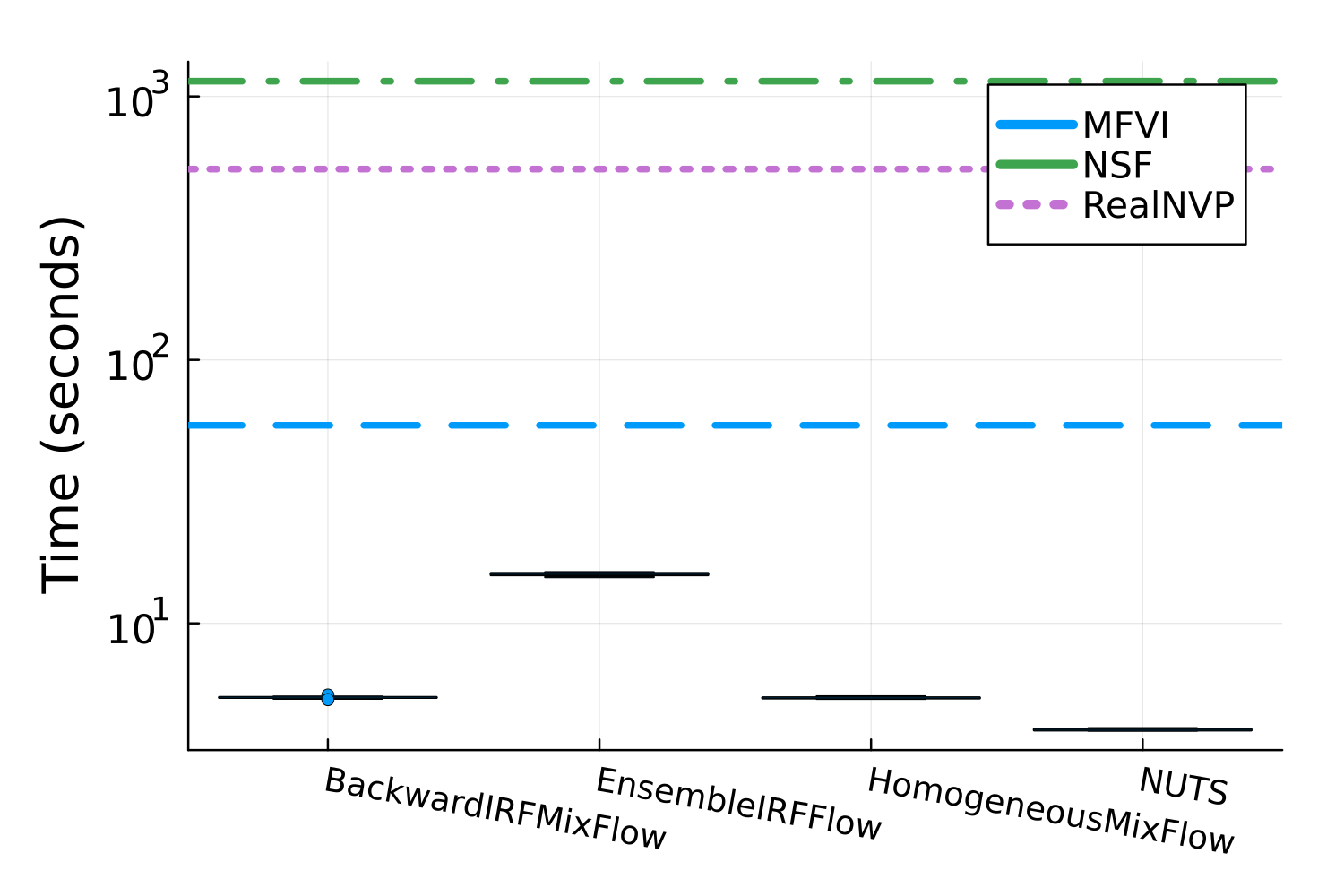}
    \includegraphics[width=0.24\columnwidth, trim=0 0 100 0, clip]{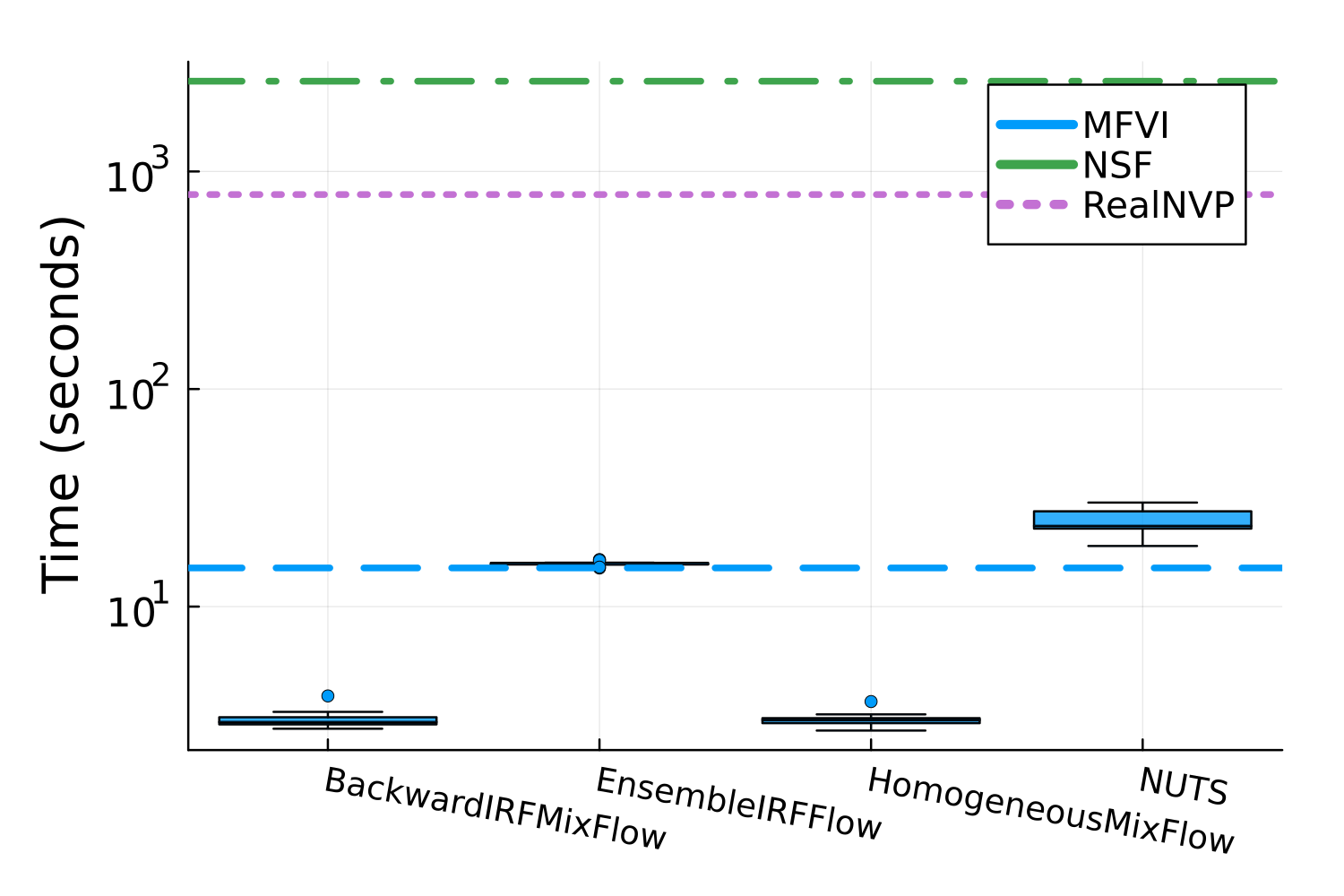}
    \includegraphics[width=0.24\columnwidth, trim=0 0 100 0, clip]{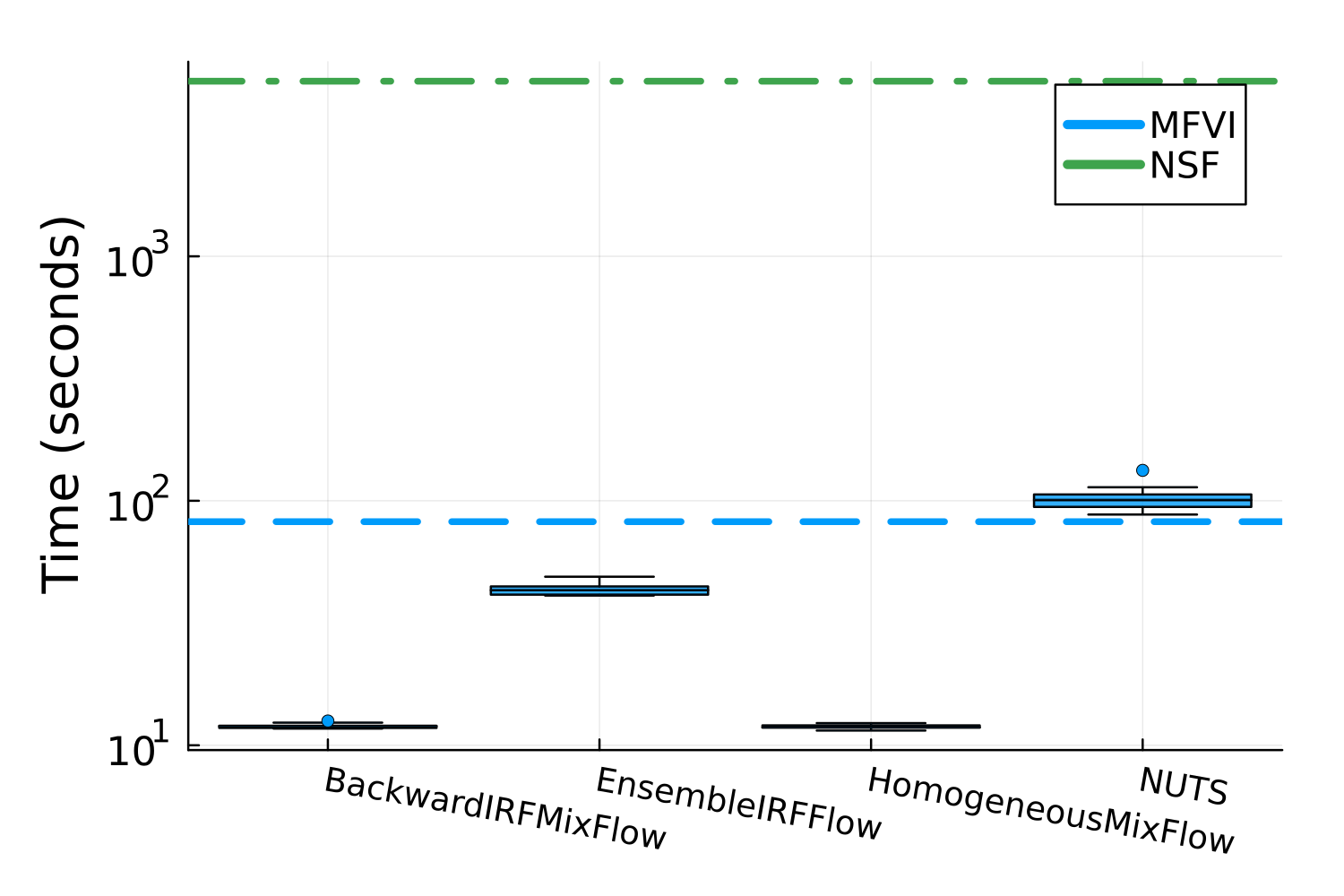}
    \includegraphics[width=0.24\columnwidth, trim=0 0 100 0, clip]{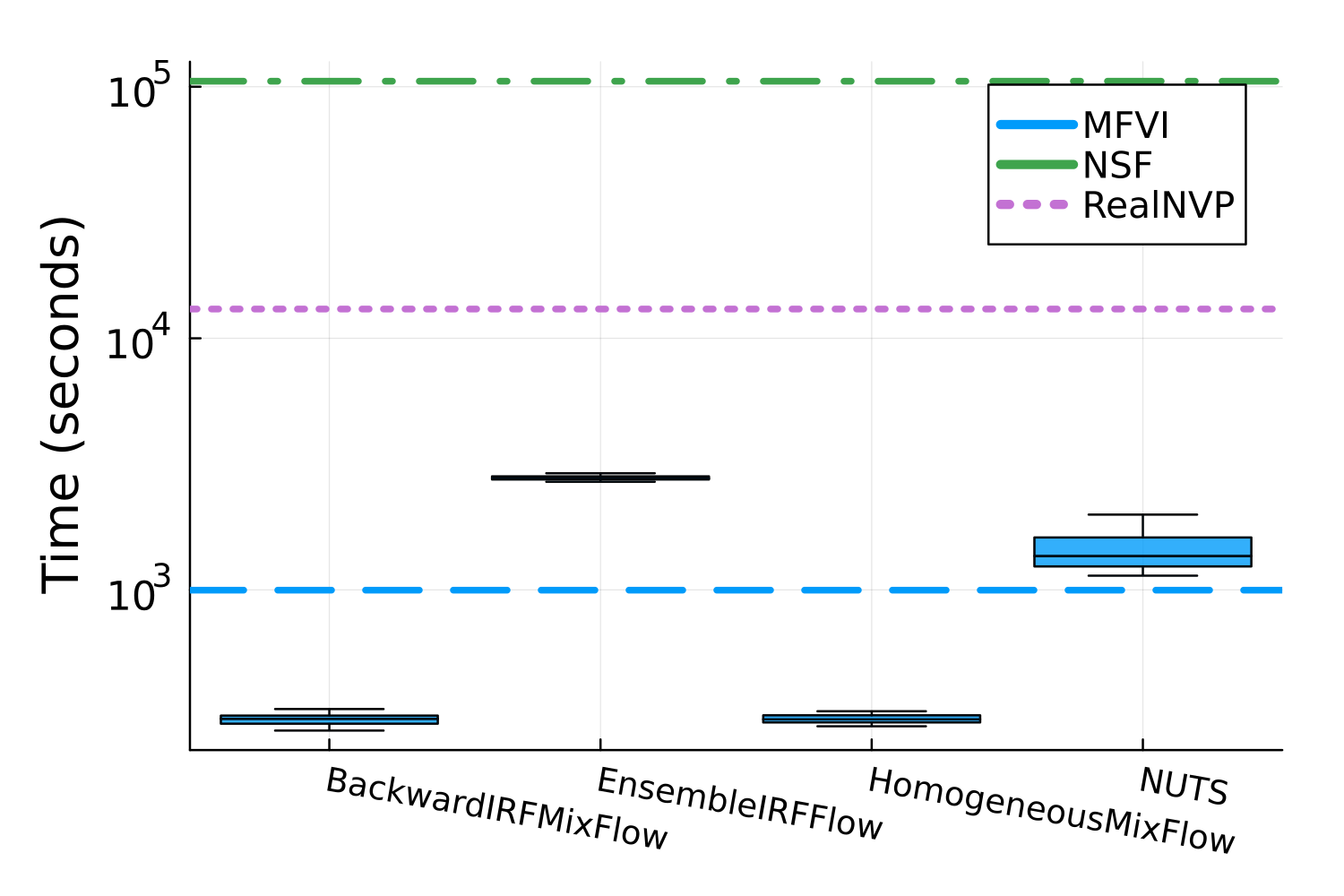}
    \caption{
        (d) Computation time (in seconds) for each method. To ensure a consistent environment, all timing results were obtained by rerunning
        the methods on the same local machine (hardware details provided in \cref{apdx:expt}). 
        \texttt{MFVI}, \texttt{NSF}, and \texttt{RealNVP} were each run once, as their execution times are deterministic given
        the flow architecture and optimization settings. For IRF flows and \texttt{NUTS}, timing statistics are based on 10 independent runs.
}\label{fig:real_cost}
	\end{subfigure}
\caption{Results on real-data benchmarks (columns, from left to right): 
    \texttt{TReg}($d=4$), \texttt{Brownian}($d = 32$), \texttt{SparseReg} ($d = 83$), and \texttt{LGCP} ($d = 1600$). 
}
\end{figure}

As in the synthetic experiments, our exact flows match---or modestly improve upon---the best-tuned \texttt{RealNVP} and \texttt{NSF} 
in both ELBO (\cref{fig:real_elbo}) and $\log Z$ accuracy (\cref{fig:real_logz}), and outperform the mean-field baseline by a wide margin. 
The per-sample importance-sampling ESS shows the same advantage
(\cref{fig:real_ess}). 
Crucially, normalizing flow training is orders of magnitude more expensive (\cref{fig:real_cost}), whereas the exact flows
achieve comparable accuracy at a fraction of the computational cost.

We further compare coordinate-wise posterior mean estimates (\cref{fig:real_mean} in \cref{apdx:real_expt}) and standard deviation estimates (\cref{fig:real_std}) against NUTS, 
reporting the maximum absolute error across dimensions relative to the estimated ground truth.
NUTS is initialized with independent draws from $q_0$ and run for $10{,}000$ iterations including $5000$ warm-up iterations.
IRF flows outperform NUTS on two models and are slightly worse on the other two---yet
they do so at generally faster computation time (\cref{fig:real_cost}). Note that the goal of 
this work is not to outperform MCMC, but rather to construct a variational family that provides 
asymptotic exactness and similar sampling performance; IRF MixFlows meet this standard.

\vspace{-1em}
\section{Conclusion}
\vspace{-0.5em}
We introduced a general framework for building asymptotically exact variational
families from general involutive MCMC kernels. By constructing invertible,
measure-preserving maps directly from these kernels, we overcome the main
practical limitation of MixFlow \citep{Xu22mixflow} and enable the construction of 
a broad class of exact flows.
We also provided a streamlined theoretical analysis for flows based on
measure-preserving transformations and demonstrated their empirical advantages
in density approximation and importance sampling.
A promising direction is to pair our framework with recent automatic-tuning MCMC \citep{Liu24,BironLattes24,bou2024gist,bou2024incorporating},
developing truly tuning-free exact flows in practice.

\begin{ack}
The authors sincerely thank Peter Orbanz for pointing us to the IRF literature,
which provided the initial inspiration for this work, 
and Alexandre Bouchard-C\^{o}t\'{e} for suggesting applying our methods to normalizing constant estimation.
T. Campbell and Z. Xu acknowledge support from the NSERC Discovery Grant
RGPIN-2025-04208. We are also grateful for access to the ARC Sockeye computing
platform at the University of British Columbia, and the compute cluster provided
by the Digital Research Alliance of Canada.
\end{ack}

\small
\bibliographystyle{unsrtnat}
\bibliography{sources}

\begin{thebibliography}{86}
\providecommand{\natexlab}[1]{#1}
\providecommand{\url}[1]{\texttt{#1}}
\expandafter\ifx\csname urlstyle\endcsname\relax
  \providecommand{\doi}[1]{doi: #1}\else
  \providecommand{\doi}{doi: \begingroup \urlstyle{rm}\Url}\fi

\bibitem[Jordan et~al.(1999)Jordan, Ghahramani, Jaakkola, and Saul]{Jordan99}
Michael Jordan, Zoubin Ghahramani, Tommi Jaakkola, and Lawrence Saul.
\newblock An introduction to variational methods for graphical models.
\newblock \emph{Machine Learning}, 37:\penalty0 183--233, 1999.

\bibitem[Wainwright and Jordan(2008)]{Wainwright08}
Martin Wainwright and Michael Jordan.
\newblock Graphical models, exponential families, and variational inference.
\newblock \emph{Foundations and Trends in Machine Learning}, 1\penalty0
  (1--2):\penalty0 1--305, 2008.

\bibitem[Blei et~al.(2017)Blei, Kucukelbir, and McAuliffe]{Blei17}
David Blei, Alp Kucukelbir, and Jon McAuliffe.
\newblock Variational inference: a review for statisticians.
\newblock \emph{Journal of the American Statistical Association}, 112\penalty0
  (518):\penalty0 859--877, 2017.

\bibitem[Rezende and Mohamed(2015)]{Rezende15}
Danilo Rezende and Shakir Mohamed.
\newblock Variational inference with normalizing flows.
\newblock In \emph{International Conference on Machine Learning}, 2015.

\bibitem[Ranganath et~al.(2016)Ranganath, Tran, and Blei]{Ranganath16}
Rajesh Ranganath, Dustin Tran, and David Blei.
\newblock Hierarchical variational models.
\newblock In \emph{International Conference on Machine Learning}, 2016.

\bibitem[Papamakarios et~al.(2021)Papamakarios, Nalisnick, Rezende, Mohamed,
  and Lakshminarayanan]{Papamakarios21}
George Papamakarios, Eric Nalisnick, Danilo~Jimenez Rezende, Shakir Mohamed,
  and Balaji Lakshminarayanan.
\newblock Normalizing flows for probabilistic modeling and inference.
\newblock \emph{Journal of Machine Learning Research}, 22:\penalty0 1--64,
  2021.

\bibitem[Margossian and Saul(2024)]{margossian2024variational}
Charles Margossian and Lawrence Saul.
\newblock Variational inference in location-scale families: Exact recovery of
  the mean and correlation matrix.
\newblock In \emph{Advances in Neural Information Processing Systems}, 2024.

\bibitem[Kullback and Leibler(1951)]{Kullback51}
Solomon Kullback and Richard Leibler.
\newblock On information and sufficiency.
\newblock \emph{The Annals of Mathematical Statistics}, 22\penalty0
  (1):\penalty0 79--86, 1951.

\bibitem[Kahn and Marshall(1953)]{kahn1953methods}
Herman Kahn and Andy~W Marshall.
\newblock Methods of reducing sample size in {M}onte {C}arlo computations.
\newblock \emph{Journal of the Operations Research Society of America},
  1\penalty0 (5):\penalty0 263--278, 1953.

\bibitem[Kahn(1955)]{kahn1955use}
Herman Kahn.
\newblock Use of different {M}onte {C}arlo sampling techniques.
\newblock Technical report, Rand Corporation, 1955.

\bibitem[Guo et~al.(2016)Guo, Wang, Fan, Broderick, and Dunson]{Guo16}
Fangjian Guo, Xiangyu Wang, Kai Fan, Tamara Broderick, and David Dunson.
\newblock Boosting variational inference.
\newblock In \emph{Advances in Neural Information Processing Systems}, 2016.

\bibitem[Miller et~al.(2017)Miller, Foti, and Adams]{Miller17}
Andrew Miller, Nicholas Foti, and Ryan Adams.
\newblock Variational boosting: iteratively refining posterior approximations.
\newblock In \emph{International Conference on Machine Learning}, 2017.

\bibitem[Wang(2016)]{Wang16}
Xiangyu Wang.
\newblock Boosting variational inference: theory and examples.
\newblock Master's thesis, Duke University, 2016.

\bibitem[Locatello et~al.(2018{\natexlab{a}})Locatello, Khanna, Ghosh, and
  R\"atsch]{Locatello18}
Francesco Locatello, Rajiv Khanna, Joydeep Ghosh, and Gunnar R\"atsch.
\newblock Boosting variational inference: an optimization perspective.
\newblock In \emph{International Conference on Artificial Intelligence and
  Statistics}, 2018{\natexlab{a}}.

\bibitem[Locatello et~al.(2018{\natexlab{b}})Locatello, Dresdner, Khanna,
  Valera, and R\"atsch]{Locatello18b}
Francesco Locatello, Gideon Dresdner, Rajiv Khanna, Isabel Valera, and Gunnar
  R\"atsch.
\newblock Boosting black box variational inference.
\newblock In \emph{Advances in Neural Information Processing Systems},
  2018{\natexlab{b}}.

\bibitem[Campbell and Li(2019)]{Campbell19b}
Trevor Campbell and Xinglong Li.
\newblock Universal boosting variational inference.
\newblock In \emph{Advances in Neural Information Processing Systems}, 2019.

\bibitem[Kobyzev et~al.(2021)Kobyzev, Prince, and Brubaker]{Kobyzev21}
Ivan Kobyzev, Simon Prince, and Marcus Brubaker.
\newblock Normalizing flows: an introduction and review of current methods.
\newblock \emph{IEEE Transactions on Pattern Analysis and Machine
  Intelligence}, 43\penalty0 (11):\penalty0 3964--3979, 2021.

\bibitem[Agrawal et~al.(2020)Agrawal, Sheldon, and Domke]{Agrawal20}
Abhinav Agrawal, Daniel Sheldon, and Justin Domke.
\newblock Advances in black-box {VI}: Normalizing flows, importance weighting,
  and optimization.
\newblock In \emph{Advances in Neural Information Processing Systems}, 2020.

\bibitem[Dinh et~al.(2017)Dinh, Sohl-Dickstein, and Bengio]{Dinh17}
Laurent Dinh, Jascha Sohl-Dickstein, and Samy Bengio.
\newblock Density estimation using {R}eal {NVP}.
\newblock In \emph{International Conference on Learning Representations}, 2017.

\bibitem[van~den Berg et~al.(2018)van~den Berg, Hasenclever, Tomczak, and
  Welling]{vansylvester}
Rianne van~den Berg, Leonard Hasenclever, Jakub Tomczak, and Max Welling.
\newblock Sylvester normalizing flows for variational inference.
\newblock In \emph{Conference on Uncertainty in Artificial Intelligence}, 2018.

\bibitem[Koehler et~al.(2021)Koehler, Mehta, and
  Risteski]{koehler2021representational}
Frederic Koehler, Viraj Mehta, and Andrej Risteski.
\newblock Representational aspects of depth and conditioning in normalizing
  flows.
\newblock In \emph{International Conference on Machine Learning}, 2021.

\bibitem[Kong and Chaudhuri(2020)]{kong2020expressive}
Zhifeng Kong and Kamalika Chaudhuri.
\newblock The expressive power of a class of normalizing flow models.
\newblock In \emph{International Conference on Artificial Intelligence and
  Statistics}, 2020.

\bibitem[Robert and Casella(2004)]{Robert04}
Christian Robert and George Casella.
\newblock \emph{Monte Carlo Statistical Methods}.
\newblock Springer, 2nd edition, 2004.

\bibitem[Robert and Casella(2011)]{Robert11}
Christian Robert and George Casella.
\newblock A short history of {M}arkov chain {M}onte {C}arlo: subjective
  recollections from incomplete data.
\newblock \emph{Statistical Science}, 26\penalty0 (1):\penalty0 102--115, 2011.

\bibitem[Gelman et~al.(2013)Gelman, Carlin, Stern, Dunson, Vehtari, and
  Rubin]{Gelman13}
Andrew Gelman, John Carlin, Hal Stern, David Dunson, Aki Vehtari, and Donald
  Rubin.
\newblock \emph{Bayesian data analysis}.
\newblock CRC Press, 3rd edition, 2013.

\bibitem[Salimans et~al.(2015)Salimans, Kingma, and Welling]{Salimans15}
Tim Salimans, Diederik Kingma, and Max Welling.
\newblock {M}arkov chain {M}onte {C}arlo and variational inference: bridging
  the gap.
\newblock In \emph{International Conference on Machine Learning}, 2015.

\bibitem[Wolf et~al.(2016)Wolf, Karl, and van~der Smagt]{Wolf16}
Christopher Wolf, Maximilian Karl, and Patrick van~der Smagt.
\newblock Variational inference with {H}amiltonian {M}onte {C}arlo.
\newblock \emph{arXiv:1609.08203}, 2016.

\bibitem[Geffner and Domke(2021)]{Geffner21}
Tomas Geffner and Justin Domke.
\newblock {MCMC} variational inference via uncorrected {H}amiltonian annealing.
\newblock In \emph{Advances in Neural Information Processing Systems}, 2021.

\bibitem[Zhang et~al.(2021)Zhang, Hsu, Li, Finn, and Grosse]{Zhang21}
Guodong Zhang, Kyle Hsu, Jianing Li, Chelsea Finn, and Roger Grosse.
\newblock Differentiable annealed importance sampling and the perils of
  gradient noise.
\newblock In \emph{Advances in Neural Information Processing Systems}, 2021.

\bibitem[Thin et~al.(2021)Thin, Kotelevskii, Durmus, Panov, Moulines, and
  Doucet]{Thin21}
Achille Thin, Nikita Kotelevskii, Alain Durmus, Maxim Panov, Eric Moulines, and
  Arnaud Doucet.
\newblock {M}onte {C}arlo variational auto-encoders.
\newblock In \emph{International Conference on Machine Learning}, 2021.

\bibitem[Jankowiak and Phan(2021)]{Jankowiak21}
Martin Jankowiak and Du~Phan.
\newblock Surrogate likelihoods for variational annealed importance sampling.
\newblock \emph{arXiv:2112.12194}, 2021.

\bibitem[Geffner and Domke(2023)]{geffner2023langevin}
Tomas Geffner and Justin Domke.
\newblock Langevin diffusion variational inference.
\newblock In \emph{International Conference on Artificial Intelligence and
  Statistics}, 2023.

\bibitem[Arbel et~al.(2021)Arbel, Matthews, and Doucet]{Arbel21}
Michael Arbel, Alexander Matthews, and Arnaud Doucet.
\newblock Annealed flow transport {M}onte {C}arlo.
\newblock In \emph{International Conference on Machine Learning}, 2021.

\bibitem[Matthews et~al.(2022)Matthews, Arbel, Rezende, and
  Doucet]{matthews2022continual}
Alex Matthews, Michael Arbel, Danilo~Jimenez Rezende, and Arnaud Doucet.
\newblock Continual repeated annealed flow transport {M}onte {C}arlo.
\newblock In \emph{International Conference on Machine Learning}, 2022.

\bibitem[Naesseth et~al.(2018)Naesseth, Linderman, Ranganath, and
  Blei]{naesseth2018variational}
Christian Naesseth, Scott Linderman, Rajesh Ranganath, and David Blei.
\newblock Variational sequential {M}onte {C}arlo.
\newblock In \emph{International Conference on Artificial Intelligence and
  Statistics}, 2018.

\bibitem[Maddison et~al.(2017)Maddison, Lawson, Tucker, Heess, Norouzi, Mnih,
  Doucet, and Teh]{maddison2017filtering}
Chris~J Maddison, John Lawson, George Tucker, Nicolas Heess, Mohammad Norouzi,
  Andriy Mnih, Arnaud Doucet, and Yee Teh.
\newblock Filtering variational objectives.
\newblock \emph{Advances in Neural Information Processing Systems}, 2017.

\bibitem[Zenn and Bamler(2024)]{zenn2024differentiable}
Johannes Zenn and Robert Bamler.
\newblock Differentiable annealed importance sampling minimizes the symmetrized
  {K}ullback-{L}eibler divergence between initial and target distribution.
\newblock In \emph{Forty-first International Conference on Machine Learning
  (ICML)}, 2024.

\bibitem[Kim et~al.(2025)Kim, Xu, Gardner, and Campbell]{Kim25}
Kyurae Kim, Zuheng Xu, Jacob Gardner, and Trevor Campbell.
\newblock Tuning sequential {M}onte {C}arlo samplers via greedy incremental
  divergence minimization.
\newblock In \emph{International Conference on Machine Learning}, 2025.

\bibitem[Xu et~al.(2022)Xu, Chen, and Campbell]{Xu22mixflow}
Zuheng Xu, Naitong Chen, and Trevor Campbell.
\newblock Mix{F}lows: principled variational inference via mixed flows.
\newblock In \emph{International Conference on Machine Learning}, 2022.

\bibitem[Neal(2012)]{Neal12}
Radford Neal.
\newblock How to view an {MCMC} simulation as permutation, with applications to
  parallel simulation and improved importance sampling.
\newblock \emph{arXiv:1205.0070}, 2012.

\bibitem[Diluvi et~al.(2024)Diluvi, Bloem-Reddy, and
  Campbell]{Diluvi2024MADMix}
Gian~Carlo Diluvi, Benjamin Bloem-Reddy, and Trevor Campbell.
\newblock Mixed variational flows for discrete variables.
\newblock In \emph{International Conference on Artificial Intelligence and
  Statistics}, 2024.

\bibitem[{ver Steeg} and Galstyan(2021)]{ver2021hamiltonian}
Greg {ver Steeg} and Aram Galstyan.
\newblock Hamiltonian dynamics with non-{N}ewtonian momentum for rapid
  sampling.
\newblock In \emph{Advances in Neural Information Processing Systems}, 2021.

\bibitem[Robnik et~al.(2023)Robnik, De~Luca, Silverstein, and
  Seljak]{robnik2023microcanonical}
Jakob Robnik, G~Bruno De~Luca, Eva Silverstein, and Uro{\v{s}} Seljak.
\newblock Microcanonical {H}amiltonian {M}onte {C}arlo.
\newblock \emph{Journal of Machine Learning Research}, 24\penalty0
  (311):\penalty0 1--34, 2023.

\bibitem[Neklyudov et~al.(2021)Neklyudov, Bondesan, and
  Welling]{neklyudov2021deterministic}
Kirill Neklyudov, Roberto Bondesan, and Max Welling.
\newblock Deterministic {G}ibbs sampling via ordinary differential equations.
\newblock \emph{arXiv:2106.10188}, 2021.

\bibitem[Murray and Elliott(2012)]{Murray12}
Iain Murray and Lloyd Elliott.
\newblock Driving {M}arkov chain {M}onte {C}arlo with a dependent random
  stream.
\newblock \emph{arXiv:1204.3187}, 2012.

\bibitem[Morita(1988)]{morita1988deterministic}
Takehiko Morita.
\newblock Deterministic version lemmas in ergodic theory of random dynamical
  systems.
\newblock \emph{Hiroshima mathematical journal}, 18:\penalty0 15--29, 1988.

\bibitem[Diaconis and Freedman(1999)]{diaconis1999iterated}
Persi Diaconis and David Freedman.
\newblock Iterated random functions.
\newblock \emph{SIAM review}, 41:\penalty0 45--76, 1999.

\bibitem[Tierney(1994)]{Tierney94}
Luke Tierney.
\newblock {M}arkov chains for exploring posterior distributions.
\newblock \emph{The Annals of Statistics}, 22\penalty0 (4):\penalty0
  1701--1728, 1994.

\bibitem[Neklyudov et~al.(2020)Neklyudov, Welling, Egorov, and
  Vetrov]{Neklyudov20}
Kirill Neklyudov, Max Welling, Evgenii Egorov, and Dmitry Vetrov.
\newblock Involutive {MCMC}: a unifying framework.
\newblock In \emph{International Conference on Machine Learning}, 2020.

\bibitem[Xu and Campbell(2023)]{Xu23chaos}
Zuheng Xu and Trevor Campbell.
\newblock Embracing the chaos: analysis and diagnosis of numerical instability
  in variational flows.
\newblock In \emph{Advances in Neural Information Processing Systems}, 2023.

\bibitem[Tierney(1998)]{tierney1998note}
Luke Tierney.
\newblock A note on {Metropolis-Hastings} kernels for general state spaces.
\newblock \emph{Annals of applied probability}, 1998.

\bibitem[Liu et~al.(2024)Liu, Surjanovic, Biron-Lattes, Bouchard-C{\^o}t{\'e},
  and Campbell]{Liu24}
Tiange Liu, Nikola Surjanovic, Miguel Biron-Lattes, Alexandre
  Bouchard-C{\^o}t{\'e}, and Trevor Campbell.
\newblock {A}uto{S}tep: locally adaptive involutive {MCMC}.
\newblock \emph{arXiv:2410.18929}, 2024.

\bibitem[Biron-Lattes et~al.(2024)Biron-Lattes, Surjanovic, Syed, Campbell, and
  Bouchard-C{\^o}t{\'e}]{BironLattes24}
Miguel Biron-Lattes, Nikola Surjanovic, Saifuddin Syed, Trevor Campbell, and
  Alexandre Bouchard-C{\^o}t{\'e}.
\newblock {autoMALA}: Locally adaptive {M}etropolis-adjusted {L}angevin
  algorithm.
\newblock In \emph{International Conference on Artificial Intelligence and
  Statistics}, 2024.

\bibitem[Bou-Rabee et~al.(2024{\natexlab{a}})Bou-Rabee, Carpenter, and
  Marsden]{bou2024gist}
Nawaf Bou-Rabee, Bob Carpenter, and Milo Marsden.
\newblock {GIST: Gibbs self-tuning for locally adaptive Hamiltonian Monte
  Carlo}.
\newblock \emph{arXiv:2404.15253}, 2024{\natexlab{a}}.

\bibitem[Bou-Rabee et~al.(2024{\natexlab{b}})Bou-Rabee, Carpenter, Kleppe, and
  Marsden]{bou2024incorporating}
Nawaf Bou-Rabee, Bob Carpenter, Tore~Selland Kleppe, and Milo Marsden.
\newblock Incorporating local step-size adaptivity into the {No-U-Turn} sampler
  using {G}ibbs self tuning.
\newblock \emph{arXiv:2408.08259}, 2024{\natexlab{b}}.

\bibitem[Roberts and Rosenthal(2004)]{Roberts04}
Gareth Roberts and Jeffrey Rosenthal.
\newblock General state space {M}arkov chains and {MCMC} algorithms.
\newblock \emph{Probability Surveys}, 1:\penalty0 20--71, 2004.

\bibitem[Kakutani(1950)]{kakutani1950random}
Shizuo Kakutani.
\newblock Random ergodic theorems and {M}arkoff processes with a stable
  distribution.
\newblock In \emph{Proceedings of the Second Berkeley Symposium on Mathematical
  Statistics and Probability}, 1950.

\bibitem[Kifer(2012)]{kifer2012ergodic}
Yuri Kifer.
\newblock \emph{Ergodic theory of random transformations}, volume~10.
\newblock Springer Science \& Business Media, 2012.

\bibitem[Neal(2011)]{Neal11}
Radford Neal.
\newblock {MCMC} using {H}amiltonian dynamics.
\newblock In Steve Brooks, Andrew Gelman, Galin Jones, and Xiao-Li Meng,
  editors, \emph{Handbook of {M}arkov chain {M}onte {C}arlo}, chapter~5. CRC
  Press, 2011.

\bibitem[Duane et~al.(1987)Duane, Kennedy, Pendleton, and
  Roweth]{duane1987hybrid}
Simon Duane, Anthony Kennedy, Brian Pendleton, and Duncan Roweth.
\newblock Hybrid {M}onte {C}arlo.
\newblock \emph{Physics letters B}, 195\penalty0 (2):\penalty0 216--222, 1987.

\bibitem[Xu and Campbell(2022)]{Xu22}
Zuheng Xu and Trevor Campbell.
\newblock The computational asymptotics of variational inference and the
  {L}aplace approximation.
\newblock \emph{Statistics and Computing}, 32\penalty0 (4):\penalty0 1--37,
  2022.

\bibitem[Rossky et~al.(1978)Rossky, Doll, and Friedman]{rossky1978brownian}
Peter~J Rossky, Jimmie~D Doll, and Harold~L Friedman.
\newblock Brownian dynamics as smart {M}onte {C}arlo simulation.
\newblock \emph{The Journal of Chemical Physics}, 69\penalty0 (10):\penalty0
  4628--4633, 1978.

\bibitem[Livingstone(2021)]{livingstone2021geometric}
Samuel Livingstone.
\newblock Geometric ergodicity of the random walk {M}etropolis with
  position-dependent proposal covariance.
\newblock \emph{Mathematics}, 9\penalty0 (4), 2021.

\bibitem[Eisner et~al.(2015)Eisner, Farkas, Haase, and Nagel]{Eisner15}
Tanja Eisner, B\'alint Farkas, Markus Haase, and Rainer Nagel.
\newblock \emph{Operator Theoretic Aspects of Ergodic Theory}.
\newblock Graduate Texts in Mathematics. Springer, 2015.

\bibitem[Birkhoff(1931)]{Birkhoff31}
George Birkhoff.
\newblock Proof of the ergodic theorem.
\newblock \emph{Proceedings of the National Academy of Sciences}, 17\penalty0
  (12):\penalty0 656--660, 1931.

\bibitem[Douc et~al.(2018)Douc, Moulines, Priouret, and
  Soulier]{douc2018markov}
Randal Douc, Eric Moulines, Pierre Priouret, and Philippe Soulier.
\newblock \emph{Markov chains}.
\newblock Springer, 2018.

\bibitem[Qiao and Minematsu(2010)]{qiao2010study}
Yu~Qiao and Nobuaki Minematsu.
\newblock A study on invariance of $ f $-divergence and its application to
  speech recognition.
\newblock \emph{IEEE Transactions on Signal Processing}, 58\penalty0
  (7):\penalty0 3884--3890, 2010.

\bibitem[Margossian et~al.(2024)Margossian, Hoffman, Sountsov, Riou-Durand,
  Vehtari, and Gelman]{margossian2024nested}
Charles~C Margossian, Matthew~D Hoffman, Pavel Sountsov, Lionel Riou-Durand,
  Aki Vehtari, and Andrew Gelman.
\newblock Nested $\hat {R}$: Assessing the convergence of {M}arkov chain
  {M}onte {C}arlo when running many short chains.
\newblock \emph{Bayesian Analysis}, 1\penalty0 (1):\penalty0 1--28, 2024.

\bibitem[Sountsov et~al.(2024)Sountsov, Carroll, and
  Hoffman]{sountsov2024running}
Pavel Sountsov, Colin Carroll, and Matthew~D Hoffman.
\newblock Running {M}arkov chain {M}onte {C}arlo on modern hardware and
  software.
\newblock \emph{arXiv:2411.04260}, 2024.

\bibitem[Tupper(2005)]{Tupper05}
Paul Tupper.
\newblock Ergodicity and the numerical simulation of {H}amiltonian systems.
\newblock \emph{SIAM Journal on Applied Dynamical Systems}, 4\penalty0
  (3):\penalty0 563--587, 2005.

\bibitem[Durkan et~al.(2019)Durkan, Bekasov, Murray, and
  Papamakarios]{Durkan19}
Conor Durkan, Artur Bekasov, Iain Murray, and George Papamakarios.
\newblock Neural spline flows.
\newblock In \emph{Advances in Neural Information Processing Systems}, 2019.

\bibitem[Hoffman and Gelman(2014)]{Hoffman14}
Matthew Hoffman and Andrew Gelman.
\newblock The {N}o-{U}-{T}urn {S}ampler: adaptively setting path lengths in
  {H}amiltonian {M}onte {C}arlo.
\newblock \emph{Journal of Machine Learning Research}, 15\penalty0
  (1):\penalty0 1593--1623, 2014.

\bibitem[Kong(1992)]{kong1992note}
Augustine Kong.
\newblock A note on importance sampling using standardized weights.
\newblock \emph{University of Chicago, Dept. of Statistics, Tech. Rep},
  348:\penalty0 14, 1992.

\bibitem[Kong et~al.(1994)Kong, Liu, and Wong]{kong1994sequential}
Augustine Kong, Jun~S Liu, and Wing~Hung Wong.
\newblock Sequential imputations and {B}ayesian missing data problems.
\newblock \emph{Journal of the American statistical association}, 89\penalty0
  (425):\penalty0 278--288, 1994.

\bibitem[Liu(1996)]{Liu96}
Jun Liu.
\newblock Metropolized independent sampling with comparisons to rejection
  sampling and importance sampling.
\newblock \emph{Statistics and Computing}, 6:\penalty0 113--119, 1996.

\bibitem[Haario et~al.(2001)Haario, Saksman, and Tamminen]{haario2001adaptive}
Heikki Haario, Eero Saksman, and Johanna Tamminen.
\newblock An adaptive {M}etropolis algorithm.
\newblock \emph{Bernoulli}, pages 223--242, 2001.

\bibitem[Neal(2003)]{neal2003slice}
Radford Neal.
\newblock Slice sampling.
\newblock \emph{The Annals of Statistics}, 31\penalty0 (3):\penalty0 705--767,
  2003.

\bibitem[Agapiou et~al.(2017)Agapiou, Papaspiliopoulos, Sanz-Alonso, and
  Stuart]{agapiou2017importance}
Sergios Agapiou, Omiros Papaspiliopoulos, Daniel Sanz-Alonso, and Andrew~M
  Stuart.
\newblock Importance sampling: Intrinsic dimension and computational cost.
\newblock \emph{Statistical Science}, pages 405--431, 2017.

\bibitem[Sountsov et~al.(2020)Sountsov, Radul, and
  contributors]{inferencegym2020}
Pavel Sountsov, Alexey Radul, and contributors.
\newblock Inference gym, 2020.
\newblock URL \url{https://pypi.org/project/inference_gym}.

\bibitem[Hastings(1970)]{Hastings70}
Wilfred Hastings.
\newblock {M}onte {C}arlo sampling methods using {M}arkov chains and their
  applications.
\newblock \emph{Biometrika}, 57:\penalty0 97--109, 1970.

\bibitem[Onsager(1944)]{onsager1944crystal}
Lars Onsager.
\newblock Crystal statistics. {I}. {A} two-dimensional model with an
  order-disorder transition.
\newblock \emph{Physical review}, 65\penalty0 (3-4), 1944.

\bibitem[Stephens(1997)]{stephens1997bayesian}
Matthew Stephens.
\newblock \emph{Bayesian methods for mixtures of normal distributions}.
\newblock PhD thesis, University of Oxford, 1997.

\bibitem[George and McCulloch(1993)]{george1993variable}
Edward George and Robert McCulloch.
\newblock Variable selection via {G}ibbs sampling.
\newblock \emph{Journal of the American Statistical Association}, 88\penalty0
  (423):\penalty0 881--889, 1993.

\bibitem[Xu et~al.(2020)Xu, Ge, Tebbutt, Tarek, Trapp, and
  Ghahramani]{xu2020advancedhmc}
Kai Xu, Hong Ge, Will Tebbutt, Mohamed Tarek, Martin Trapp, and Zoubin
  Ghahramani.
\newblock {AdvancedHMC.jl}: A robust, modular and efficient implementation of
  advanced {HMC} algorithms.
\newblock In \emph{Symposium on Advances in Approximate Bayesian Inference},
  2020.

\bibitem[Chopin et~al.(2024)Chopin, Crucinio, and Korba]{chopin2024a}
Nicolas Chopin, Francesca Crucinio, and Anna Korba.
\newblock A connection between tempering and entropic mirror descent.
\newblock In \emph{International Conference on Machine Learning}, 2024.

\bibitem[Syed et~al.(2024)Syed, Bouchard-C{\^o}t{\'e}, Chern, and
  Doucet]{syed2024optimised}
Saifuddin Syed, Alexandre Bouchard-C{\^o}t{\'e}, Kevin Chern, and Arnaud
  Doucet.
\newblock Optimised annealed sequential {M}onte {C}arlo samplers.
\newblock \emph{arXiv:2408.12057}, 2024.

\end{thebibliography}

\newpage

\appendix

\tableofcontents
\newpage

\section{Additional content about involutive MCMC}

\subsection{Examples of involutive MCMC} \label{apdx:example}

Here, we illustrate how 
the generic Metropolis-Hastings (MH) algorithm \citep{Tierney94,tierney1998note}, 
random-walk Metropolis-Hastings (RWMH) \citep{Hastings70,livingstone2021geometric}, 
and Hamiltonian Monte Carlo (HMC) \citep{duane1987hybrid,Neal11}, fit into this framework 
by specifying the corresponding auxiliary distribution $\rho(\cdot|x)$ and the involution map $g$.

\bexa[MH sampler; Section B.3. of \citep{Neklyudov20}] \label{example:mhasimcmc}
The Metropolis-Hastings sampler with proposal distribution $\rho(\d x'|x)$ can be cast as an involutive MCMC method
by defining the auxiliary distribution as $\rho(\d v|x)$, and using the swap involution $g: (x, v) \mapsto (v, x)$.
\eexa

\bexa[RWMH sampler; Section 2. of \citep{Liu24}] \label{example:rwmhasimcmc}
RWMH with step size $\epsilon$ is obtained by setting
\[
g(x, v) = (x + \epsilon v, -v), \quad v\sim \rho(\d v | x) =  \Norm(0, I).
\]
\eexa

\bexa[HMC; \citep{duane1987hybrid}] \label{example:hmcasimcmc}
In the involutive formulation of HMC, the auxiliary variable $v$ corresponds to the momentum variable, 
and $\rho(v|x)$ is the momentum distribution, typically a Gaussian distribution independent of $x$.
The involution map $g$ consists of applying $k$ steps of the leapfrog integrator, followed by a momentum sign flip:
\[
    g\left(\bbmat x \\ v \ebmat\right) = \bbmat I & 0 \\ 0 & -I \ebmat L^k\left(\bbmat x \\ v \ebmat\right), 
\]
where $L: (x, v) \to (x', v')$ denotes a single leapfrog step (of step size $\epsilon$) given by
\[
v_{1 / 2} & \leftarrow v+\frac{\epsilon}{2} \nabla \log \pi(x) \\
x^{\prime} & \leftarrow x+\epsilon v_{1 / 2} \\
v^{\prime} & \leftarrow v_{1 / 2}+\frac{\epsilon}{2} \nabla \log \pi\left(x^{\prime}\right) .
\]
\eexa

\subsection{Pseudocode of involutive MCMC} \label{apdx:codeimcmc}

\begin{algorithm}[H]
    \caption{Involutive MCMC kernel $K(x'| x)$} \label{alg:involutivemcmc}
    \begin{algorithmic}[1]
\Require current state $x$, target $\pi$, auxiliary distribution $\rho(\d v|x)$, involution $g$

\State $v \sim \rho(\d v|x)$                       \Comment sample auxiliary variable
\State $(x', v') \leftarrow g(x, v)$            \Comment generate proposal via the involution
\State $\alpha \leftarrow \min \left(1, \frac{\sbpi(x', v')}{\sbpi(x,v)}J_g(x, v)\right) $ \Comment compute the acceptance probability
\LineComment Accept or reject
\State $u \sim \Unif[0,1]$
\If{$u > \alpha$}                               
\State $x' \leftarrow x$                        \Comment reject
\EndIf
\State\Return $x'$
\end{algorithmic}
\end{algorithm}

\newpage
\section{Pseudocode for IRF and inverse IRF based on involutive MCMC} \label{apdx:codeirf}

\begin{algorithm}[H]
    \caption{IRF based on involutive MCMC $f_\theta(s)$} \label{alg:IFSinvlution}
    \begin{algorithmic}[1]
\Require joint state $s = (x, v, u_v, u_a)$, random parameters $\theta = (\theta_v, \theta_a)$ 

\LineComment update uniform auxiliary variables
\State $u_v \gets \left(u_v + \theta_v \right) \mod 1$
\State $u_a \gets \left(u_a + \theta_a \right) \mod 1$

\LineComment involutive MCMC with target $\pi(x)$, auxiliary distribution $\rho(\d v|x)$, involution $g$
\State $u_v' \gets F_{\rho(\cdot|x)}(v)$
\State $\stv \gets F_{\rho(\cdot|x)}^{-1}(u_v)$ 
\State $(x', v') \leftarrow g(x, \stv)$
\State $r \gets \frac{\sbpi(x', v')}{\sbpi(x,\stv)}J_g(x, \stv)$ \Comment Compute MH ratio
\If{$u_a > r$} 
\State \Return $x, \stv, u_v', u_a$ \Comment reject and return pre-involution state
\EndIf
\State $u_a' \gets \frac{u_a}{r}$  \Comment $u_a \leq r$ implies that $u_a' \in [0, 1]$
\State\Return $x', v', u_v', u_a'$  \Comment accept and return after-involution state
\end{algorithmic}
\end{algorithm}

\begin{algorithm}[H]
    \caption{Inverse IRF based on involutive MCMC $f^{-1}_\theta(s')$} \label{alg:invIFSinvlution}
    \begin{algorithmic}[1]
\Require joint state $s' = (x', v', u_v', u_a')$, random parameters $\theta = (\theta_v, \theta_a)$ 

\LineComment recover pre- and post-involution pair 
\State $(x, \stv) \gets g(x', v')$ 

\LineComment this will either be $r$ in line 6 of \cref{alg:IFSinvlution} if accepted, or $r^{-1}$ otherwise
\State $\str \gets \frac{\sbpi(x', v')}{\sbpi(x,\stv)}J_g(x, \stv)$ 

\LineComment check accept or reject
\State $u_a \gets u_a' \cdot \str$  \Comment update $u_a$ (line 10 of \cref{alg:IFSinvlution}) as if the forward pass was an accept
\If{$u_a > 1$}      \Comment forward pass was a reject (see line 6-7 of \cref{alg:IFSinvlution}) 
\LineComment pre-involution state
\State $(x, \stv) \gets x', v'$ 
\State $u_a\gets u_a'$      
\EndIf

\LineComment inverse of line 3-4 of \cref{alg:IFSinvlution}
\State $v \gets  F_{\rho(\cdot|x)}^{-1}(u_v)$ 
\State $u_v \gets F_{\rho(\cdot|x)}(\stv)$ 

\LineComment inverse update of the uniform auxiliary variables (line 1-2 of \cref{alg:IFSinvlution})
\State $u_v \gets \left(u_v + 1 -\theta_v \right) \mod 1$
\State $u_a \gets \left(u_a + 1 -\theta_a \right) \mod 1$

\State\Return $x, v, u_v, u_a$  
\end{algorithmic}
\end{algorithm}

\newpage
\section{Measure-theoretic formulation of pushforward density} \label{apdx:changeofvar}

A fundamental formula when studying variational inference is the the change of variable formula, 
which characterizes the density of a transformed distribution.
For a diffeomorphism $f:\scX \to \scX$ on a continuous space, the density of $X = f(Y), Y\sim q_0$, is given by
\[
\forall x \in \mathscr{X}, \quad q_\lambda(x)=f
q_0(x)=\frac{q_0\left(f^{-1}(x)\right)}{J\left(f^{-1}(x)\right)},
\quad J(x)=\left|\operatorname{det} \nabla f(x)\right| .
\]
However, the assumptions of differentiability and a continuous state space can
be restrictive, as many inference problems involve discrete or hybrid spaces
(e.g., Ising model \citep{onsager1944crystal}, Bayesian Gaussian mixture model
\citep{stephens1997bayesian}, and spike-and-slab model
\citep{george1993variable}). To handle general state spaces, we adopt a
measure-theoretic formulation of the pushforward density, stated in
\cref{prop:generalpushforwarddensity}, for a generic bijection $f$.
This result is well known (see, e.g., \citealp{tierney1998note} for its use in
the general involutive MCMC framework), but we include a proof here for completeness.

\bprop \label{prop:generalpushforwarddensity}
Suppose that $f:\scX \to \scX$ is bijective. For a distribution $q \ll \pi$, for all $x \in \mathscr{X}$ :
\[
\frac{\mathrm{d}(f q)}{\mathrm{d} \pi}(x)=\frac{\mathrm{d} q}{\mathrm{~d} \pi}\left(f^{-1} x\right) \frac{\mathrm{d} f \pi}{\mathrm{~d} \pi}(x) .
\]
\eprop

\bprfof{\cref{prop:generalpushforwarddensity}}
First, note that if $q \ll \pi$, then $fq \ll f\pi$. This implies that 
\[
\frac{\d (f q)}{\d \pi}(x) = \frac{\d (f q)}{\d f\pi}(x) \frac{\d f\pi}{\d \pi}(x), \quad \forall x \in \scX.
\]
It remains to show that $\frac{\d (f q)}{\d f\pi} = \frac{\d q}{\d \pi} \circ f^{-1}$. 
It suffices to show that $\forall A \in \scB$, 
\[
\int_A \frac{\d q}{\d \pi} \circ f^{-1}\d f\pi
=
\int_A \frac{\d (f q)}{\d f\pi} \d f\pi 
=
fq(A).
\]
Note that for all $A \in \scB$, we have that
\[
\int_A \frac{\d q}{\d \pi}(f^{-1}x) f\pi(\d x) = \int_{f^{-1}(A)} \frac{\d q}{\d \pi}(x) \pi(\d x) 
= q(f^{-1} A) = fq(A),
\]
which completes the proof.
\eprfof
It is worth noting that for a Euclidean space $\scX$ equipped with the Lebesgue measure $m$, and a diffeomorphism $f$, 
$\frac{\d fm}{\d m}(x)$ is precisely the Jacobian determinant $\abs{\det \nabla f^{-1}(x)}$.

If $f$ is further $\pi$-measure-preserving, then $\frac{\d f\pi}{\d \pi} = 1$, yielding a simplified expression for the pushforward density.

\bcor \label{cor:pushforwarddensitymp}
Suppose that $f$ is bijective and $\pi$-measure-preserving. For a distribution $q \ll \pi$, for all $x \in \scX$:
\[\label{eq:pushforwarddensitymp}
\frac{\d (f q)}{\d \pi}(x) = \frac{\d q}{\d \pi}(f^{-1} x).
\]
\ecor
Aside from the generality of \cref{cor:pushforwarddensitymp} over the diffeomorphic case,
it provides an elegant formula of the pushforward density under a measure-preserving map.
We invoke \cref{cor:pushforwarddensitymp} frequently when developing and analyzing MixFlows.

Beyond extending the diffeomorphic case, \cref{cor:pushforwarddensitymp} offers
an elegant expression for the pushforward density under a measure-preserving
map. We frequently invoke this result when developing and analyzing MixFlows.
Finally, we present a specialization of \cref{cor:pushforwarddensitymp} for
diffeomorphic $f$, which provides a convenient characterization of $\pi$-measure-preservation.

\bprop\label{prop:mpjacobian}
Let $f:\scX \to \scX$ be a diffeomorphism, $\pi$ be a probability distribution on $\scX$, with density (denoted by $\pi(x)$) 
with respect to a dominating measure $\lambda$.
Then, 
\benum
\item $f$ is $\pi$-measure-preserving if and only if $f^{-1}$ is $\pi$-measure-preserving.
\item
$f$ is $\pi$-measure-preserving if and only if for $\lambda$-\aev $x \in \scX$, 
$J_f(x) :=\left|\det \nabla f^{-1}(x)\right| = \frac{\pi(x)}{\pi(f^{-1}(x))}$.
\eenum
\eprop

\bprfof{\cref{prop:mpjacobian}}
By definition, $f$ is $\pi$-preserving if and only if $f \pi(x)= \pi(x) = f^{-1}\pi(x)$.
Examining the density of the pushforward $f\pi$ via the change-of-variable formula, we have
\[
    \forall x\in \scX, \quad f\pi(x) = \pi(f^{-1}(x)) J_f(x) = \pi(x) \Leftrightarrow J_f(x) = \frac{\pi(x)}{\pi(f^{-1}(x))}.
\]
The second claim follows from the fact that $\pi = (f \circ f^{-1}) \pi = (f^{-1} \circ f) \pi$.
\eprfof

\newpage
\section{Proofs} \label{apdx:proof}

\subsection{Proof of \cref{thm:randomergodic}} \label{apdx:irf}

As introduced in the main text, the IRF $f_\theta$ induces a Markov kernel given by:
\[
\forall x \in \scX, \quad \forall B \in \scB, \quad 
P(x, B) := \int_\Theta \1_{B}(f_\theta(x)) \mu(\d \theta).
\]
This yields a simple characterization of the action of the Markov process $P$ on a distribution $q$:
\[
    (Pq)(y) := \int_{\scX} P(x, y) q(\d x) = \E\left[ f_\theta q(y) \right], 
    \quad
    \theta \sim \mu, 
    \quad  
    \text{$f_\theta q$: pushforward of $q$ under $f_\theta$}.
\]
We can further characterize the Markov kernel $R(\cdot, \cdot)$ induced by the \emph{inverse IRF} 
$f_{\theta_t}^{-1}$:
\[\label{eq:markovkernelinvirf}
\forall x \in \scX, \quad \forall A \in \scB, \quad 
R(x, A) := \int_\Theta \1_{A}(f^{-1}_\theta(x)) \mu(\d \theta).
\]
which is precisely the \emph{reversal} of $P(\cdot, \cdot)$:
\[\label{eq:Qab}
    \pi \otimes P (A \times B) = \pi \otimes R(B \times A) = \int \pi(f_\theta(A) \cap B) \mu(\d \theta), 
\]
where $\pi \otimes P (A \times B) := \int_A P(x, B) \pi(\d x)$. 
See \citet[Eq. (4.5)]{kakutani1950random} for the detailed derivation.
Notice that if $P$ is reversible wrt $\pi$, i.e., $\pi \otimes P = \pi \otimes R$,
both the IRF $f_\theta$ and its inverse $f_\theta^{-1}$ induce the same Markov process $P$. 
In other words, $P = R$.
From \cref{eq:Qab}, we can see that a sufficient and necessary condition so that 
$P = R$ is that 
\[
\int \pi(f_\theta(A) \cap B) \mu(\d \theta) =  \int \pi(f_\theta^{-1}(A) \cap B) \mu(\d \theta).
\]

\bprfof{\cref{thm:randomergodic}}
From \cref{eq:irfpushforward}, we see that $P$ must admit $\pi$ as a stationary distribution.
\citet[Theorem 5.2.6]{douc2018markov} further states that if $\pi$ is the unique invariant probability measure of $P$, 
then the Markov process $P$ is ergodic.
Therefore, the LLM of ergodic Markov process \citep[Theorem 5.29]{douc2018markov} guarantees \cref{eq:LLN}, 
and the random ergodic theorem \citep[Cor. 2.2.]{kifer2012ergodic} ensures \cref{eq:randomergodic}.

Then as discussed above, \citet[Theorem 3.]{kakutani1950random} show that \cref{assump:niceirf} holds for $f_\theta$
and its induced Markov process $P$ if and only if \cref{assump:niceirf} holds for the inverse IRF $f_\theta^{-1}$ and its induced $R$.
Therefore, the same convergence holds for the inverse IRF.
\eprfof

\subsection{Convergence of the homogeneous MixFlow} \label{apdx:homomixflowconv}

\bdefn[Ergodic map {\citep[pp.~73, 105]{Eisner15}}]\label{def:ergodic}
$f :\scX\to\scX$ is \emph{ergodic} for $\pi$ if for all measurable sets 
$A\subseteq \scX$, $f(A) = A$ implies that $\pi(A) \in \{0,1\}$.  
\edefn

The most notable implication of a $\pi$-e.m.p $f$ is that the long-run average of
repeated applications of $f$ converges to the expectation under $\pi$, a result
known as the Birkhoff ergodic theorem [\citealp{Birkhoff31};
\citealp[p.~212]{Eisner15}]. The full statement is given in
\cref{thm:pointwiseergodic}.

\bthm[{Ergodic Theorem [\citealp{Birkhoff31}; \citealp[p.~212]{Eisner15}]}]\label{thm:pointwiseergodic}
Suppose $f : \scX \to \scX$ is measure-preserving and ergodic for $\pi$,
and $\phi \in L^1(\pi)$. Then 
\[
\lim_{T\to\infty} \frac{1}{T}\sum_{t=1}^{T} \phi(f^tx) = \int \phi\d \pi, 
\qquad \text{$\pi$-\aev $x\in\scX$}.
\]
\ethm

\blem[Scheffé's Lemma] \label{lemma:scheffe}
Let $\phi_n$ be a sequence of integrable functions on a measure space $(\scX, \scB, \pi)$ that convergences $\pi$-a.s. to $\phi$. Then
\[
    \int |\phi_n(x) - \phi(x)| \pi(\d x) \to 0, \quad n \to \infty,
\]
if and only if 
\[
    \int |\phi_n(x)| \pi(\d x)  \to \int |\phi(x)| \pi(\d x), \quad n \to \infty.
\]
\elem

\bprfof{\cref{thm:convergenceofmixflow}}

Note that the Jacobian of the $\pi$-e.m.p $f$ is $\pi(x)/\pi(f^{-1}(x))$ by \cref{prop:mpjacobian}, allowing the
density of $\bar{q}_T$ to be expressed as:
\[
    \sbq_T(x) = \frac{1}{T}\sum_{t=1}^{T} f^t q_0(x) 
    = 
    \pi(x)\cdot \frac{1}{T}\sum_{t=1}^{T} \frac{q_0}{\pi}(f^{-t}(x)), 
    \quad \forall x\in \scX.
\]
The pointwise density convergence is the direct consequence of \cref{eq:convofdensityratio}.
Specifically, provided $q_0 \ll \pi$, we have $q_0/\pi \in L^1(\pi)$, so the
Birkhoff ergodic theorem [\citealp{Birkhoff31}; \citealp[p.~212]{Eisner15}]
(see \cref{thm:pointwiseergodic}) ensures:
\[\label{eq:convofdensityratio}
    \frac{1}{T}\sum_{t=1}^{T} \frac{q_0}{\pi}(f^{-t}(x)) \to 1, 
    \qquad \pi-\text{\aev } \, x\in\scX, 
    \qquad \text{as } T\to \infty.
\]

The total variation convergence is then by the direct application of the \emph{Scheffé's lemma} \cref{lemma:scheffe}. 
Notice that 
\[
    \TV(\sbq_T, \pi) = \int \left|\frac{\sbq_T}{\pi}(x) - 1\right|\pi(\d x) = \int \left|\frac{1}{T}\sum_{t=1}^T \frac{q_0}{\pi}(f^{-t}(x)) - 1\right| \pi(\d x).
\]
To apply \cref{lemma:scheffe}, we set $\phi_t(x) := \frac{1}{T}\sum_{t=1}^T \frac{q_0}{\pi}(f_\theta^{-t}(x))$, 
and set $\phi(x) := 1$. Because $q_0 \ll \pi$, all $\phi_n$'s are $\pi$-integrable. Then, for all $n \in \nats$,
we obtain that
\[
    \int |\phi_n(x)| \pi(\d x) 
    &=  \int \phi_n(x) \pi(\d x) \\
    &= \frac{1}{T} \sum_{t=1}^T \int \frac{q_0}{\pi}(f_\theta^{-t} x) \pi(\d x) \\
    &= \int q_0(\d x)
    \quad (\text{as } f_\theta \pi = \pi) \\
    & = 1 = \int |\phi(x)| \pi(\d x),
\] 
yielding the second convergence in \cref{lemma:scheffe}.
\eprfof

\subsection{Convergence of the IRF MixFlow} \label{apdx:irfflowconv}

The proof of \cref{thm:tvconvergence} proceeds by first establishing the
convergence in expectation
\[\label{eq:irfexpectedtv}
\E\left[\TV\left(  \overrightarrow{q_T}, \pi\right)\right] \longrightarrow 0 
\quad
\text{ as } \,T \to \infty,
\]
where 
the expectation is taken over the cached \iid stream $(\theta_t)_{t\in\nats}\sim\mathbb{P}$;
convergence in probability then follows from Markov's inequality.

For notational simplicity, besides the kernels $P$ (\cref{eq:markovkernel}) and $R$ (\cref{eq:markovkernelinvirf}), 
we also use $P$ and $R$ to denote the corresponding operators acting on bounded measurable functions:
\[
    (Pu)(x) := \int_\Theta u(f_\theta(x)) \mu(\d\theta) = \int u(y) P(x, \d y), 
    \qquad 
    (Ru)(x) := \int_\Theta u(f_\theta^{-1}(x)) \mu(\d\theta) = \int u(y) R(x, \d y).
\]
Since each $f_\theta$ is $\pi$-m.p., $P$ admits $\pi$ as a stationary distribution:
for all $A \in \scB$, $\int P(x, A) \pi(\d x) = \int_\Theta \pi(f_\theta^{-1}(A)) \mu(\d\theta) = \pi(A)$, and similarly for $R$. 
Moreover, $R$ is the $L^2(\pi)$-adjoint operator of $P$: for bounded measurable $\phi, \psi$,
we have
\[\label{eq:PRduality}
    \langle \phi, P\psi \rangle_\pi
    = \int_\Theta\int_{\scX} \phi(x)\psi(f_\theta(x)) \pi(\d x)\mu(\d\theta)
    = \int_\Theta\int_{\scX} \phi(f_\theta^{-1}(y))\psi(y) \pi(\d y)\mu(\d\theta)
    = \langle R\phi, \psi \rangle_\pi.
\]
Iterating forward yields that, for all $k \in \nats$, $\langle \phi, P^k\psi \rangle_\pi = \langle R^k\phi, \psi \rangle_\pi$. 
Note that this is the operator form of \cref{eq:Qab}.

The proof relies on a Ces\`aro-averaged mixing property of $P$.
For $m \in \nats$, define the Ces\`aro-averaged kernel and its averaged total variation distance to $\pi$:
\[\label{eq:betam}
    \overline{P}_m(x, \cdot) := \frac{1}{m}\sum_{k=1}^m P^k(x, \cdot),
    \qquad 
    \beta_m := \int_{\scX} \TV\left(\overline{P}_m(x, \cdot), \pi\right) \pi(\d x).
\]
The Ces\`aro average is essential here: if $P$ is periodic, $\TV(P^m(x,\cdot),\pi)$ need not vanish as $m \to \infty$, 
whereas its Ces\`aro average does (\cref{prop:cesaromixing}). 
This is why \cref{thm:tvconvergence} does not require the aperiodicity that is typical of MCMC convergence theory.

\bprop \label{prop:cesaromixing}
Suppose \cref{assump:niceirf} holds. Then $\beta_m \in [0, 1]$ 
for all $m \in \nats$, and 
$\beta_m \to 0$ as $m \to \infty$.
\eprop

\bprfof{\cref{prop:cesaromixing}}

As shown above, $\pi$ is a stationary distribution of $P$, and $P$ is $\pi$-irreducible by \cref{assump:niceirf}. 
Let $d \geq 1$ denote the period of $P$, and for $x \in \scX$ and $n \in \nats$, 
let $W_n(x, \cdot) := \frac{1}{d}\sum_{k=n}^{n+d-1}P^k(x, \cdot)$ be the average of $d$ consecutive transition kernels starting from step $n$.
By \citet[Theorem~4]{Roberts04} when $d = 1$ (in which case $W_n(x, \cdot) = P^n(x, \cdot)$) 
and \citet[Corollary~6]{Roberts04} when $d \geq 2$, for $\pi$-\aev $x \in \scX$,
\[\label{eq:windowconv}
    \TV\left(W_n(x, \cdot), \pi\right) \to 0, \qquad n \to \infty.
\]

Now fix $x$ such that \cref{eq:windowconv} holds, and fix $m \geq d$. 
Write $m = Jd + r$ with $J := \lfloor m/d \rfloor \geq 1$ and $0 \leq r < d$. 
We partition the steps $\{1, \dots, m\}$ into $J$ consecutive blocks of length $d$, 
namely $\{jd+1, \dots, (j+1)d\}$ for $j = 0, \dots, J-1$, followed by the $r$ remaining steps $\{Jd+1, \dots, m\}$. 
The average of $P^k(x, \cdot)$ over the $j$-th block is $W_{jd+1}(x, \cdot)$, so
\[
    \sum_{k=1}^m P^k(x, \cdot) = d\sum_{j=0}^{J-1} W_{jd+1}(x, \cdot) + \sum_{k=Jd+1}^{m} P^k(x, \cdot),
\]
where the last sum is empty if $r = 0$. 
Dividing by $m$ and subtracting $\pi = \frac{Jd}{m}\pi + \frac{r}{m}\pi$ gives
\[
    \overline{P}_m(x, \cdot) - \pi 
    = \frac{d}{m}\sum_{j=0}^{J-1}\left(W_{jd+1}(x, \cdot) - \pi\right) + \frac{1}{m}\sum_{k=Jd+1}^{m} \left(P^{k}(x, \cdot) - \pi\right).
\]
Evaluating both sides at a set $A \in \scB$, applying the triangle inequality, taking the supremum over $A$, 
and using $\TV(\cdot, \pi) \le 1$ for the $r$ remainder terms, we obtain
\[
    \TV\left(\overline{P}_m(x, \cdot), \pi\right) 
    \leq \frac{Jd}{m}\cdot \frac{1}{J}\sum_{j=0}^{J-1}\TV\left(W_{jd+1}(x, \cdot), \pi\right) + \frac{r}{m}
    \leq \frac{1}{J}\sum_{j=0}^{J-1}\TV\left(W_{jd+1}(x, \cdot), \pi\right) + \frac{d}{m},
\]
where the second inequality uses $Jd \leq m$ and $r < d$.
By \cref{eq:windowconv}, the sequence $a_j := \TV(W_{jd+1}(x, \cdot), \pi)$ converges to $0$ as $j \to \infty$, 
and hence so do its Ces\`aro means $\frac{1}{J}\sum_{j=0}^{J-1} a_j$.
Since $J = \lfloor m/d \rfloor \to \infty$ and $d/m \to 0$ as $m \to \infty$, both terms on the right-hand side vanish as $m \to \infty$. 
Hence $\TV(\overline{P}_m(x, \cdot), \pi) \to 0$ for $\pi$-\aev $x$.

Since $0 \leq \TV(\overline{P}_m(x, \cdot), \pi) \leq 1$, the dominated convergence theorem yields $\beta_m \to 0$.
\eprfof

We now prove \cref{thm:tvconvergence}. Define
\[
    g := \frac{\d q_0}{\d \pi} - 1,
    \qquad 
    F_t := f_{\theta_t}\circ \cdots \circ f_{\theta_1},
    \qquad 
    F_t^{-1} = f_{\theta_1}^{-1}\circ \cdots \circ f_{\theta_t}^{-1},
\]
and, for a measurable function $u$, the averaging operator (which depends on the realized stream $(\theta_t)_{t\in \nats}$)
\[\label{eq:ATdef}
    A_T u := \frac{1}{T}\sum_{t=1}^T u \circ F_t^{-1}.
\]
Since $q_0 \ll \pi$ is a probability measure, $g \in L^1(\pi)$ and $\int g \d\pi = 0$.
Since each $F_t$ is a $\pi$-m.p.\ bijection, so is $F_t^{-1}$; hence $u\circ F_t^{-1}$ has the same distribution under $\pi$ as $u$. 
In particular, for every $u \in L^p(\pi)$, $p \in \{1, 2\}$,
\[\label{eq:normpreserve}
    \|u\circ F_t^{-1}\|_p = \|u\|_p, 
    \qquad 
    \int u\circ F_t^{-1} \d \pi = \int u \d\pi.
\]
Consequently, $A_T$ is nonexpansive on $L^1(\pi)$ for every stream: 
$\|A_T u\|_1 \leq \frac{1}{T}\sum_{t=1}^T\|u\circ F_t^{-1}\|_1 = \|u\|_1$.

\bprfof{\cref{thm:tvconvergence}}

By \cref{cor:pushforwarddensitymp}, $\frac{\d (F_t q_0)}{\d \pi} = \frac{\d q_0}{\d \pi}\circ F_t^{-1} = 1 + g \circ F_t^{-1}$. 
Therefore, the density of $\overrightarrow{q_T}$ can be expressed as:
\[
    \frac{\d \overrightarrow{q_T}}{\d \pi} 
    = \frac{1}{T}\sum_{t=1}^T \left(1 + g\circ F_t^{-1}\right)
    = 1 + A_T g.
\]
By the definition of TV distance between two probability measures, we have that 
\[\label{eq:tvidentityirf}
    \TV\left(\overrightarrow{q_T}, \pi\right) 
    = \frac{1}{2}\int_{\scX} \left| \frac{\d \overrightarrow{q_T}}{\d \pi}(x) -1 \right|\pi(\d x)
    = \frac{1}{2}\int_{\scX} \left|\frac{1}{T}\sum_{t=1}^T g\left(F_t^{-1}(x)\right)\right|\pi(\d x)
    = \frac{1}{2}\|A_T g\|_1.
\]

As outlined at the beginning of this section, it suffices to show that 
$\E\left[ \TV\left(\overrightarrow{q_T}, \pi\right) \right] = \frac{1}{2}\E\|A_T g\|_1 \to 0$ as $T\to \infty$. 
To achieve this, we split $g$ into a bounded part $g_b$ and a tail part $g - g_b$, and control them separately. 
The bounded part is controlled in $L^2(\pi)$: we show that $\E\|A_T g_b\|_2^2 \to 0$ as $T \to \infty$ provided $b$ grows sufficiently slowly with $T$, which suffices since
\[
    \E\|A_T g_b\|_1 \leq \sqrt{\E\|A_T g_b\|_2^2}.
\]
The tail part is controlled in $L^1(\pi)$: its contribution is bounded by $\|g - g_b\|_1$ uniformly in $T$, which vanishes as $b \to \infty$.

Now we proceed with the concrete derivation.
For $b > 0$, define the clipped and recentered function
\[
    \tilde{g}_b := \max\{-b, \min\{g, b\}\}, 
    \qquad 
    g_b := \tilde{g}_b - \int \tilde{g}_b \d \pi.
\]
Note that $\int g_b \d\pi = 0$, and $|g_b| \leq 2b$ pointwise.
Then for every $b > 0$ and $T \in \nats$,
by linearity and $L^1$-nonexpansiveness of $A_T$ (i.e., $\|A_T u\|_1 \leq \|u\|_1$), we have
\[
    \|A_T g\|_1 \leq \|A_T(g - g_b)\|_1 + \|A_T g_b\|_1 \leq \|g - g_b\|_1 + \|A_T g_b\|_1,
\]
which, combined with \cref{eq:tvidentityirf}, yields
\[
    \E\left[\TV\left(\overrightarrow{q_T}, \pi\right)\right] 
    & \leq \frac{1}{2}\|g - g_b\|_1 + \frac{1}{2} \E\|A_T g_b\|_1
    \\
    & \leq \frac{1}{2}\|g - g_b\|_1 + \frac{1}{2}\sqrt{\E\|A_T g_b\|^2_2}.
\label{eq:irfstep1}
\]
The last inequality follows from Jensen's inequality, applied first under $\pi$ and then under $\mathbb{P}$.

To bound the tail $\|g - g_b\|_1$, note that $\int g \d\pi = 0$ implies that $\int \tilde g_b \d\pi = -\int (g - \tilde g_b)\d\pi$, so
\[
    g - g_b = (g - \tilde{g}_b) - \int (g - \tilde{g}_b)\d\pi.
\]
By the triangle inequality, 
\[\label{eq:clipirf}
    \|g - g_b\|_1 
    \leq 2\|g - \tilde g_b\|_1 
    = 2\int (|g| - b)_+ \d\pi
    \to 0, 
    \qquad \text{as } b \to \infty,
\]
where the equality holds because $|g - \tilde g_b| = (|g| - b)_+$ pointwise, 
and the limit follows from dominated convergence, since $(|g| - b)_+ \leq |g| \in L^1(\pi)$ and $(|g| - b)_+ \to 0$ pointwise.

It remains to bound the last term in \cref{eq:irfstep1}.
For the rest of the proof, define $v_t := g_b\circ F_t^{-1}$. 
By \cref{eq:normpreserve}, $\int v_t \d \pi = 0$ and $\|v_t\|_2 = \|g_b\|_2$; 
moreover, $|v_t(x)| = |g_b(F_t^{-1}(x))| \leq 2b$ for every $x \in \scX$.
Expanding the square,
\[
    \E\|A_T g_b\|_2^2 
    & = 
    \E\left\|\frac{1}{T}\sum_{t=1}^T g_b\circ F_t^{-1}\right\|_2^2
    \\
    & = 
    \frac{1}{T^2}\sum_{i=1}^T\sum_{j=1}^T \E\langle v_i, v_j\rangle_\pi
    \\
    & = \frac{\|g_b\|_2^2}{T} + \frac{2}{T^2}\sum_{i=1}^{T-1}\sum_{k=1}^{T-i}\E\langle v_i, v_{i+k}\rangle_\pi.
\label{eq:l2expansion}
\]
We now compute the cross terms. Let $\scF_i := \sigma(\theta_1, \dots, \theta_i)$. 
Since $F_{i+1}^{-1} = F_i^{-1}\circ f_{\theta_{i+1}}^{-1}$, we have $v_{i+1}(x) = v_i(f^{-1}_{\theta_{i+1}}(x))$, 
where $v_i$ is $\scF_i$-measurable and $\theta_{i+1}\sim\mu$ is independent of $\scF_i$. Hence
\[
    \E\left[v_{i+1}(x)\mid \scF_i\right] = \int_\Theta v_i(f_\theta^{-1}(x))\mu(\d\theta) = (R v_i)(x),
\]
and iterating with the tower property gives $\E[v_{i+k}(x) \mid \scF_i] = (R^k v_i)(x)$ for all $k\in\nats$.
By the conditional Fubini theorem and the duality \cref{eq:PRduality},
\[
    \E\langle v_i, v_{i+k}\rangle_\pi 
    = \E\int_{\scX} v_i(x)\,\E\left[v_{i+k}(x) \mid \scF_i\right]\pi(\d x)
    = \E\langle v_i, R^k v_i\rangle_\pi
    = \E\langle P^k v_i, v_i\rangle_\pi.
\]
For any $i \in \{1, \dots, T-1\}$, by the definition of $\overline{P}_m$ in \cref{eq:betam}, 
the inner sum of \cref{eq:l2expansion} can be rewritten as:
\[
    \sum_{k=1}^{T-i}\E\langle v_i, v_{i+k}\rangle_\pi 
    = \E\sum_{k=1}^{T-i}\langle P^k v_i, v_i\rangle_\pi 
    = (T-i)\,\E\int_{\scX} v_i(x)\,(\overline{P}_{T-i} v_i)(x)\,\pi(\d x),
\]
where $(\overline{P}_{T-i} v_i)(x) := \int v_i(y)\overline{P}_{T-i}(x, \d y)$.

Since $\int v_i \d\pi = 0$ and $|v_i| \le 2b$ pointwise, for every $x \in \scX$ and every realization of the stream,
\[
    \left|(\overline{P}_{T-i} v_i)(x)\right| 
    = \left|\int_{\scX} v_i(y)\left(\overline{P}_{T-i}(x, \d y) - \pi(\d y)\right)\right| 
    \leq 4b\,\TV\left(\overline{P}_{T-i}(x, \cdot), \pi\right).
\]
Multiplying by $|v_i(x)| \leq 2b$ and integrating against $\pi$ yields
\[
    \left|\sum_{k=1}^{T-i}\E\langle v_i, v_{i+k}\rangle_\pi\right| \leq 8b^2 (T-i)\, \beta_{T-i},
\]
where $\beta_{T-i} := \int_{\scX} \TV\left(\overline{P}_{T-i}(x, \cdot), \pi\right) \pi(\d x)$ as defined in \cref{eq:betam}.
Substituting this bound into \cref{eq:l2expansion} and reindexing with $m = T - i$,
\[\label{eq:l2bound}
    \E\|A_T g_b\|_2^2 
    \leq \frac{\|g_b\|_2^2}{T} + \frac{16b^2}{T^2}\sum_{i=1}^{T-1}(T-i)\,\beta_{T-i}
    = \frac{\|g_b\|_2^2}{T} + \frac{16b^2}{T^2}\sum_{m=1}^{T-1}m\,\beta_m.
\]
Since $\|g_b\|_2^2 \leq 4b^2$, \cref{eq:l2bound} yields that for every $b > 0$ and $T \in \nats$,
\[\label{eq:l2boundfinal}
    \E\|A_T g_b\|_2^2 \leq 16 b^2 \delta_T,
    \qquad 
    \delta_T := \frac{1}{T} + \frac{1}{T^2}\sum_{m=1}^{T-1} m\,\beta_m.
\]
Moreover, $\delta_T \to 0$ as $T \to \infty$, 
since $\frac{1}{T^2}\sum_{m=1}^{T-1} m\,\beta_m \leq \frac{1}{T}\sum_{m=1}^{T-1}\beta_m$ 
and $\beta_m \to 0$ as $m\to \infty$ (\cref{prop:cesaromixing}).

We now conclude the proof. 
Since \cref{eq:irfstep1} holds for arbitrary $b > 0$ and $T \in \nats$, 
we may choose $b = b(T)$ depending on $T$ such that $b(T) \to\infty$ as $T\to \infty$.
By \cref{eq:clipirf}, $\|g - g_{b(T)}\|_1 \to 0$ as $T \to \infty$ (since $b(T) \to \infty$); 
by \cref{eq:l2boundfinal}, $\E\|A_T g_{b(T)}\|_2^2 \to 0$ as $T \to \infty$ as long as $b(T)^2 \delta_T \to 0$. 
Note that both conditions hold for $b(T) := \delta_T^{-1/4}$, since $\delta_T \to 0$ and $b(T)^2 \delta_T = \delta_T^{1/2} \to 0$. 
With this choice, both terms on the right-hand side of \cref{eq:irfstep1} vanish as $T \to \infty$, which proves \cref{eq:irfexpectedtv}. 
The convergence in probability \cref{eq:condconv} then follows from Markov's inequality.
\eprfof

\subsection{Convergence of the backward IRF MixFlow} \label{apdx:bwdirfflowconv}

\bprfof{\cref{thm:bwdirfconvergence}}
The pointwise density convergence is the direct consequence of
\cref{eq:densitybwdirf} via \cref{thm:randomergodic}.
The total variation convergence is then established using identical strategy as
the proof of \cref{thm:convergenceofmixflow}
via Scheffé's lemma \cref{lemma:scheffe}.
\eprfof

\subsection{Convergence of the ensemble IRF MixFlow} \label{apdx:ensembleflow}

\bprfof{\cref{thm:tvensemble}}
By the definition of the total variation, 
\[
    \TV\left(\stq^{(M)}_T, \pi\right) 
    &= \int \left| \frac{\stq^{(M)}_T}{\pi}(x) - 1\right| \pi(\d x) \\
    &= \int \left| \frac{1}{M}\sum_{m=1}^M \frac{q_0}{\pi}\left(f_{\theta_1^{(m)}}^{-1} \circ \cdots \circ f_{\theta_T^{(m)}}^{-1}(x)\right) - 1\right| \pi(\d x).
\]
By the triangle inequality,
\[\label{eq:tvensemble_triangle_ineq}
    \leq 
    \int 
    \left| \frac{1}{M}\sum_{m=1}^M \frac{q_0}{\pi}\left(f_{\theta_1^{(m)}}^{-1} \circ \cdots \circ f_{\theta_T^{(m)}}^{-1}(x)\right) 
    - R^T\left(\frac{q_0}{\pi}\right)(x)\right| 
    \pi(\d x) 
    + 
    \int \left| \int \frac{q_0}{\pi}(y) R^T \delta_x(\d y) - 1\right| \pi(\d x).
\]
We derive upper bounds for two terms on the right-hand side separately.

For the first term, taking the expectation with respect to the randomness of $\theta \sim \mu$, and interchange the order of integrations,
\[
    &
    \E
    \left[
    \int 
    \left| \frac{1}{M}\sum_{m=1}^M \frac{q_0}{\pi}\left(f_{\theta_1^{(m)}}^{-1} \circ \cdots \circ f_{\theta_T^{(m)}}^{-1}(x)\right) 
    - R^T\left(\frac{q_0}{\pi}\right)(x)\right| 
    \pi(\d x) 
    \right]
    \\ 
    &
    = \E
    \left[
    \int 
    \left| \frac{1}{M}\sum_{m=1}^M \frac{q_0}{\pi}\left(f_{\theta_1^{(m)}}^{-1} \circ \cdots \circ f_{\theta_T^{(m)}}^{-1}(X)\right) 
    - R^T\left(\frac{q_0}{\pi}\right)(X)\right| 
    \mu \left(\d \theta_{1:T}^{(1:M)}\right) 
    \right], 
    \quad X \sim \pi
\]
Notice that $\forall x \in \scX, \quad \left\{f_{\theta_1^{(m)}}^{-1} \circ \cdots \circ f_{\theta_T^{(m)}}^{-1}(x)\right\}_{m=1}^M \distiid R^T \delta_x$, 
where the randomness comes from the independent realization of $\theta$s, where $R$ is the induced the Markov process of $f^{-1}_\theta$.
Therefore, applying Jensen's inequality yields
\[
    \leq
    \frac{1}{\sqrt{M}} 
    \E\left[
        \sqrt{ 
            \Var_{\theta_{1:T}}
           \left[
               \frac{q_0}{\pi}\left(f_{\theta_1}^{-1} \circ \cdots \circ f_{\theta_T}^{-1}(X)\right) \mid X
           \right]
        }
    \right],
\]

For the second term of \cref{eq:tvensemble_triangle_ineq}, 
since $\frac{q_0}{\pi}$ is globally bounded by constant $B < \infty$, we have that
\[
    &\int \left| \int \frac{q_0}{\pi}(y) R^T\delta_x(\d y) - 1\right| \pi(\d x)  \\
    &= \int \left| \int \frac{q_0}{\pi}(y) R^T\delta_x(\d y) - \int \frac{q_0}{\pi}(y) \pi(\d y)\right| \pi(\d x)  \\
    & \leq B \int \TV(R^T\delta_x, \pi) \pi(\d x) \\
    & = B\cdot \E\left[\TV(R^T \delta_X, \pi)\right], \quad X \sim \pi.
\]
This completes the proof.

\eprfof

\subsection{Proof of \cref{prop:measurepreserving}}

\bprfof{\cref{prop:measurepreserving}}

We first verify that the map defined in \cref{alg:IFSinvlution} is
$\bar{\pi}$-measure-preserving, invoking the second part of
\cref{prop:measurepreserving}. The algorithm has four steps (see
\cref{sec:constructirf}); we compute the Jacobian of each step.
Steps 3-4 involve a discrete accept/reject decision, so we treat the two
branches separately---within a branch the transformation is a diffeomorphism,
making the Jacobian well defined.

\benum
\item Step 1 describes constant shifts applied to uniform random variables, which preserves $\Unif_{[0, 1]}(\d u_v)$ and $\Unif_{[0, 1]}(\d u_a)$ with Jacobian $1$.
\item Step 2 is the CDF/inverse-CDF transformation of $\rho(\cdot|x)$. As long as the CDF $F(\cdot|x)$ is well-defined, this step describes a diffeomorphism 
    in $\scV \times [0, 1]$.
    The corresponding Jacobian is given by:
    \[
        \frac{\rho(\stv|x)}{\rho(v|x)} 
    \]
\item We analyze step 3 and 4 together. 
    In the rejection branch, 
    no additional transformation is applied, so the Jacobian is $1$.
    In the acceptance branch, step 3 involves the involution mapping, with Jacobian $\abs{\frac{\partial g(x, \stv)}{\partial x, \stv}}^{-1}$, 
    and step 4 rescale $u_a$ by the MH-ratio $r$, yielding a combined Jacobian with step 3 $\frac{\sbpi(x', v')}{\sbpi(x,\stv)}$.
\eenum
Hence, in the rejection branch, the combined jacobian of step 1-4 evaluated on $s' = (x, \stv, u_v', u_a)$ is 
\[
    \frac{\rho(\stv|x)}{\rho(v|x)} = \frac{\sbpi(x, \stv, u_v', u_a)}{\sbpi(x, v, u_v, u_a)}.
\]
In the acceptance branch, the combined jacobian of step 1-4 evaluated on $s' = (x', v', u_v', u_a')$ is 
\[
    \frac{\rho(\stv|x)}{\rho(v|x)} \frac{\sbpi(x', v')}{\sbpi(x,\stv)} = \frac{\sbpi(x', v', u_v', u_a')}{\sbpi(x, v, u_v, u_a)}.
\]
Both satisfy the criterion of \cref{prop:measurepreserving}; the map is therefore $\sbpi$-measure-preserving.

Finally, we show uniqueness of the invariant distribution. 
By \citet[Corollary 9.2.16]{douc2018markov}, an irreducible kernel has at most one invariant distribution. 
Because each $f_\theta$ preserves $\bar{\pi}$, the induced Markov kernel $P$ must admit $\sbpi$ as an invariant distribution.
If $P$ is irreducible, then $\sbpi$ is its unique invariant distribution. 

\eprfof

\newpage
\section{Additional experimental details} \label{apdx:expt}

For all homogeneous MixFlows variants, the uniform-shift parameters were fixed to 
$\theta_v = \pi/8$ and $\theta_a = \pi/7$.
For \texttt{NUTS} benchmarks, we use the Julia package \texttt{AdvancedHMC.jl} 
\citep{xu2020advancedhmc} with default settings throughout. 
The normalizing flow architectures were implemented as follows. In
\texttt{RealNVP}, the affine coupling layers consist of two separate multilayer
perceptrons (MLPs)---one for scaling and one for shifting---each with three
fully connected layers and LeakyReLU activations.
For \texttt{Neural Spline Flows (NSF)}, we set the spline bandwidth to $B = 30$, and used $K = 11$ knots.
For synthetic examples, the hidden dimension in each MLP was set to 32 for \texttt{RealNVP} and 64 for \texttt{NSF}.
For real-data examples, the hidden dimension was set to $\min(d, 64)$, where 
$d$ is the dimensionality of the target posterior distribution.

Experiments are conducted on the following platforms: a local machine equipped
with an AMD Ryzen 9 5900X CPU and 64 GB of RAM, 
the ARC Sockeye computing platform at the University of British Columbia, and the high-performance 
compute cluster provided by the Digital Research Alliance of Canada.
Code for reproducing the main experimental results is available at: \url{https://github.com/zuhengxu/MixFlow.jl.git}.

\subsection{Synthetic experiments} \label{apdx:synthetic_expt}

The four target distributions used in this experiment are as follows:
\benum
\item  the banana distribution \citep{haario2001adaptive}:
\[
y = \begin{bmatrix}y_1 \\ y_2\end{bmatrix} \sim 
\Norm\left( 0, \begin{bmatrix}100 & 0 \\ 0 & 1\end{bmatrix} \right), \quad 
x = \begin{bmatrix} y_1 \\ y_2 + by_1^2 - 100b \end{bmatrix}, \quad b = 0.1;
\]
\item  Neals' funnel \citep{neal2003slice}:
\[
x_1 \sim \Norm\left( 0, \sigma^2 \right), \quad 
x_2 \mid x_1 \sim \Norm\left( 0, \exp\left(\frac{x_1}{2}\right) \right), \quad \sigma^2 = 36;
\]
\item  a cross-shaped distribution: in particular, a Gaussian mixture of the form
\[
x &\sim \frac{1}{4}\Norm\left( \begin{bmatrix} 0 \\ 2 \end{bmatrix}, 
				\begin{bmatrix} 0.15^2 & 0 \\ 0 & 1 \end{bmatrix} \right) + 
	\frac{1}{4}\Norm\left( \begin{bmatrix} -2 \\ 0 \end{bmatrix}, 
				\begin{bmatrix} 1 & 0 \\ 0 & 0.15^2  \end{bmatrix} \right)\\ 
	&+ \frac{1}{4}\Norm\left( \begin{bmatrix} 2 \\ 0 \end{bmatrix}, 
	\begin{bmatrix} 1 & 0 \\ 0 & 0.15^2  \end{bmatrix} \right) + 
	\frac{1}{4}\Norm\left( \begin{bmatrix} 0 \\ -2 \end{bmatrix}, 
	\begin{bmatrix} 0.15^2 & 0 \\ 0 & 1 \end{bmatrix} \right);
\]
\item  and a warped Gaussian distribution 
\[
y = \begin{bmatrix}y_1 \\ y_2\end{bmatrix} \sim 
\Norm\left( 0, \begin{bmatrix} 1 & 0 \\ 0 & 0.12^2 \end{bmatrix} \right), 
\quad x = \begin{bmatrix} \|y\|_2\cos\left( \mathrm{atan2}\left( y_2, y_1 \right) - \frac{1}{2}\|y\|_2 \right) \\ 
\|y\|_2\sin\left( \mathrm{atan2}\left( y_2, y_1 \right) - \frac{1}{2}\|y\|_2 \right) \end{bmatrix},
\]
where $\mathrm{atan2}(y,x)$ is the angle, in radians, between the positive $x$ axis
and the ray to the point $(x,y)$.
\eenum

\newpage
\subsubsection{Relative performance of homogeneous, IRF, and backward IRF MixFlows} \label{apdx:comparemixflows}

\begin{figure}[H]
    \begin{subfigure}{\columnwidth}
		\includegraphics[width=0.24\columnwidth, trim=200 150 150 150, clip]{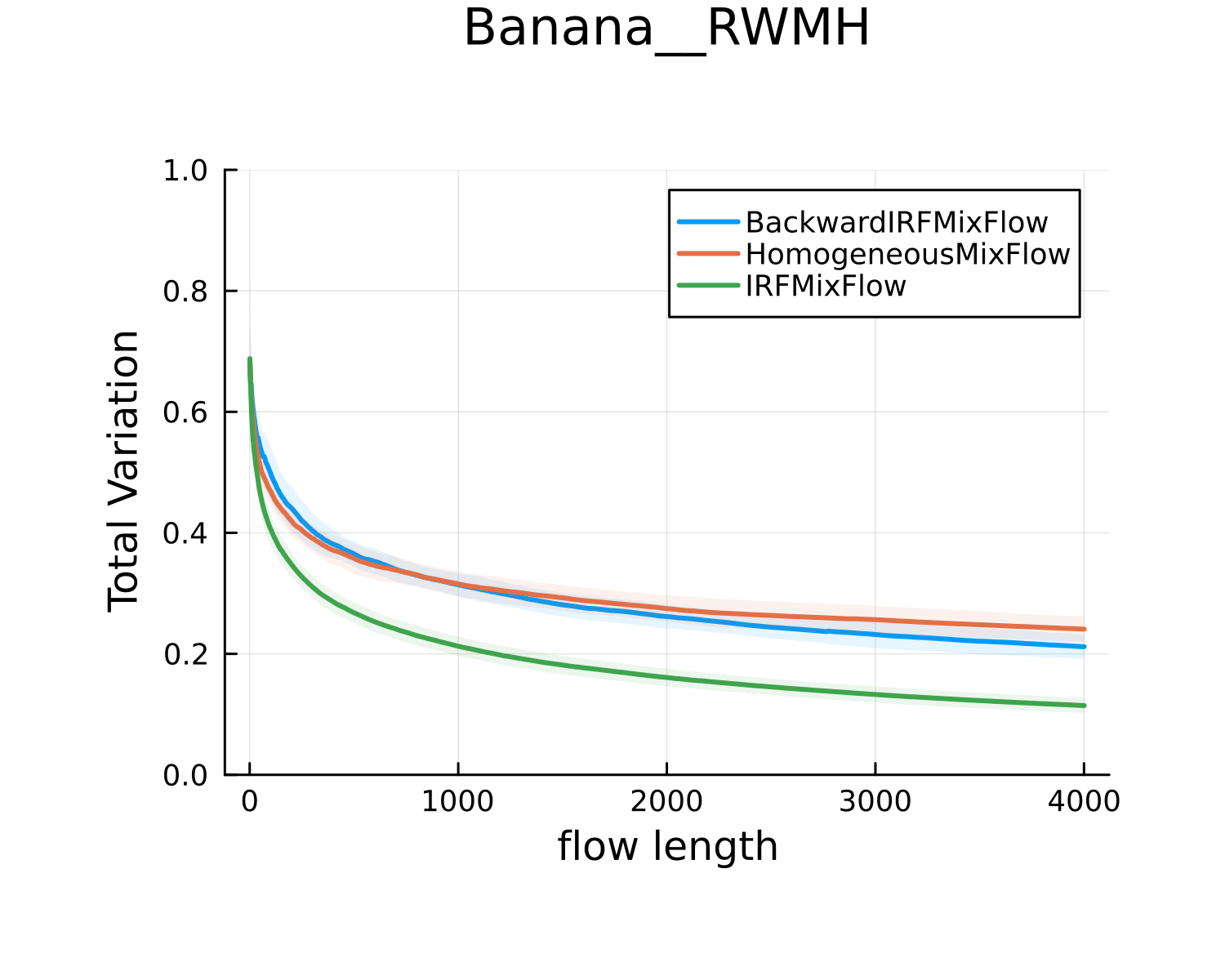}
		\includegraphics[width=0.24\columnwidth, trim=200 150 150 150, clip]{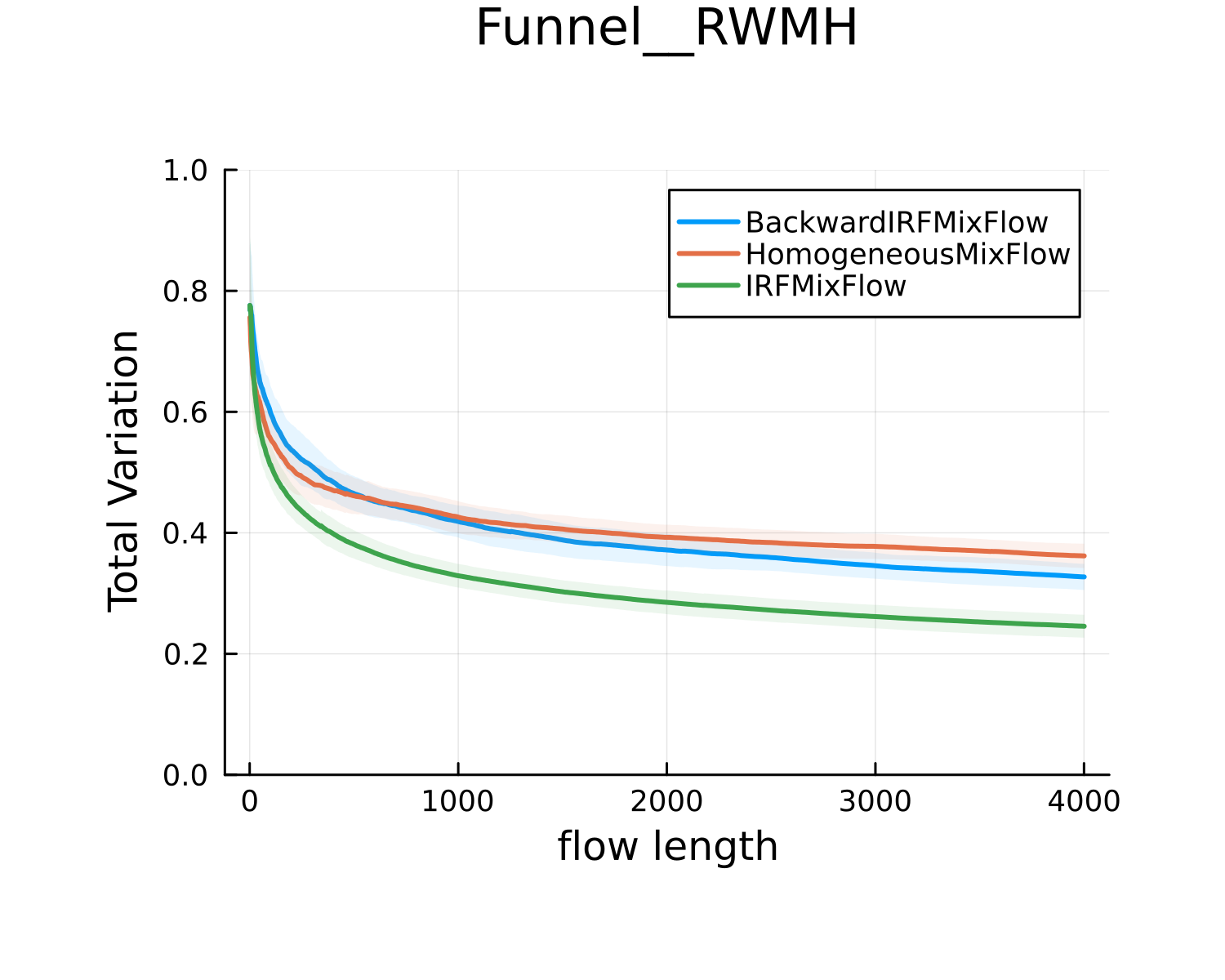}
		\includegraphics[width=0.24\columnwidth, trim=200 150 150 150, clip]{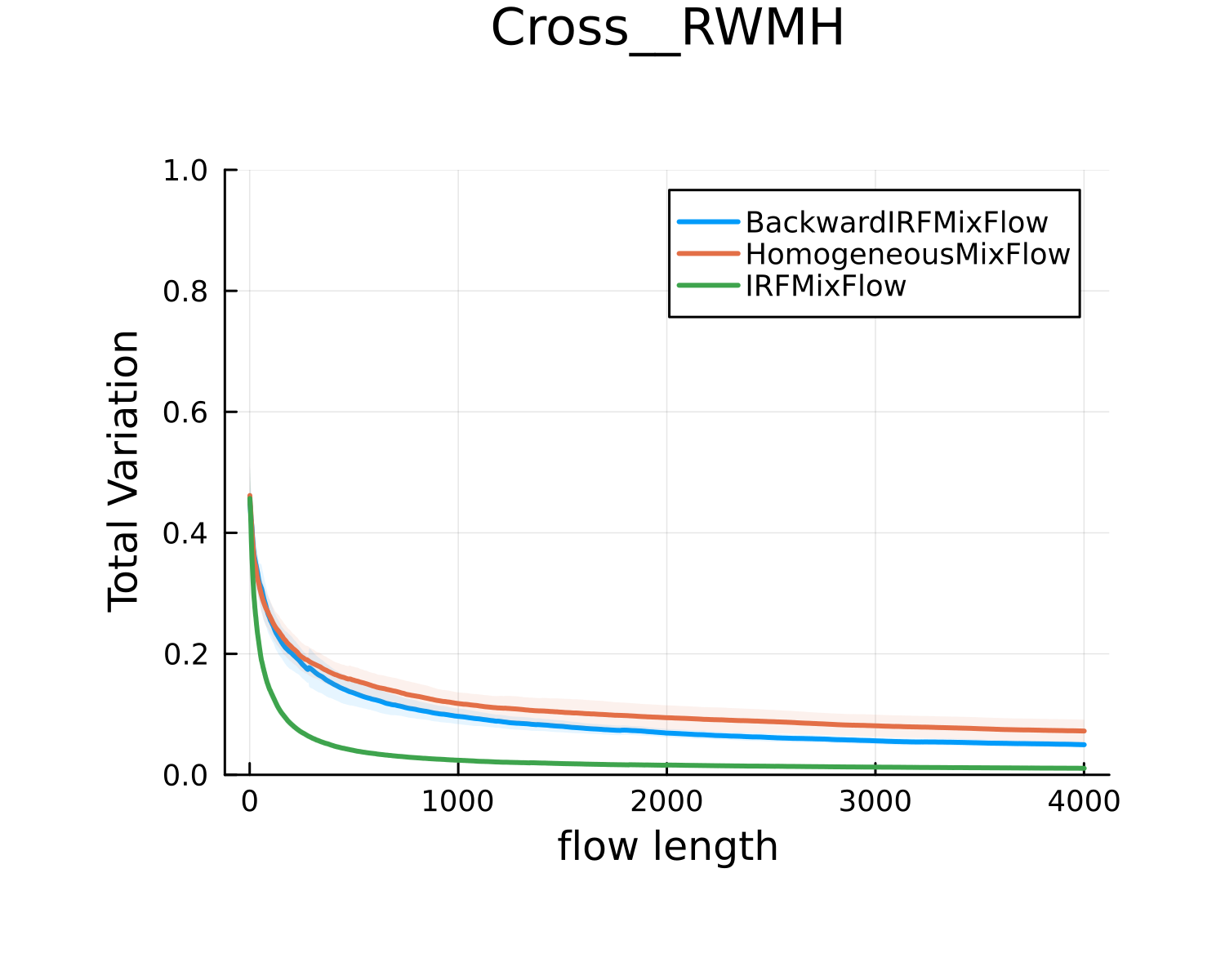}
		\includegraphics[width=0.24\columnwidth, trim=200 150 150 150, clip]{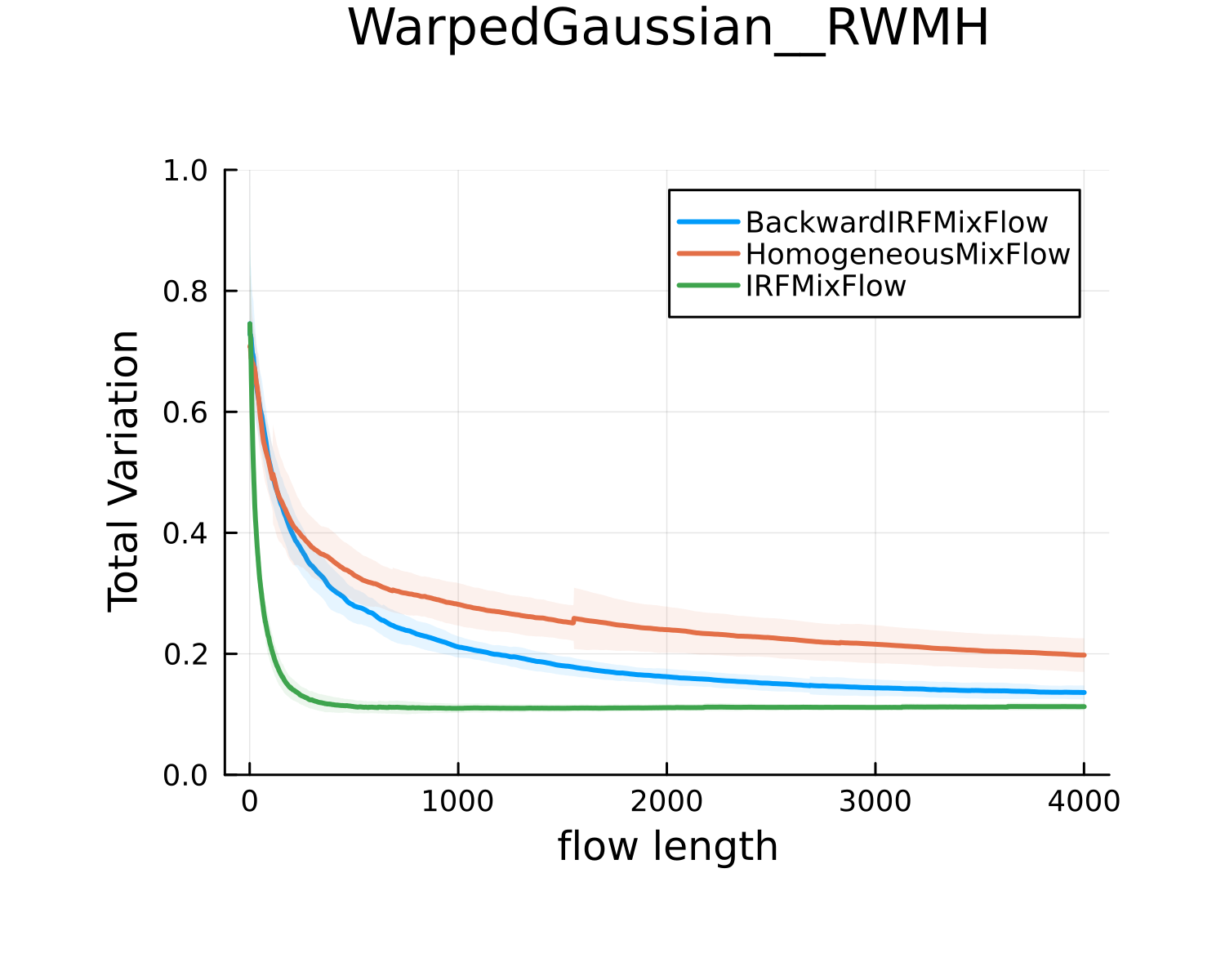}
		\caption{
            Total-variation error for homogeneous, IRF, and backward IRF MixFlows built on RWMH kernels, plotted against flow length T
        for the most performant step sizes among $\{0.05, 0.2, 1.0\}$. Each
    curve is the mean over 32 independent runs; shaded bands ($\pm 1$ SD) show
run-to-run variability. }\label{fig:compare_mixflow}
	\end{subfigure}

    \begin{subfigure}{\columnwidth}
    \includegraphics[width =0.24\columnwidth]{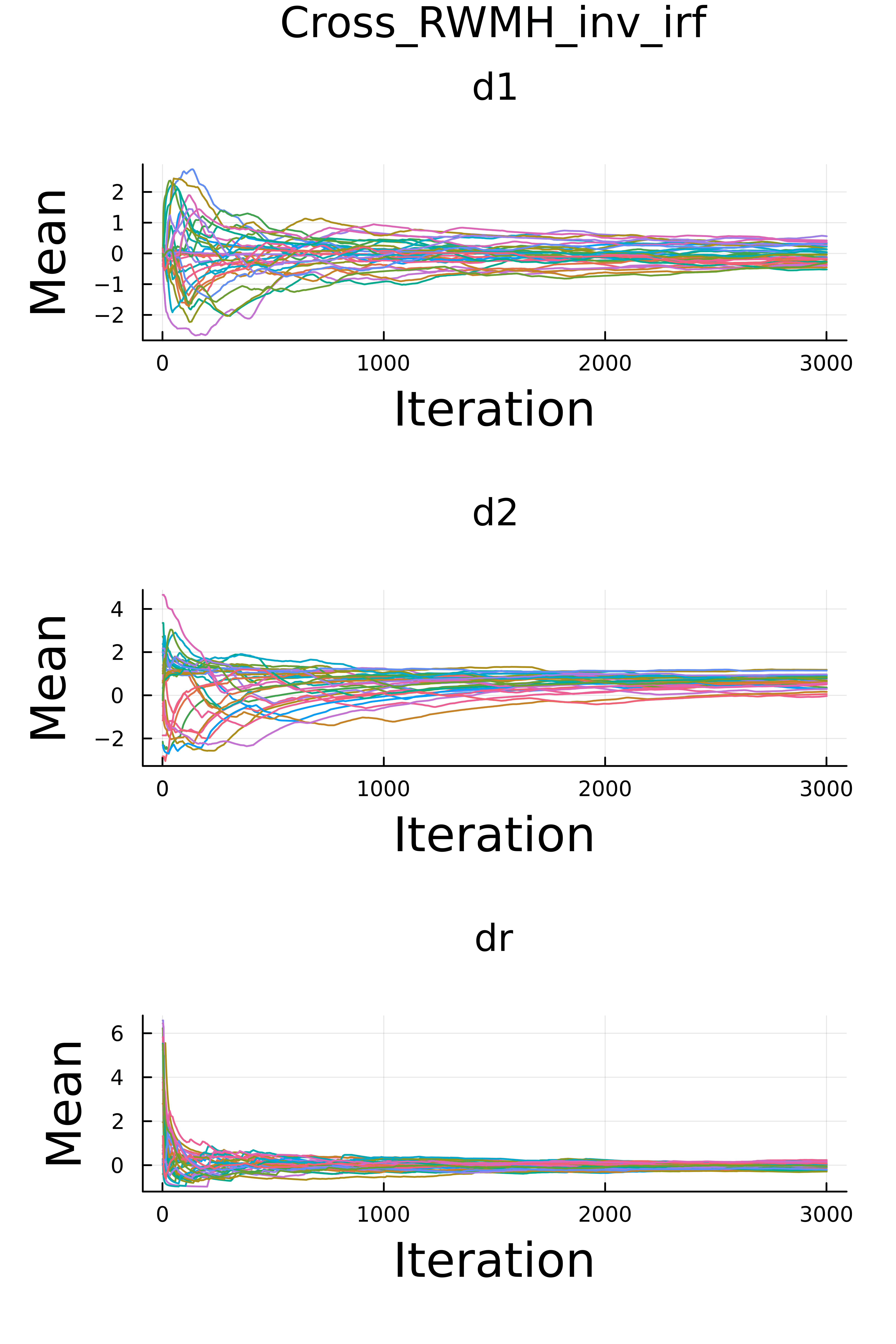}
    \includegraphics[width =0.24\columnwidth]{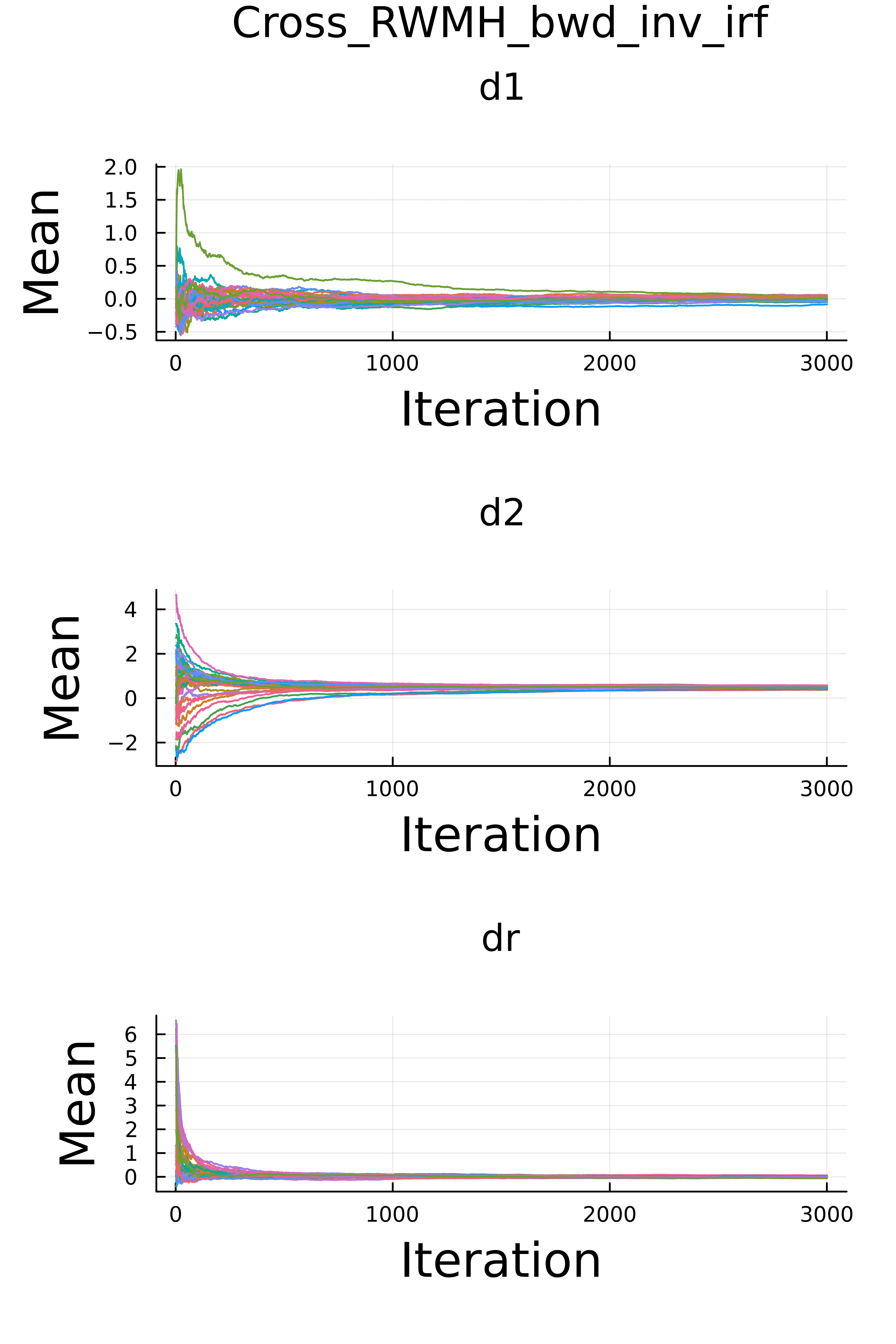}
    \includegraphics[width =0.24\columnwidth]{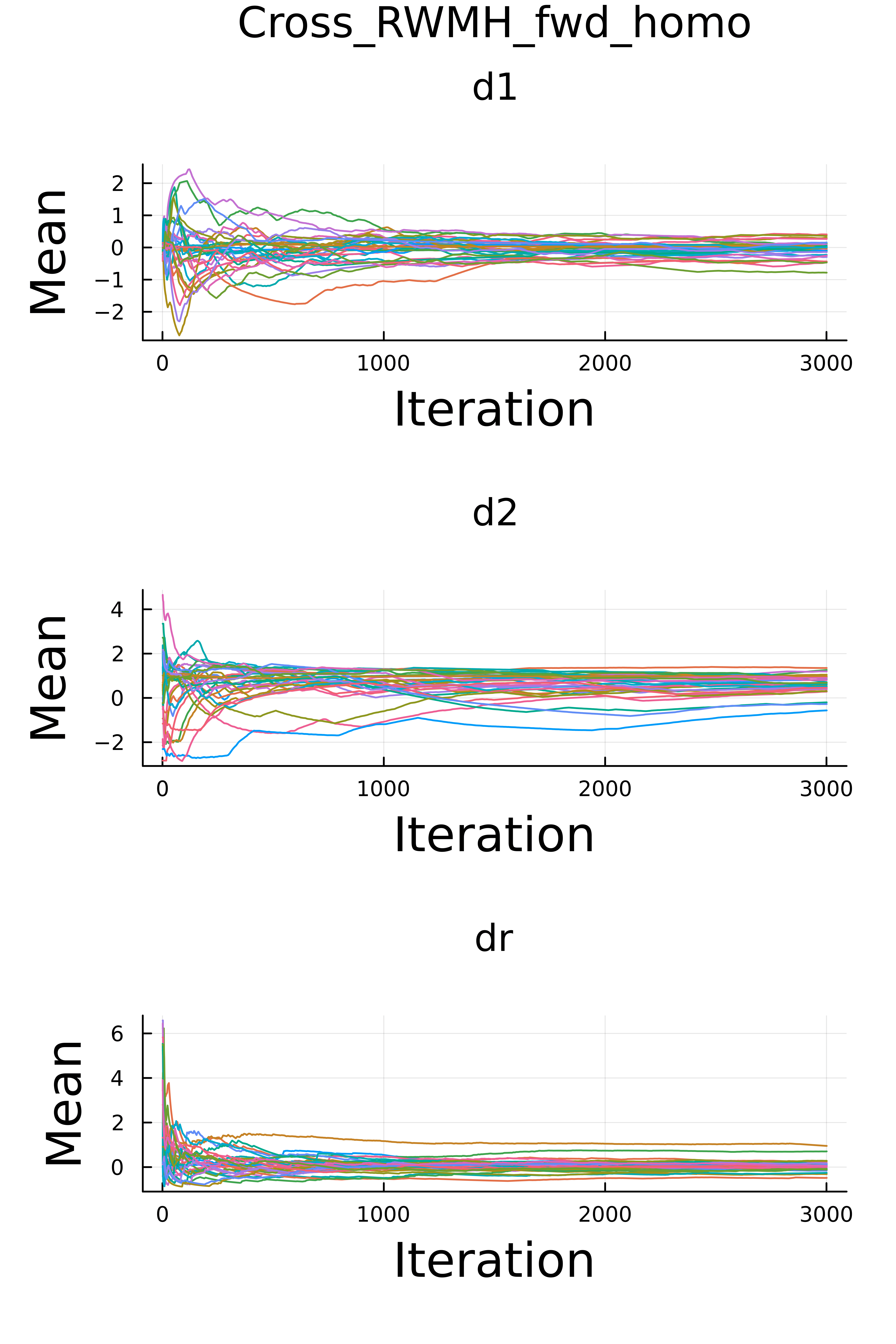}
    \includegraphics[width =0.24\columnwidth]{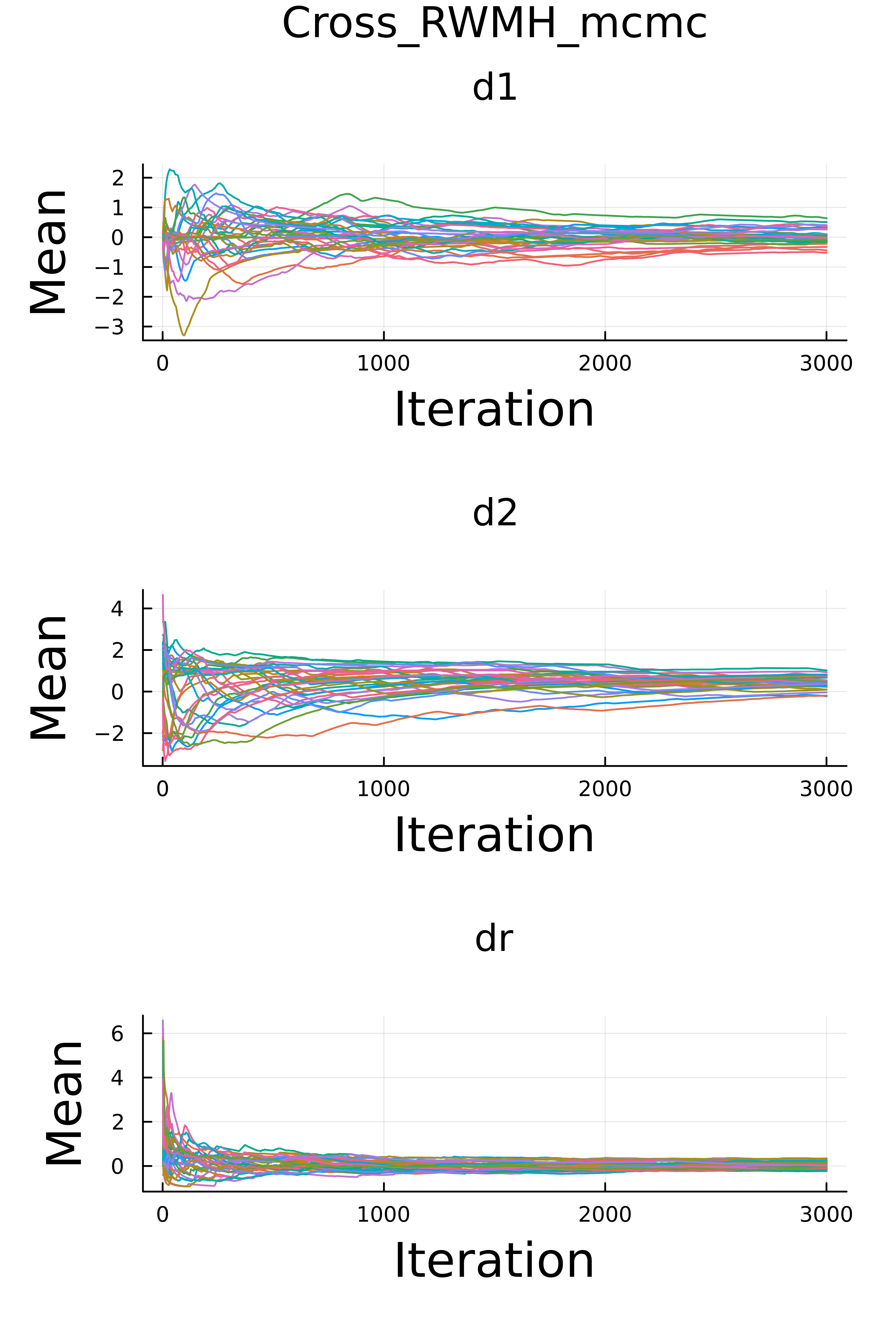}
		\caption{
            Running mean estimates over $3000$ iterates from different IRF and
            MCMC dynamics based on RWMH, evaluated on the Cross distribution
            across 32 independent runs. 
            Each line represents the trajectory of a single run.
            From top to bottom, the rows show the running mean of the test functions $(x_1, x_2) \mapsto x_1$, $(x_1, x_2) \mapsto x_2$, and 
            $(x_1, x_2) \mapsto \frac{q_0}{\pi}(x_1, x_2)$.
            From left to right, the columns correspond to the dynamic of inverse IRF $f_\theta^{-1}$, the backward process of the inverse IRF, 
            time-homogeneous dynamics $f_{\theta^\star}$, and the standard RWMH MCMC.
    }\label{fig:runningmean}
    \end{subfigure}

    \begin{subfigure}{\columnwidth}
		\includegraphics[width=0.24\columnwidth, trim=250 250 250 250, clip]{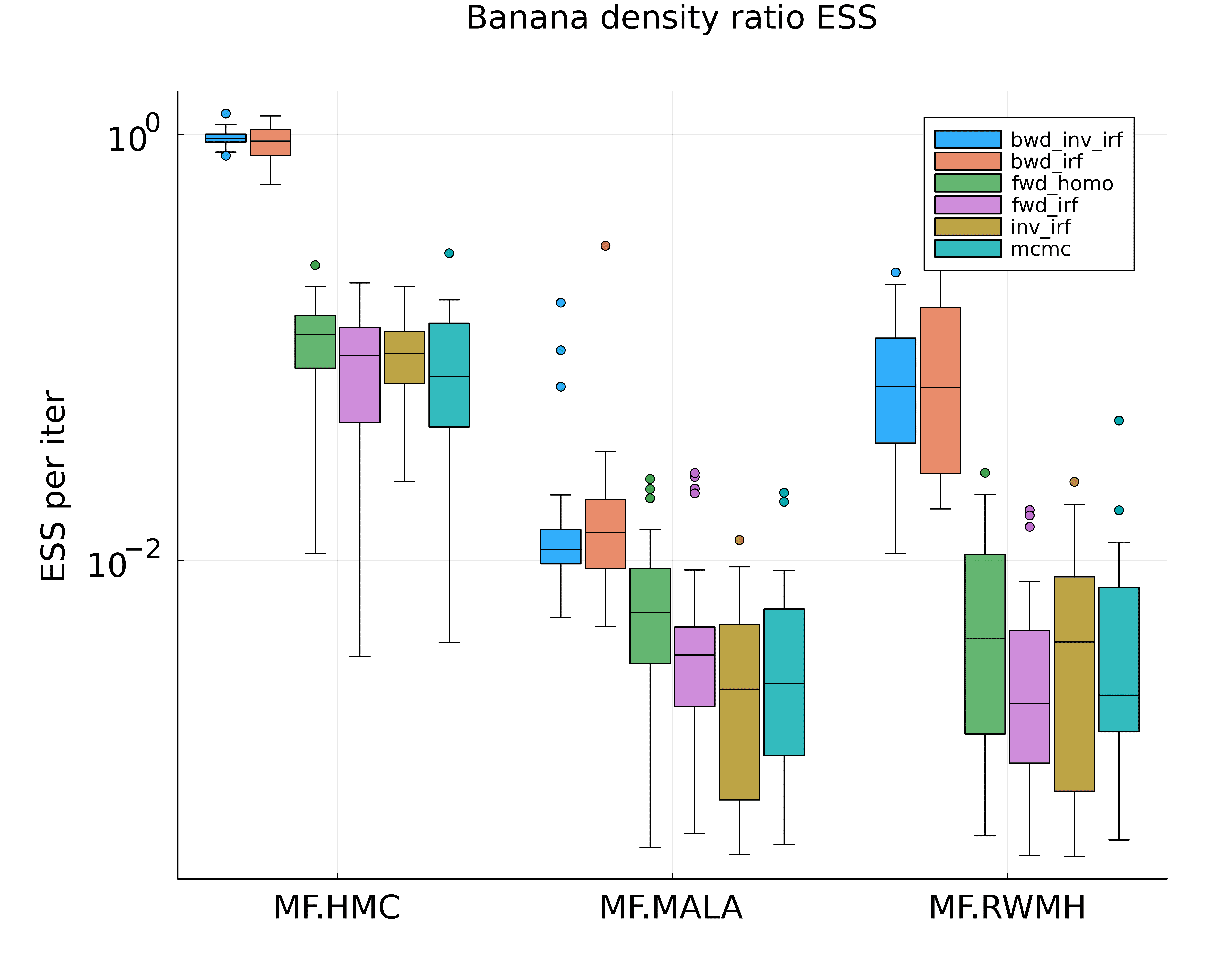}
		\includegraphics[width=0.24\columnwidth, trim=250 250 250 250, clip]{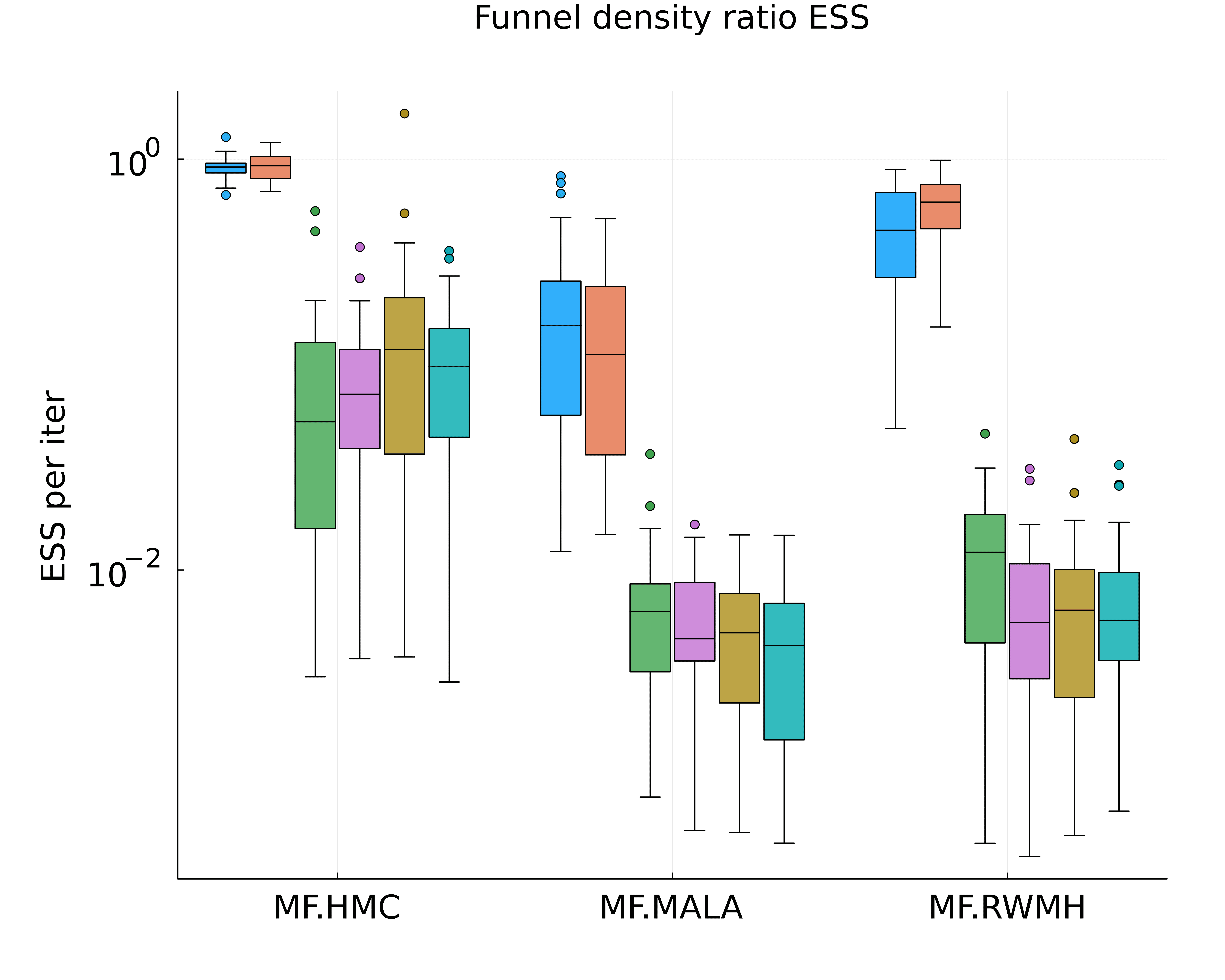}
		\includegraphics[width=0.24\columnwidth, trim=250 250 250 250, clip]{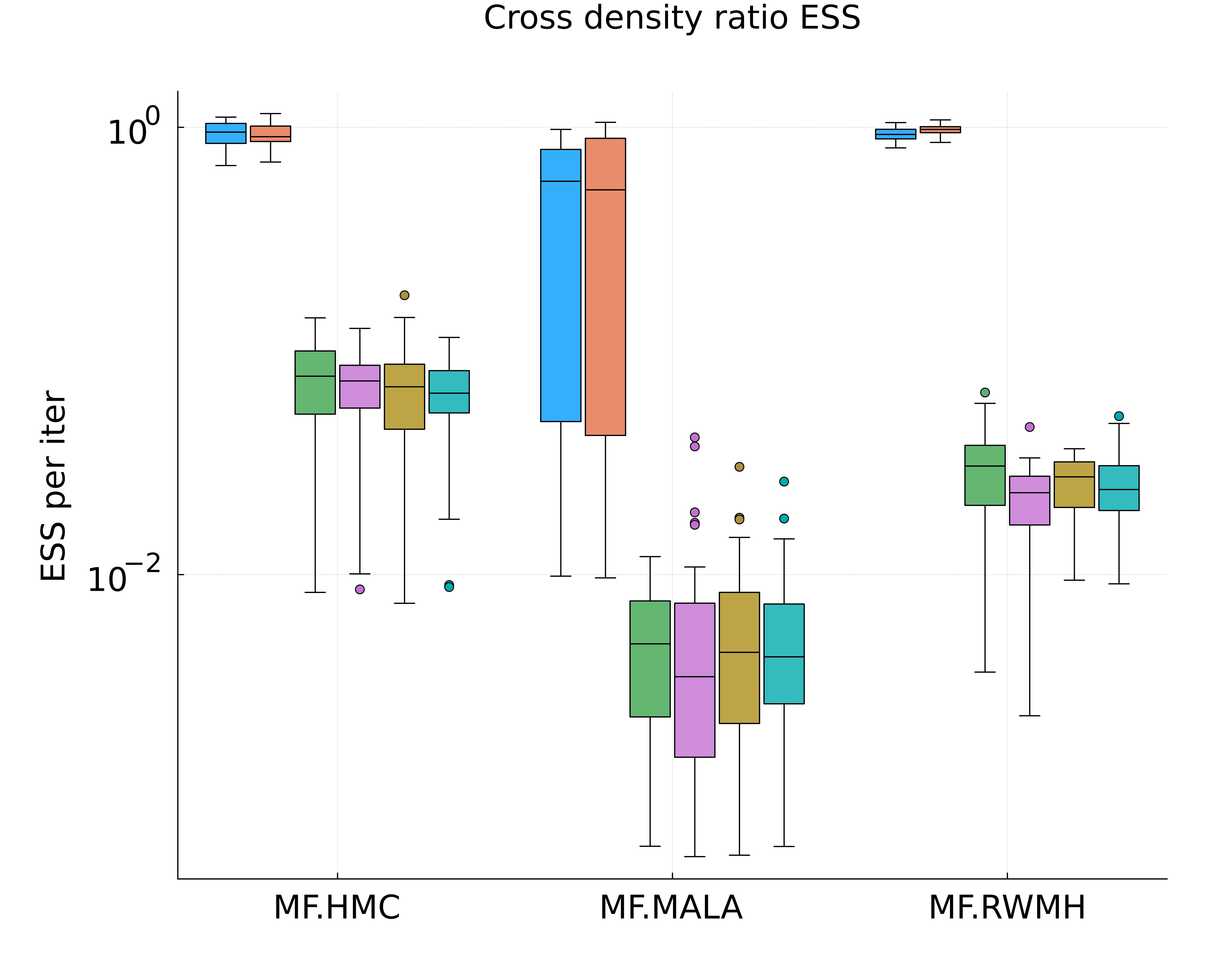}
		\includegraphics[width=0.24\columnwidth, trim=250 250 250 250, clip]{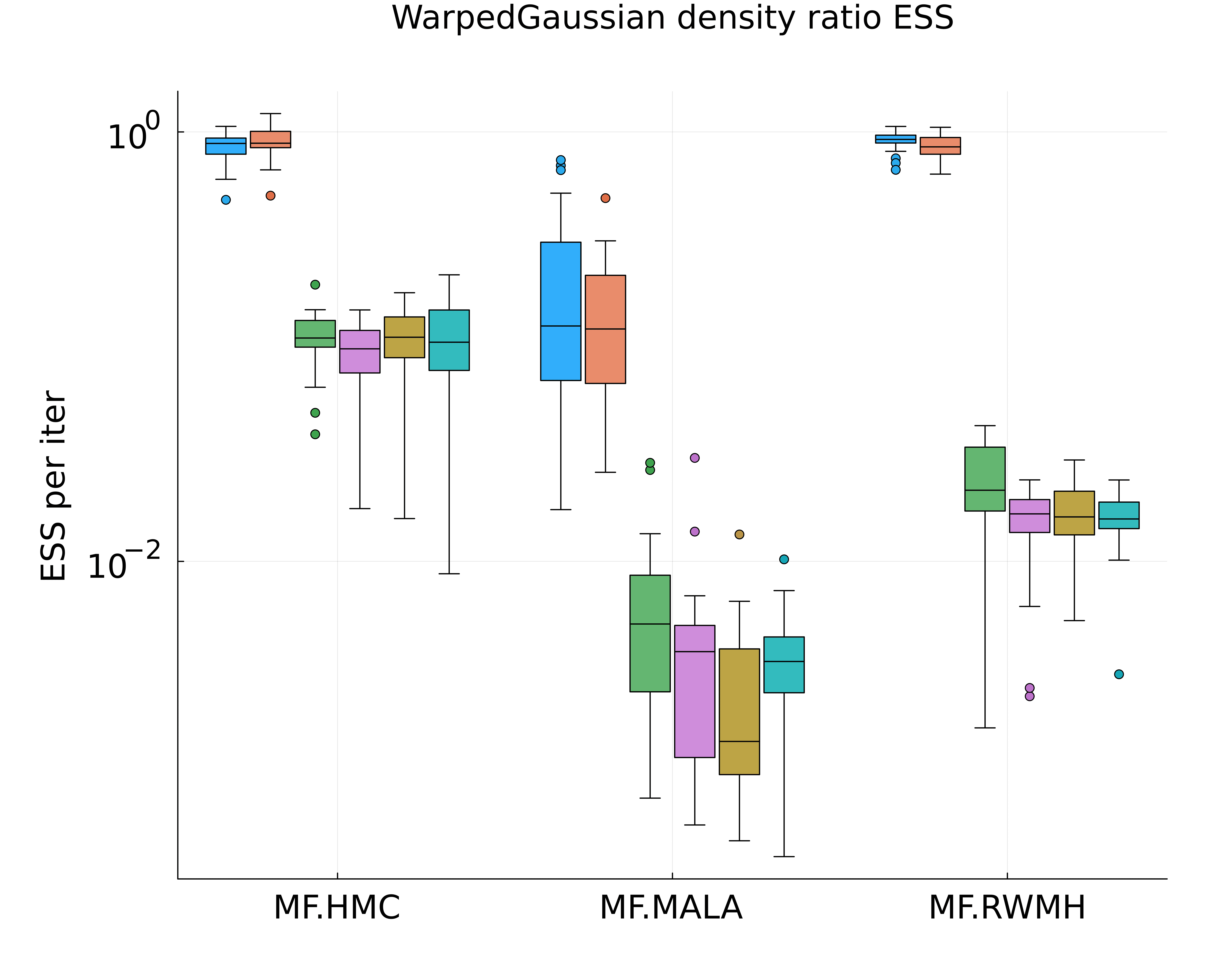}
    \caption{
        Per-sample MCMC effective sample size (ESS) estimates on the test function $\frac{q_0}{\pi}$, 
        computed from trajectories generated by various IRF and MCMC dynamics based on HMC, MALA, and RWMH kernels.
        The trajectory lengths are set to $300$ for HMC-based dynamics, $2000$ for MALA, and $4000$ for RWMH.
        Each ESS value is computed from a single trajectory, and the boxplots summarize the ESS estimates over 32 independent runs per method.
        The per-sample ESS for \iid samples will be $1$.
}\label{fig:densityratio_ess}
	\end{subfigure}

\caption{Results showing difference between homogeneous, IRF, and backward IRF MixFlows}
\end{figure}

\cref{fig:compare_mixflow} compares the total variation (TV) errors of
homogeneous, IRF, and backward IRF MixFlows constructed from RWMH kernels.
Overall, homogeneous and backward IRF MixFlows perform similarly, though the
latter exhibits slightly improved accuracy at longer flow lengths. IRF MixFlow
consistently outperforms both, achieving faster TV convergence and lower
variability across runs.
As discussed in \cref{sec:weakerassumption}, this improvement stems from
differences in the convergence behavior of the series $\frac{1}{K} \sum_{k=1}^K
\frac{q_0}{\pi}\left(T_k(x)\right)$, where $T_k$ represents the sequence of
transformations used in the density computation of each MixFlow variant.

\cref{fig:runningmean} further illustrates this effect by showing running mean
estimates over 3000 iterations for the Cross distribution. From top to bottom,
each row shows the mean of the test functions $(x_1, x_2) \mapsto x_1$, $(x_1,
x_2) \mapsto x_2$, and $(x_1, x_2) \mapsto \frac{q_0}{\pi}(x_1, x_2)$. From
left to right, the columns correspond to the inverse IRF $f_\theta^{-1}$
(backward IRF MixFlow), the backward process of the inverse IRF (IRF MixFlow),
the time-homogeneous flow $f_{\theta^\star}$ (homogeneous MixFlow), and
standard RWMH MCMC. The backward process exhibits significantly faster
convergence in all cases, consistent with the superior TV performance of IRF
MixFlows under equal flow lengths. This advantage arises from reduced
autocorrelation in the backward iterates.

\cref{fig:densityratio_ess} reports the per-sample MCMC effective sample size
(ESS) for the test function $\frac{q_0}{\pi}$, estimated from trajectories
generated using various IRF and MCMC dynamics based on HMC, MALA, and RWMH.
This metric captures the degree of autocorrelation in $\frac{q_0}{\pi}(T_k(x))$
across iterations. Backward process dynamics consistently yield ESS values
orders of magnitude higher than other methods---often approaching the ideal of
independent sampling, with relative ESS close to 1 in some cases.

\subsubsection{Ensemble IRF MixFlows: scaling up $M$ or $T$} \label{apdx:ensembles}

\begin{figure}[H]
    \begin{subfigure}{\columnwidth}
		\includegraphics[width=0.24\columnwidth, trim=250 130 250 150, clip]{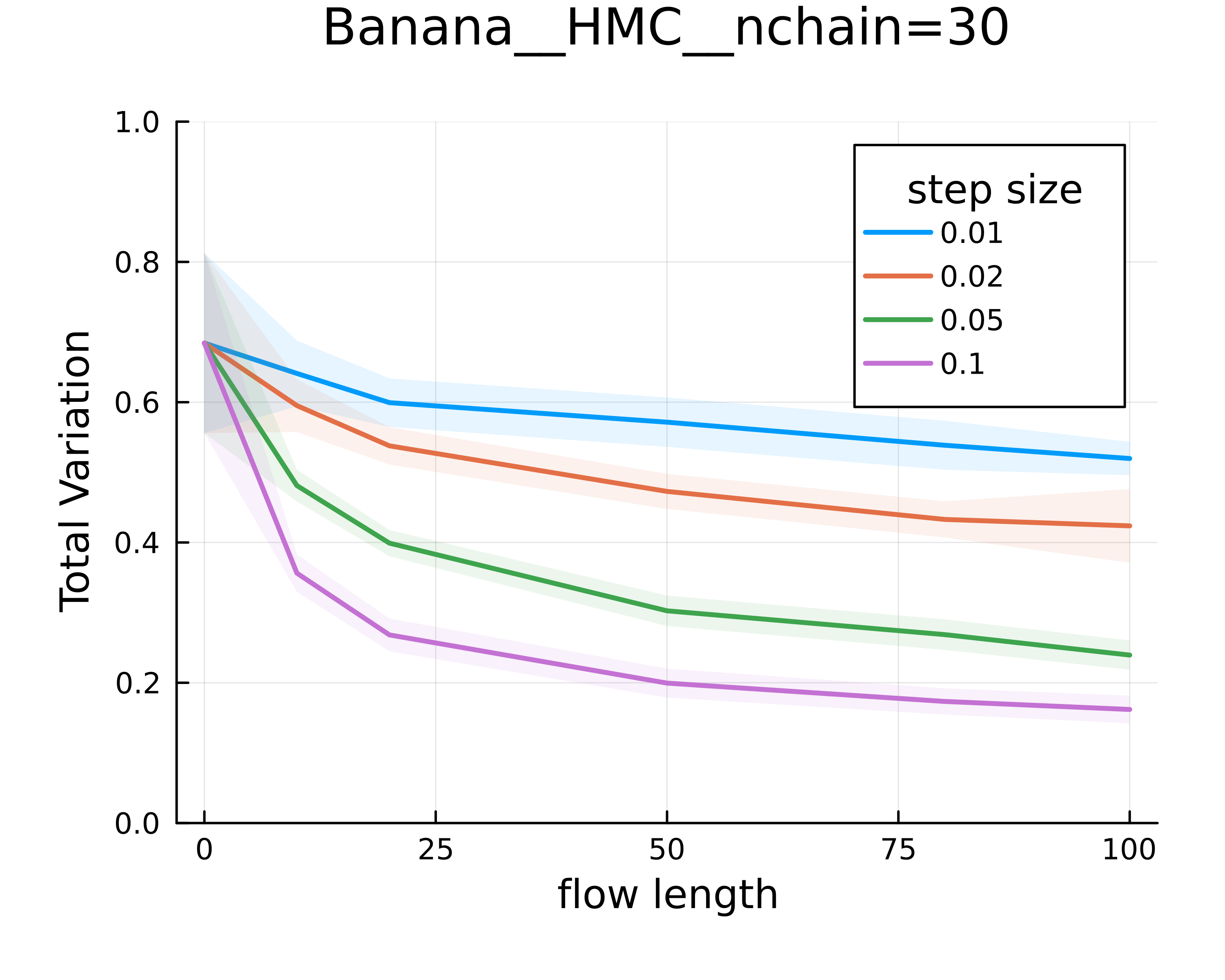}
		\includegraphics[width=0.24\columnwidth, trim=250 130 250 150, clip]{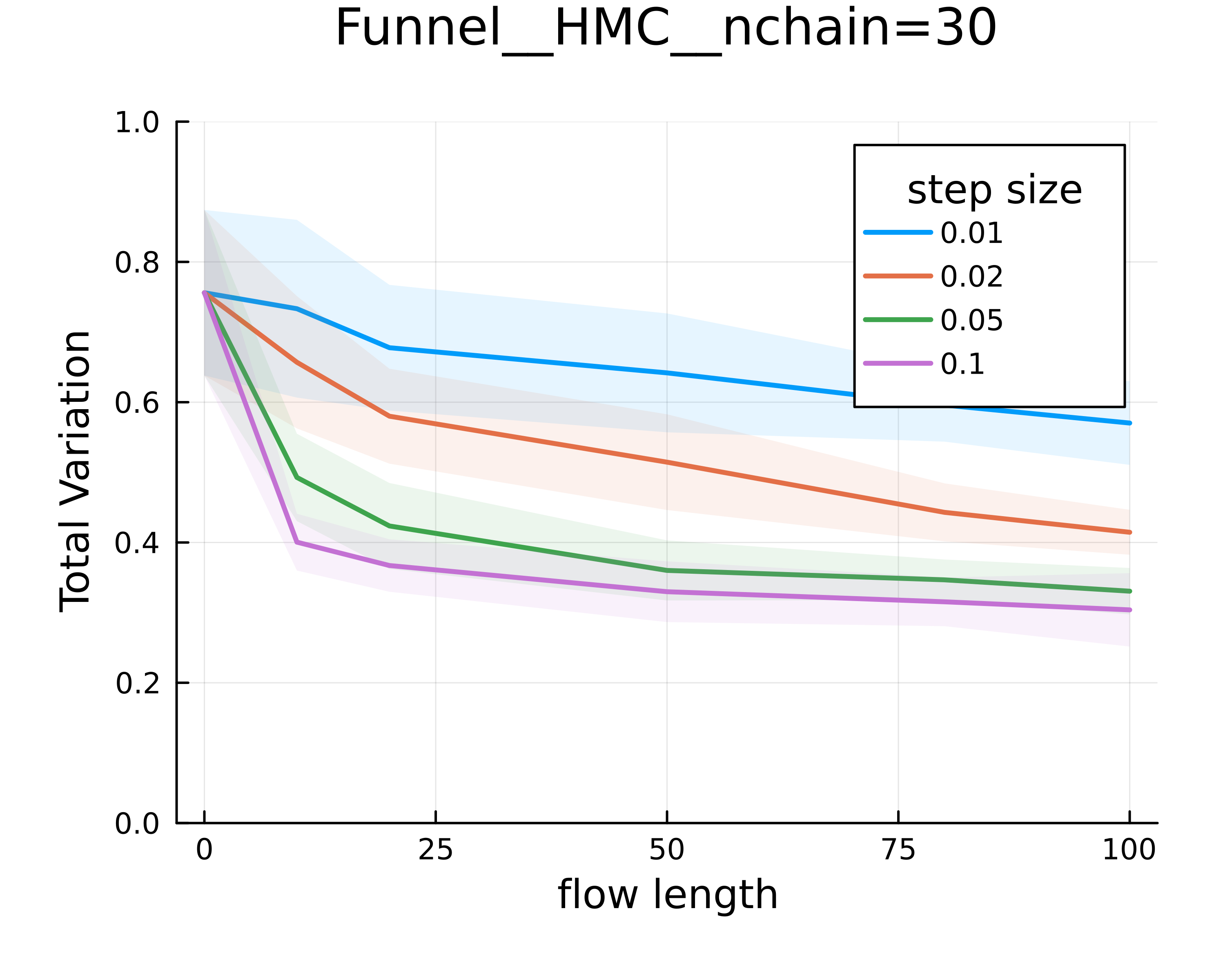}
		\includegraphics[width=0.24\columnwidth, trim=250 130 250 150, clip]{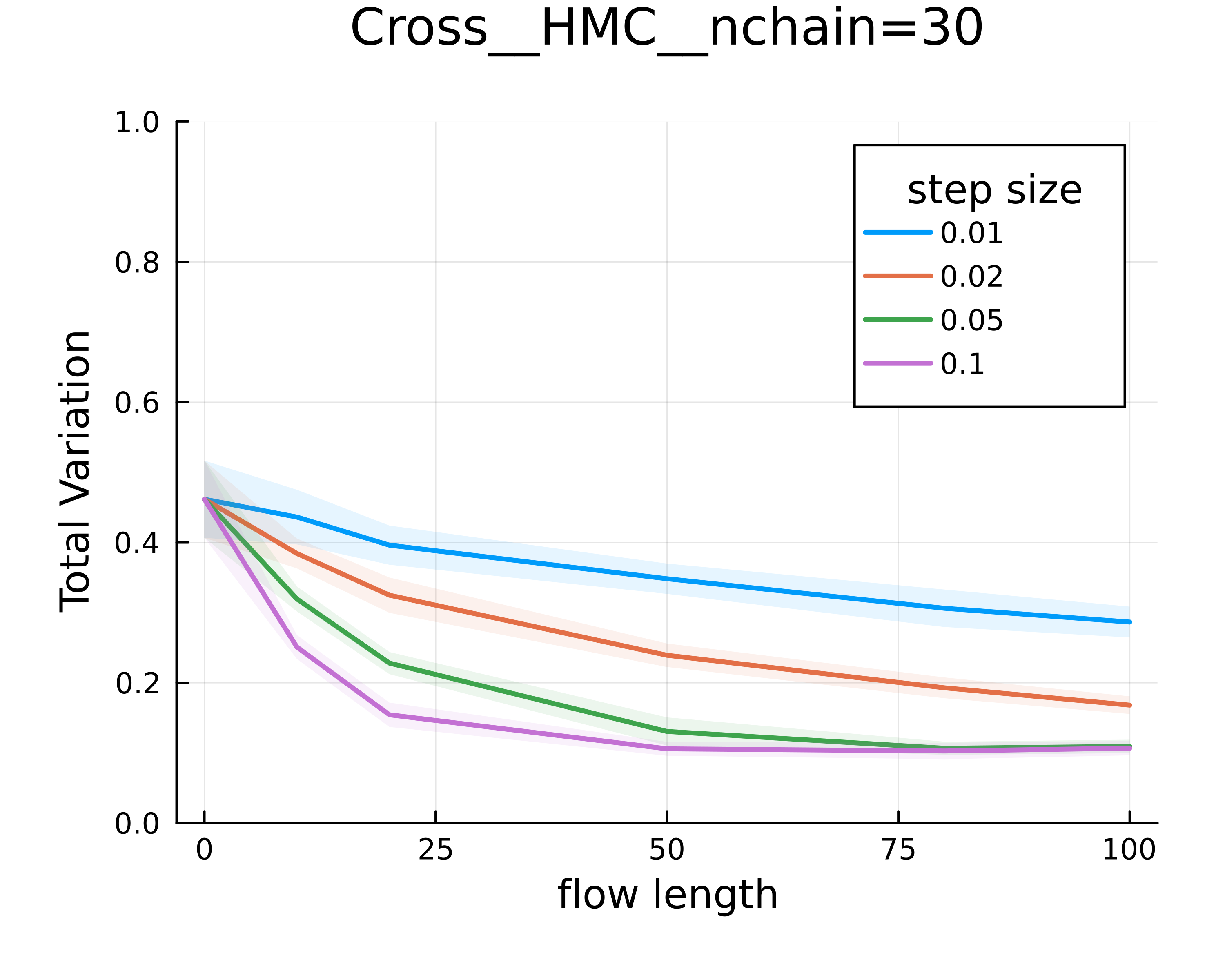}
		\includegraphics[width=0.24\columnwidth, trim=250 130 250 150, clip]{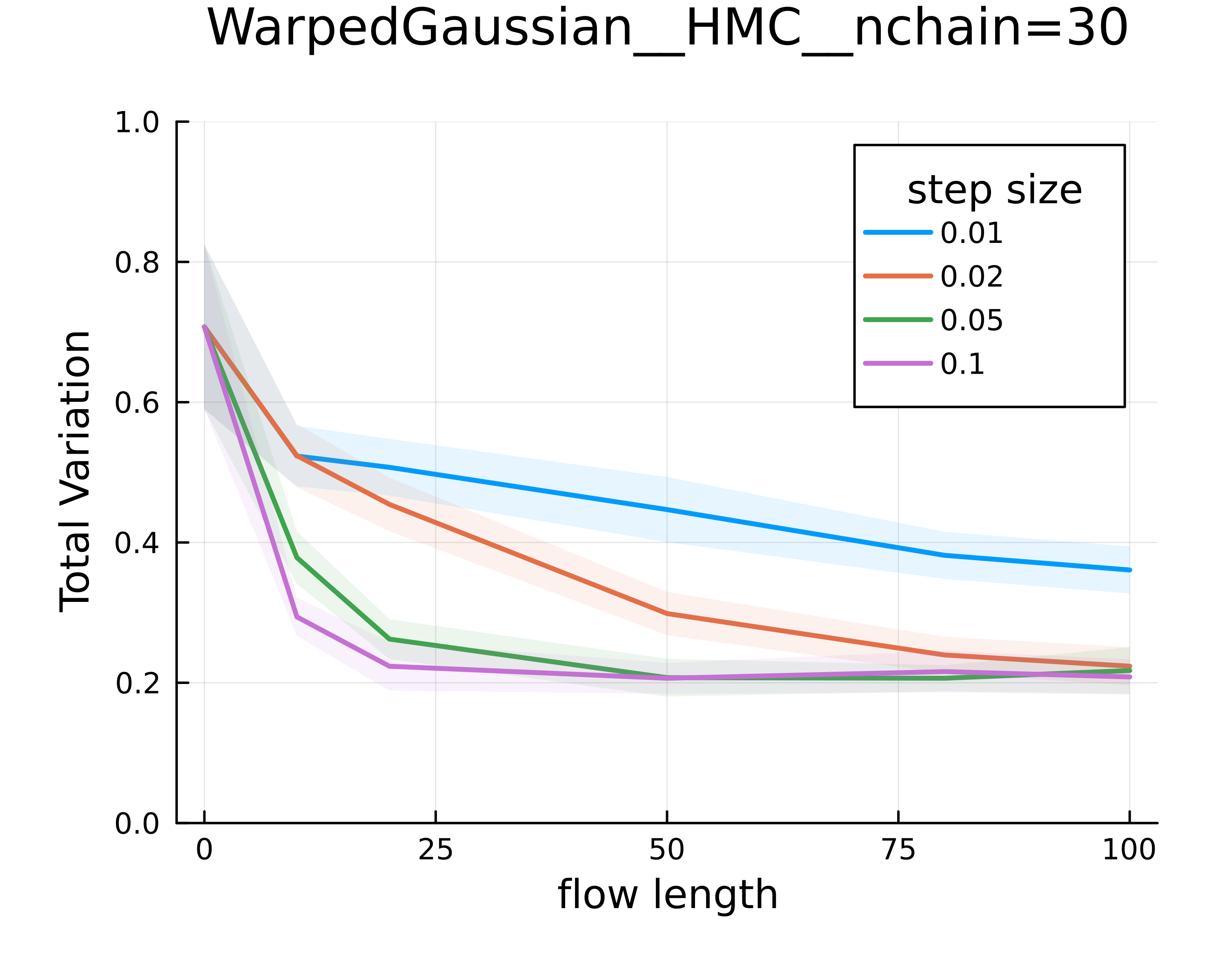}
		\caption{Fix ensemble size $M = 30$ increase flow length.}\label{fig:ensemble_increase_T}
	\end{subfigure}
    \begin{subfigure}{\columnwidth}
		\includegraphics[width=0.24\columnwidth, trim=250 130 250 150, clip]{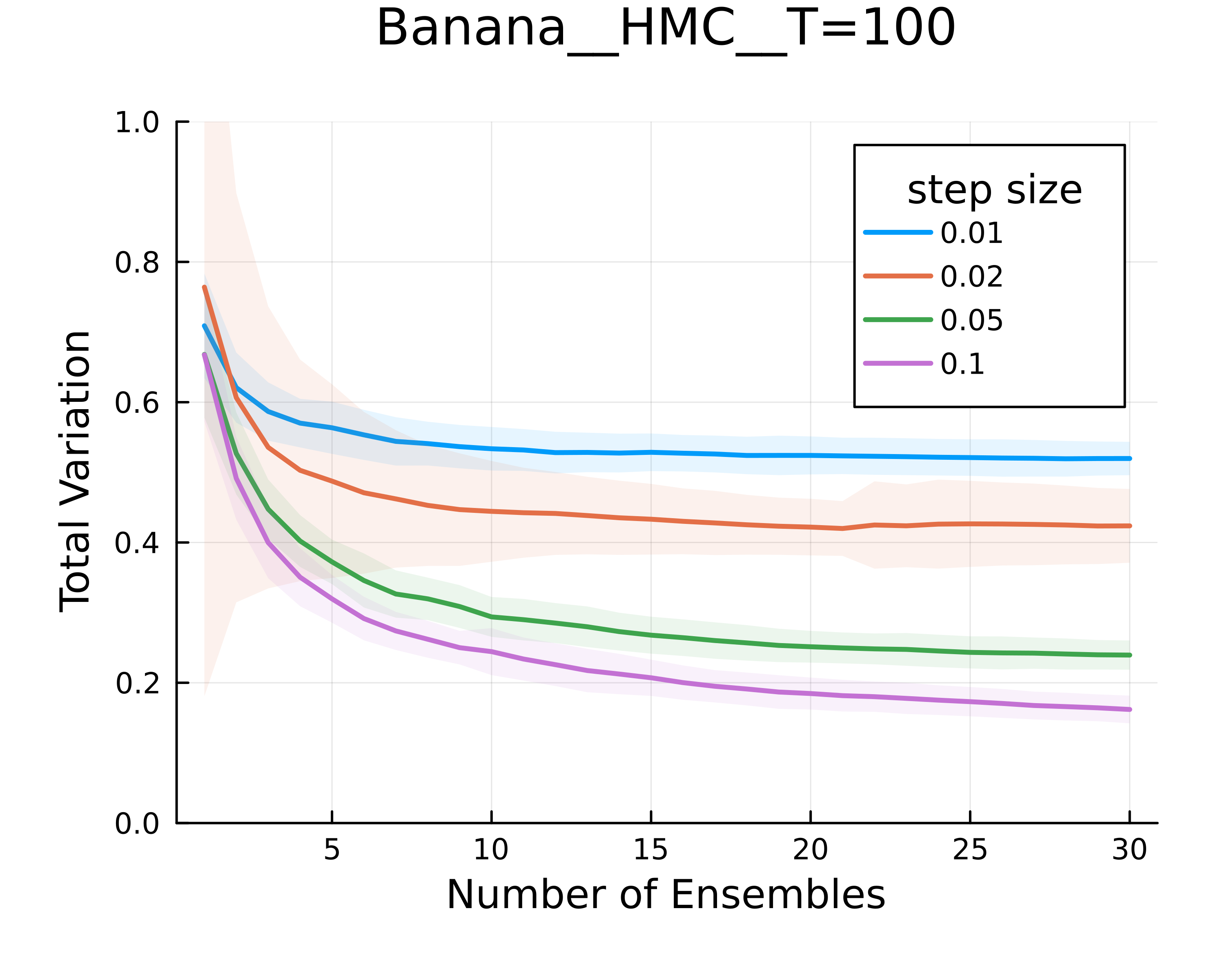}
		\includegraphics[width=0.24\columnwidth, trim=250 130 250 150, clip]{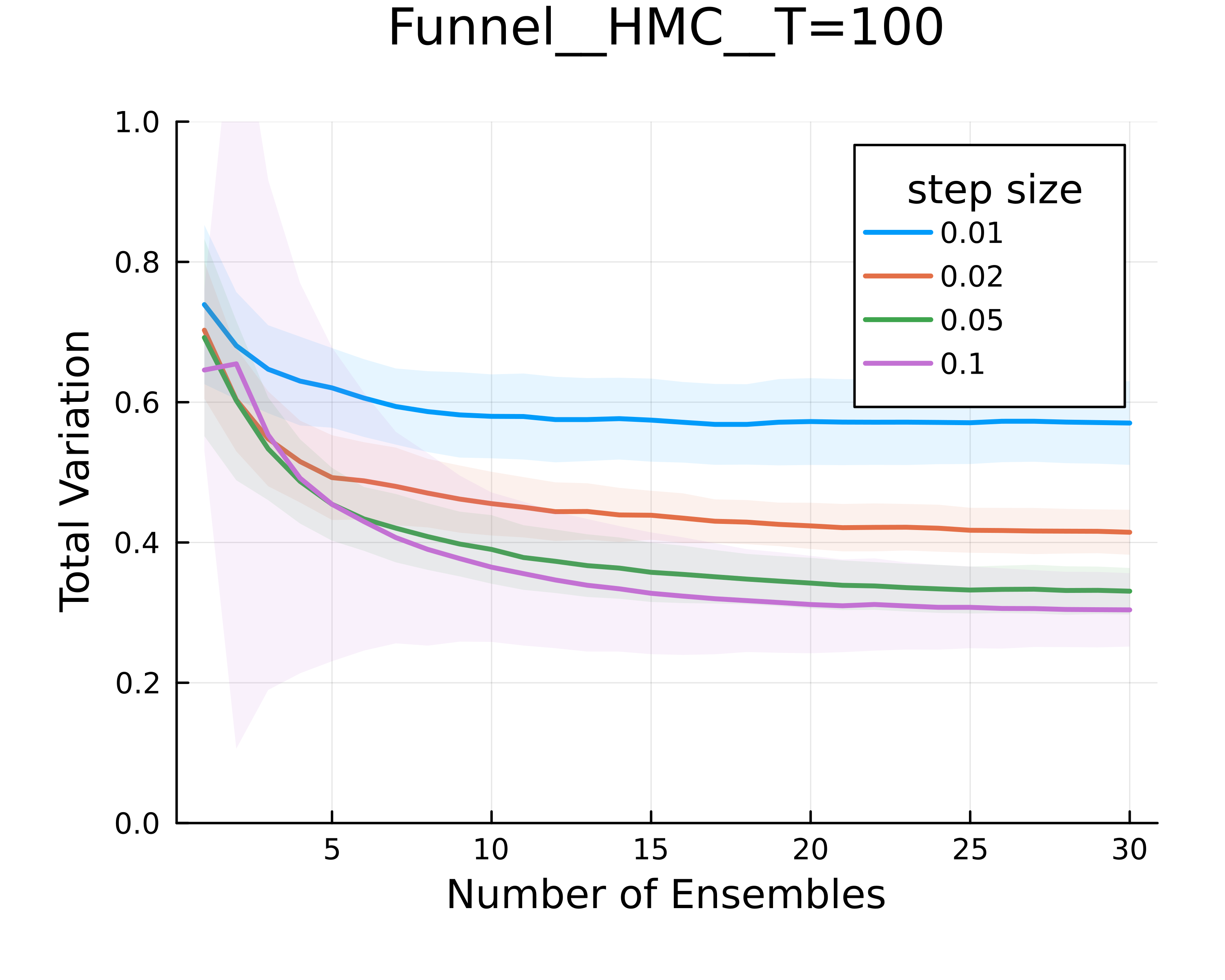}
		\includegraphics[width=0.24\columnwidth, trim=250 130 250 150, clip]{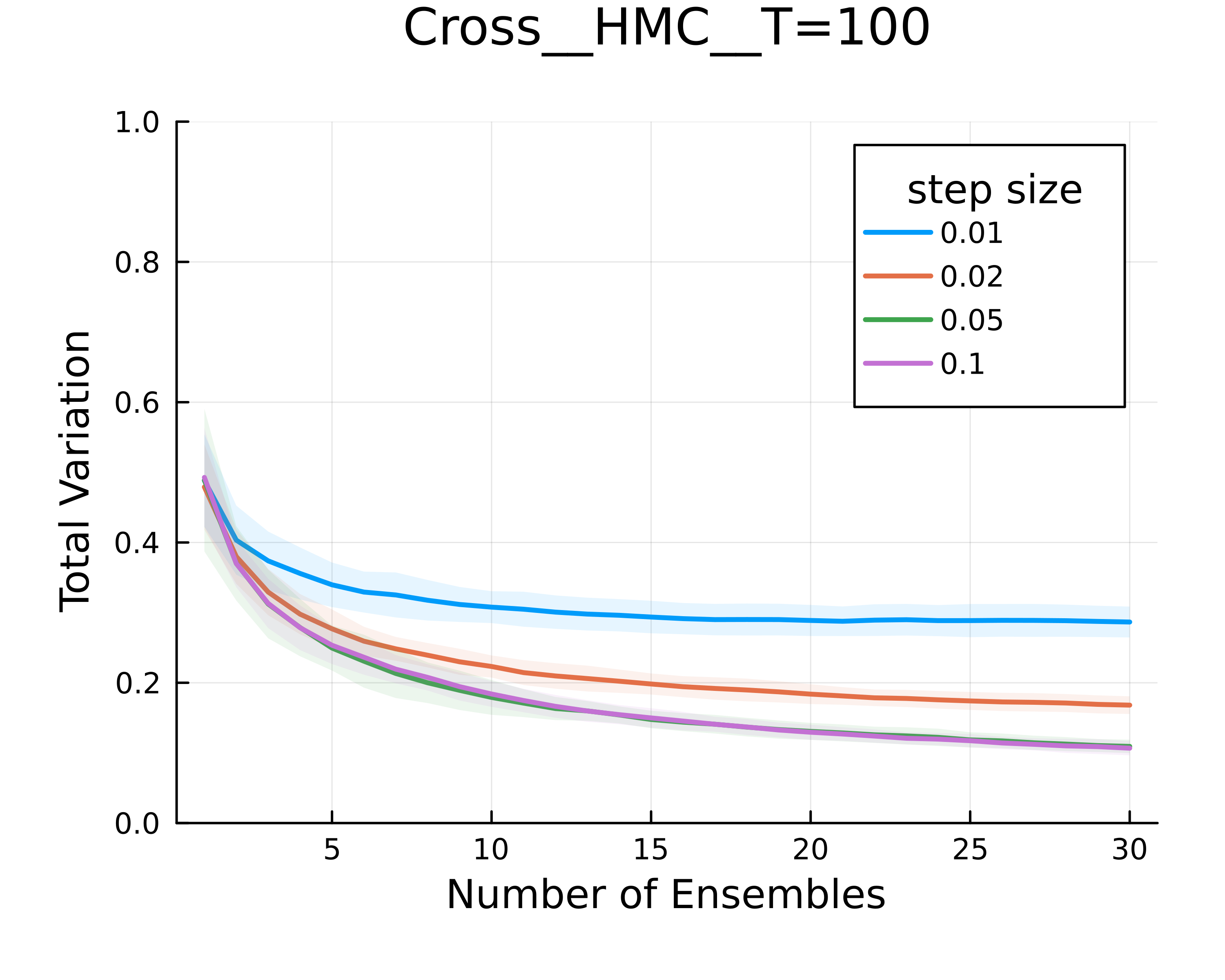}
		\includegraphics[width=0.24\columnwidth, trim=250 130 250 150, clip]{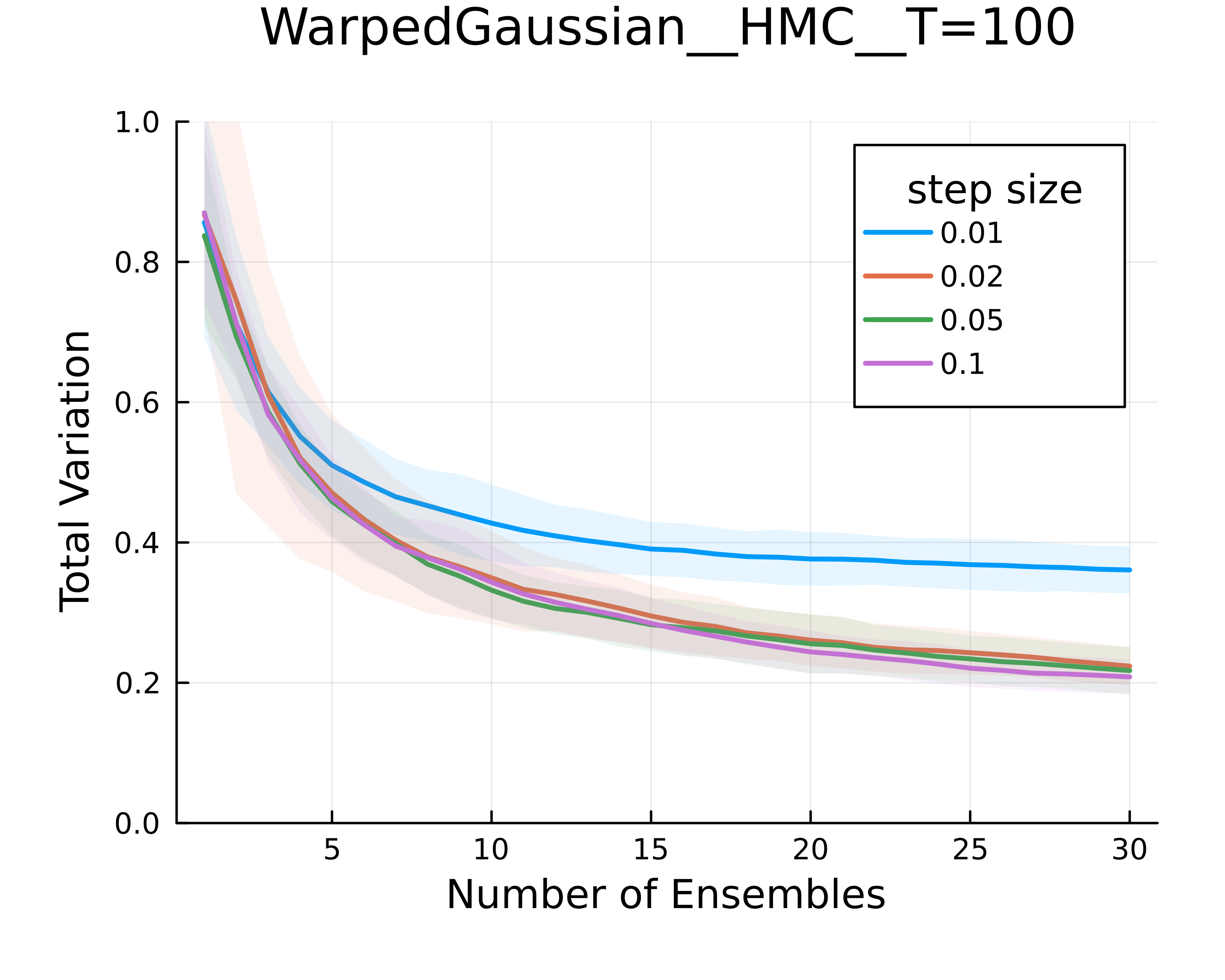}
		\caption{Fix flow length $T = 100$ increase ensemble size.}\label{fig:ensemble_increase_M}
	\end{subfigure}
\caption{TV error of ensemble IRF MixFlows based on HMC over increasing ensemble size $M$ and flow length $T$.
    Each curve is the mean over 32 independent runs; shaded bands ($\pm 1$ SD) show run-to-run variability. }
\end{figure}

\newpage
\subsubsection{Additional results for synthetic examples}\label{apdx:additional_synthetic_results}

\begin{figure}[H]
    \begin{subfigure}{\columnwidth}
        \includegraphics[width=0.33\columnwidth, trim=0 20 50 0, clip]{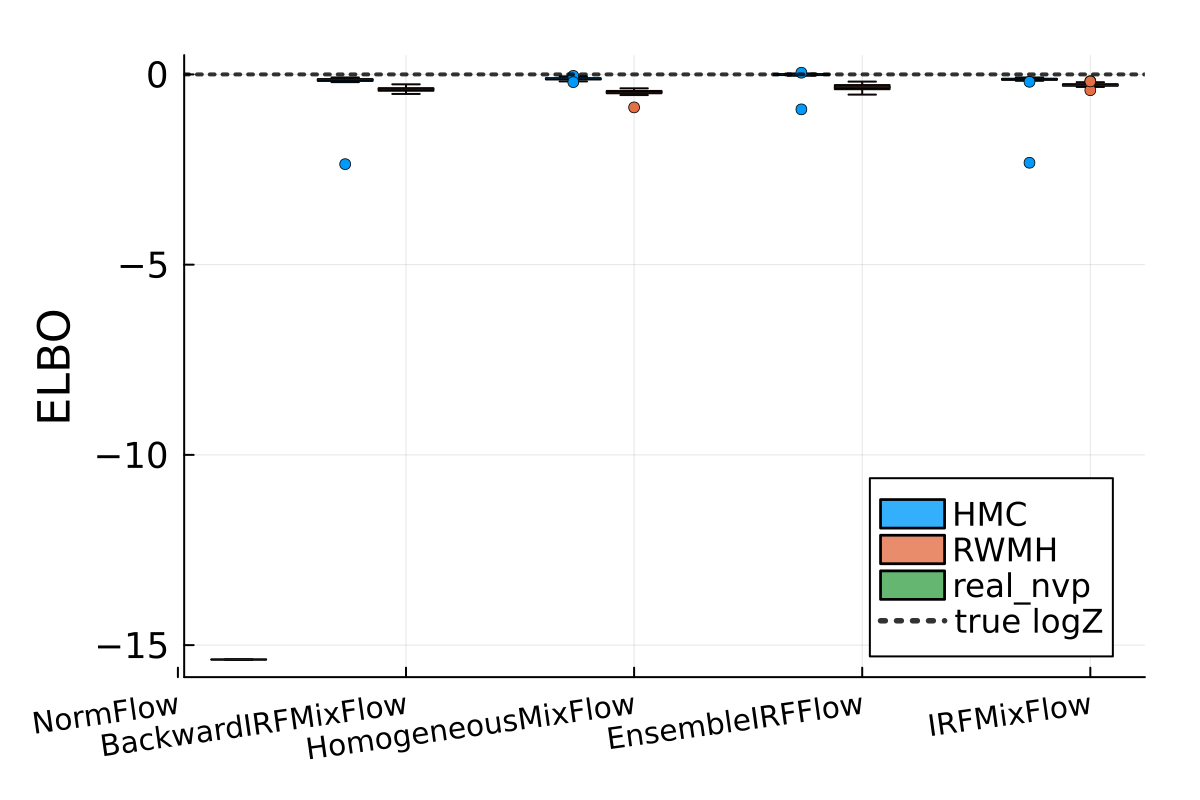}
        \includegraphics[width=0.33\columnwidth, trim=0 20 50 0, clip]{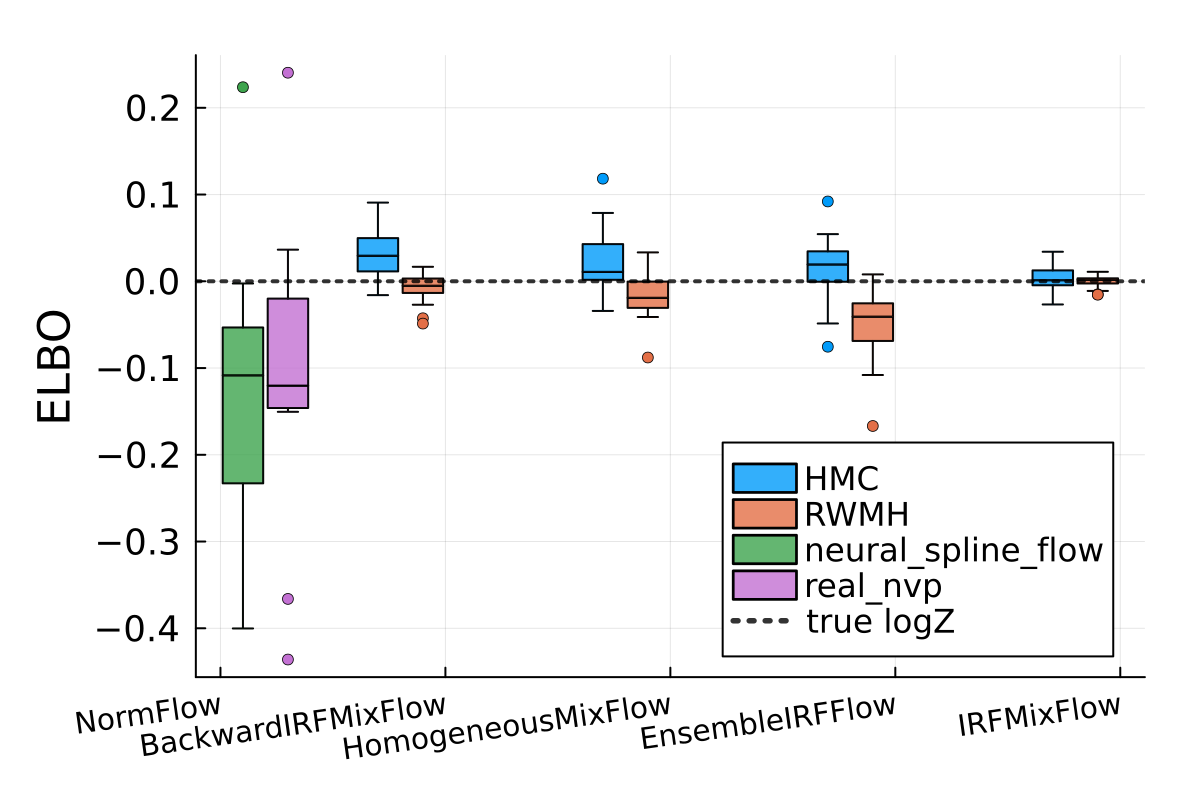}
        \includegraphics[width=0.33\columnwidth, trim=0 20 50 0, clip]{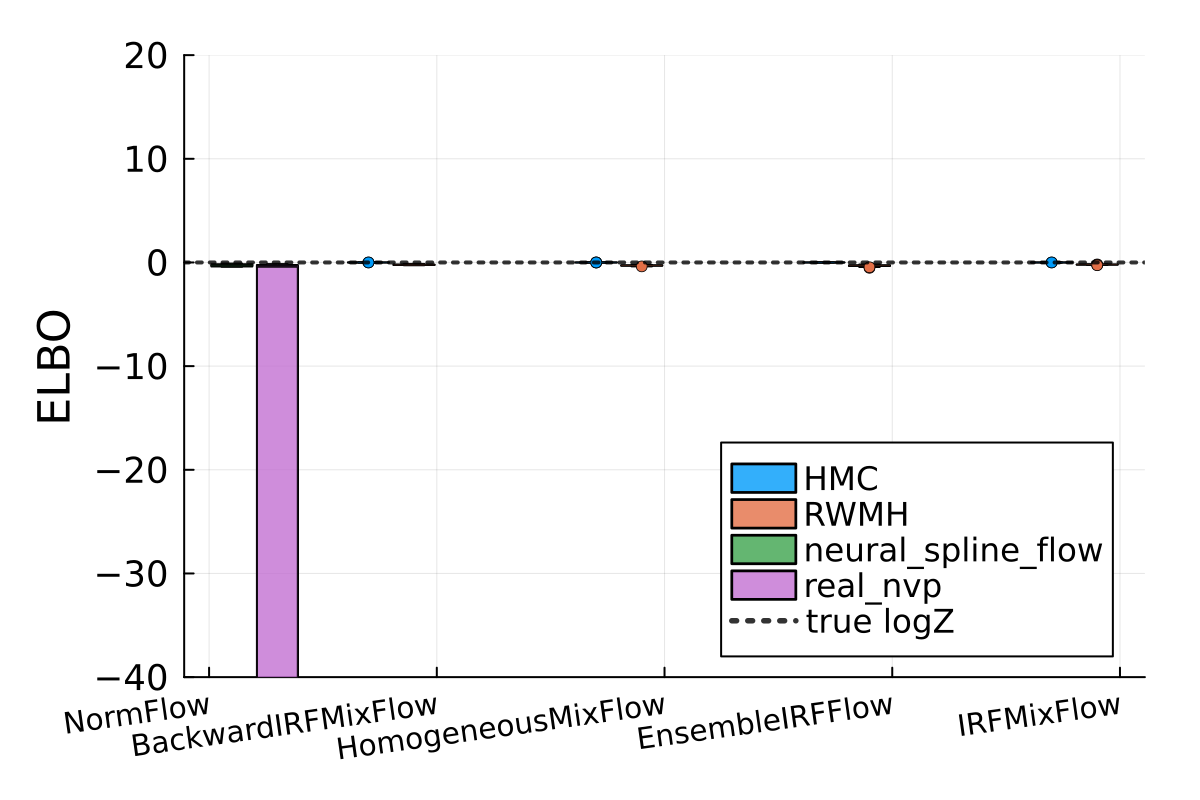}
		\caption{(a) ELBO: (from left to right) funnel, cross, warped Gaussian.}\label{fig:syn_elbo_no_banana}
	\end{subfigure}
    \begin{subfigure}{\columnwidth}
        \includegraphics[width=0.24\columnwidth, trim=0 20 50 0, clip]{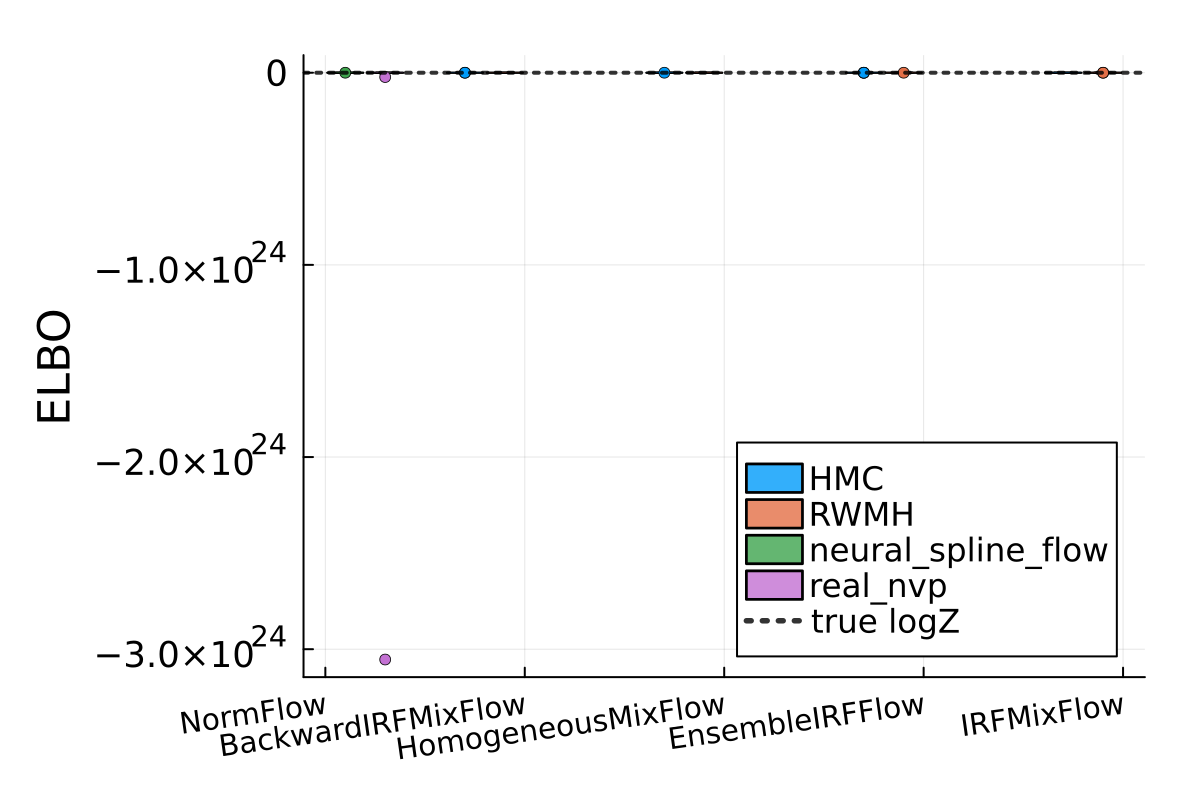}
        \includegraphics[width=0.24\columnwidth, trim=0 20 50 0, clip]{sync_evaluation_Funnel_elbo_raw.png}
        \includegraphics[width=0.24\columnwidth, trim=0 20 50 0, clip]{sync_evaluation_Cross_elbo_raw.png}
        \includegraphics[width=0.24\columnwidth, trim=0 20 50 0, clip]{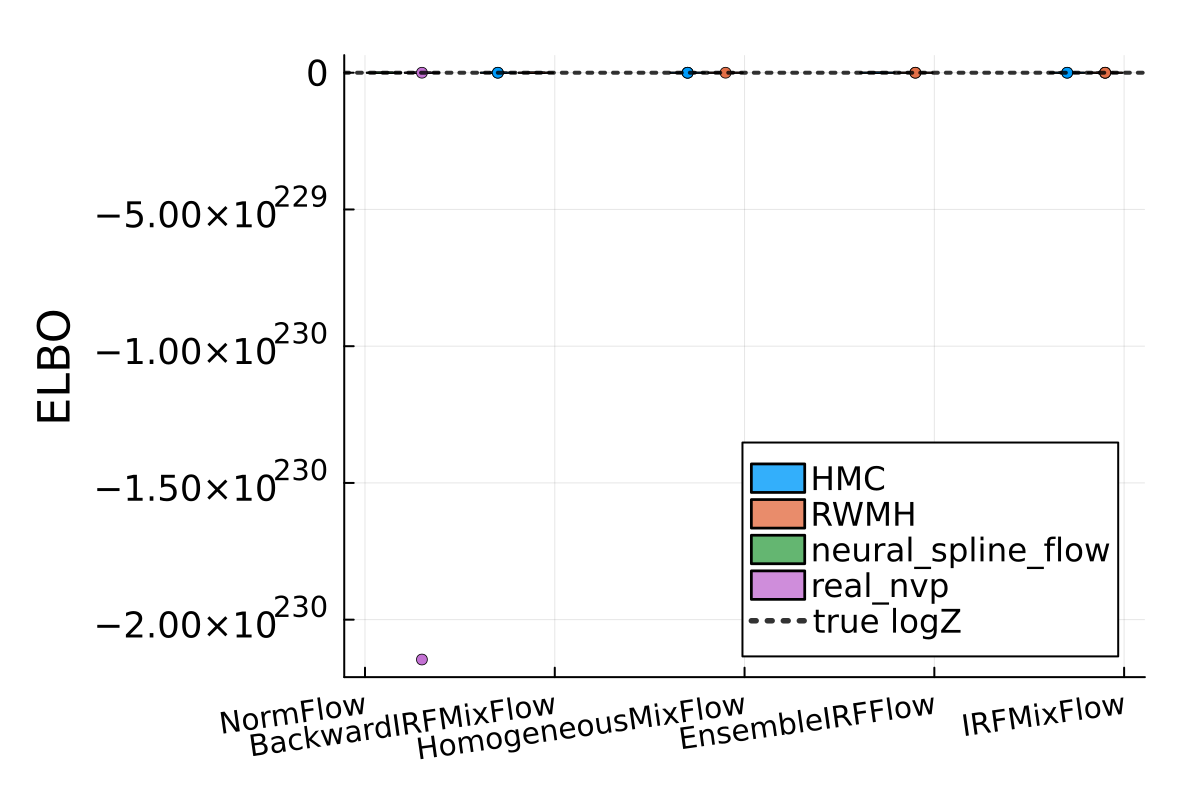}
		\caption{(b) ELBO in full range: (from left to right) Banana, funnel, cross, warped Gaussian.}\label{fig:syn_elbo_raw}
	\end{subfigure}
    \begin{subfigure}{\columnwidth}
        \includegraphics[width=0.33\columnwidth, trim=0 20 50 0, clip]{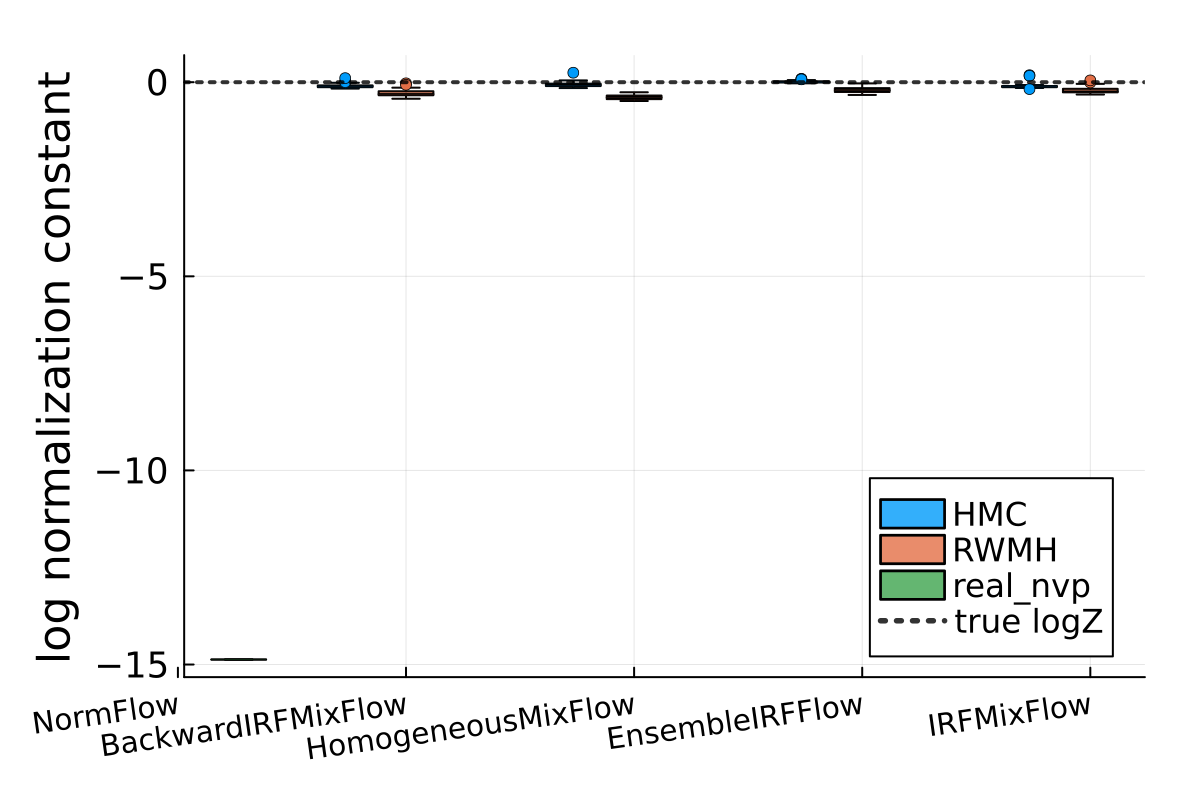}
        \includegraphics[width=0.33\columnwidth, trim=0 20 50 0, clip]{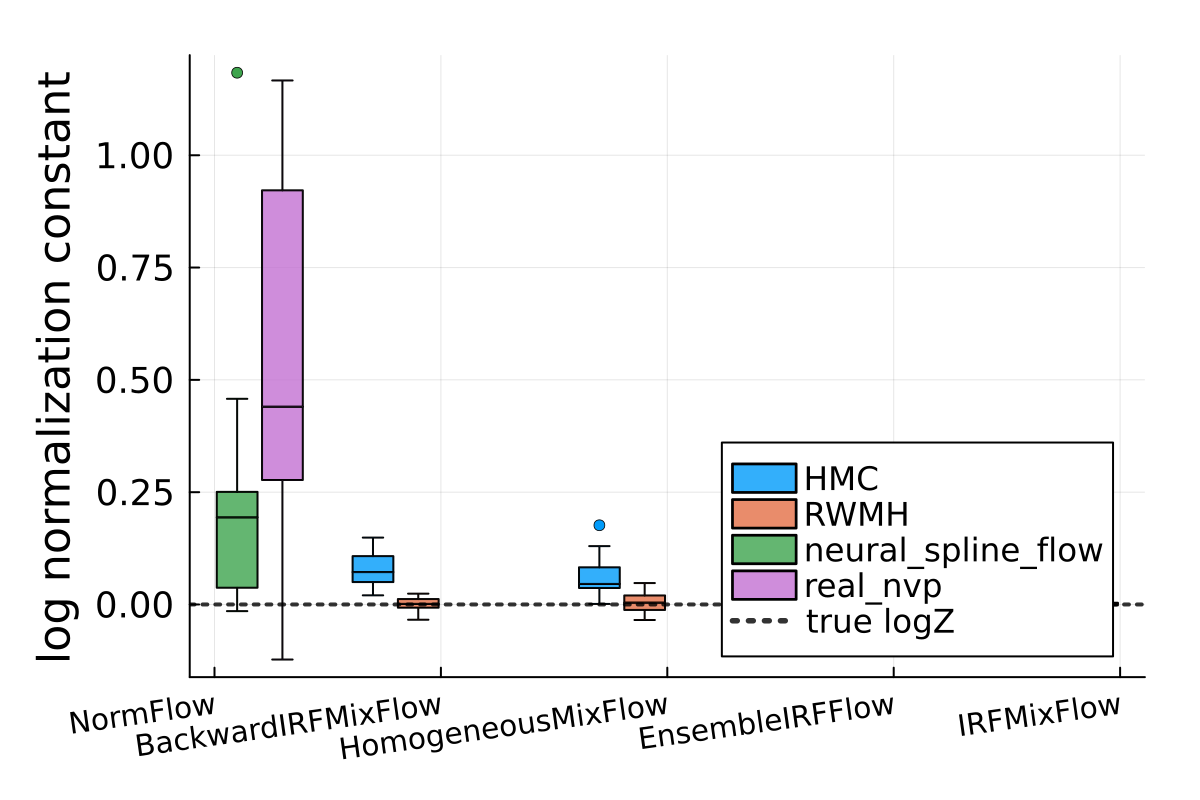}
        \includegraphics[width=0.33\columnwidth, trim=0 20 50 0, clip]{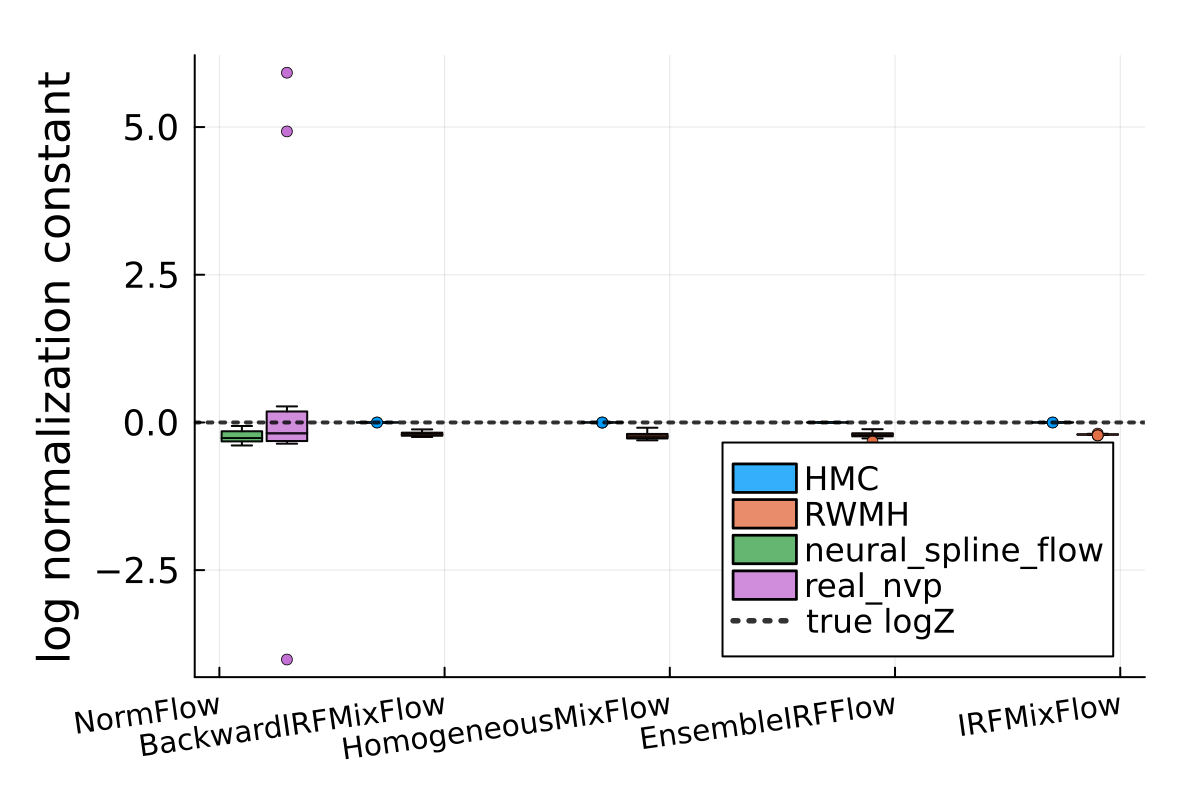}
		\caption{(c) $\log Z$ estimates: (from left to right) funnel, cross, warped Gaussian. }\label{fig:syn_logz_no_banana}
	\end{subfigure}
    \begin{subfigure}{\columnwidth}
        \includegraphics[width=0.33\columnwidth, trim=0 20 50 0, clip]{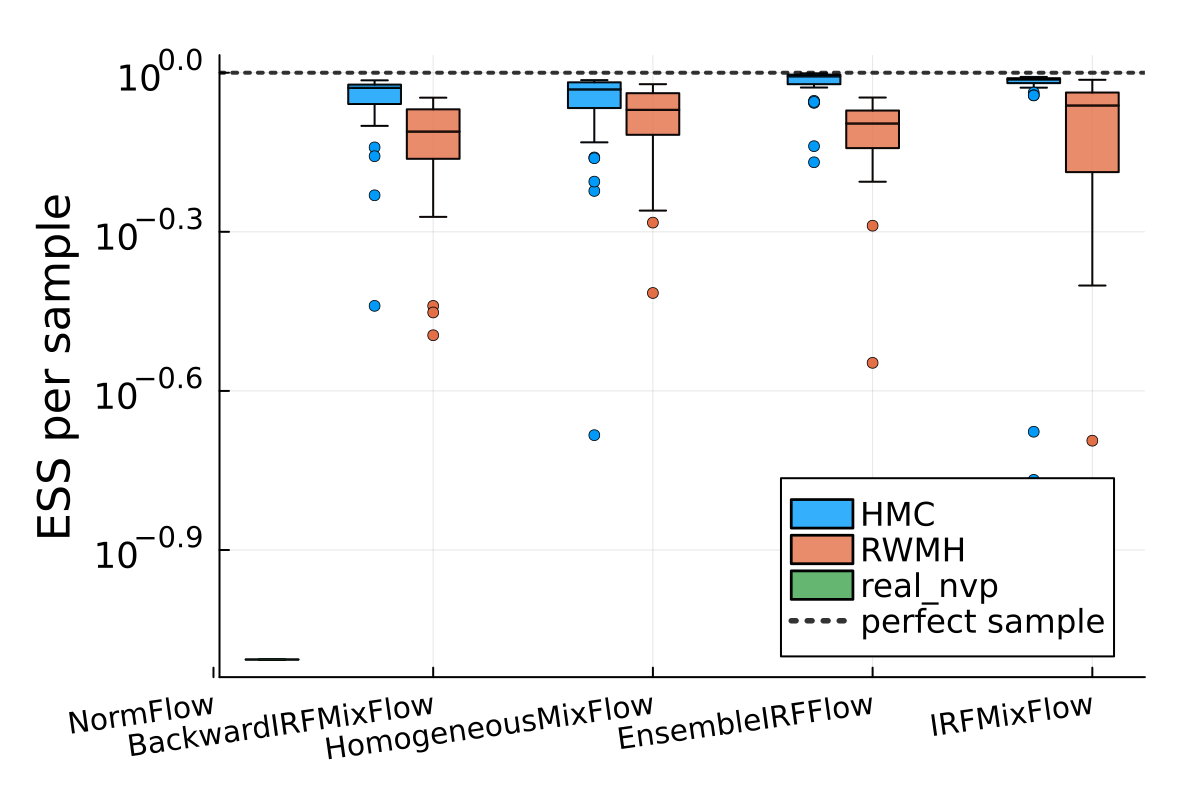}
        \includegraphics[width=0.33\columnwidth, trim=0 20 50 0, clip]{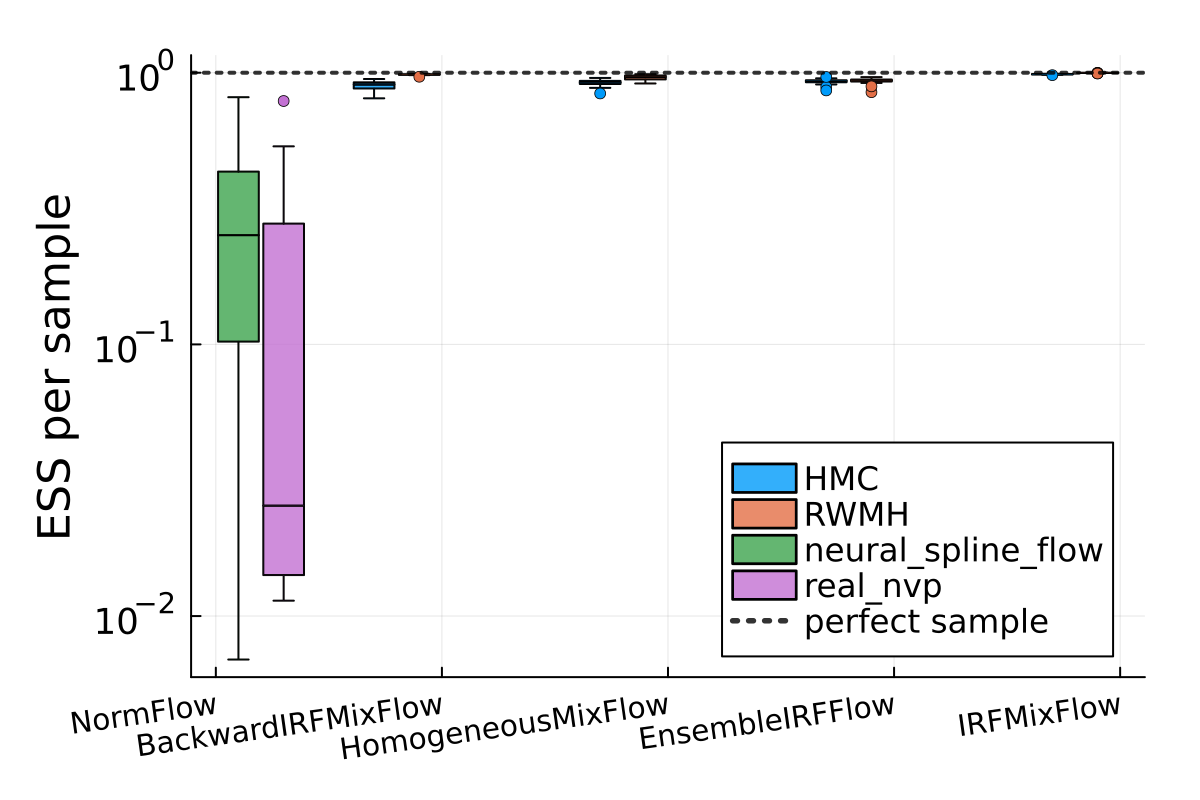}
        \includegraphics[width=0.33\columnwidth, trim=0 20 50 0, clip]{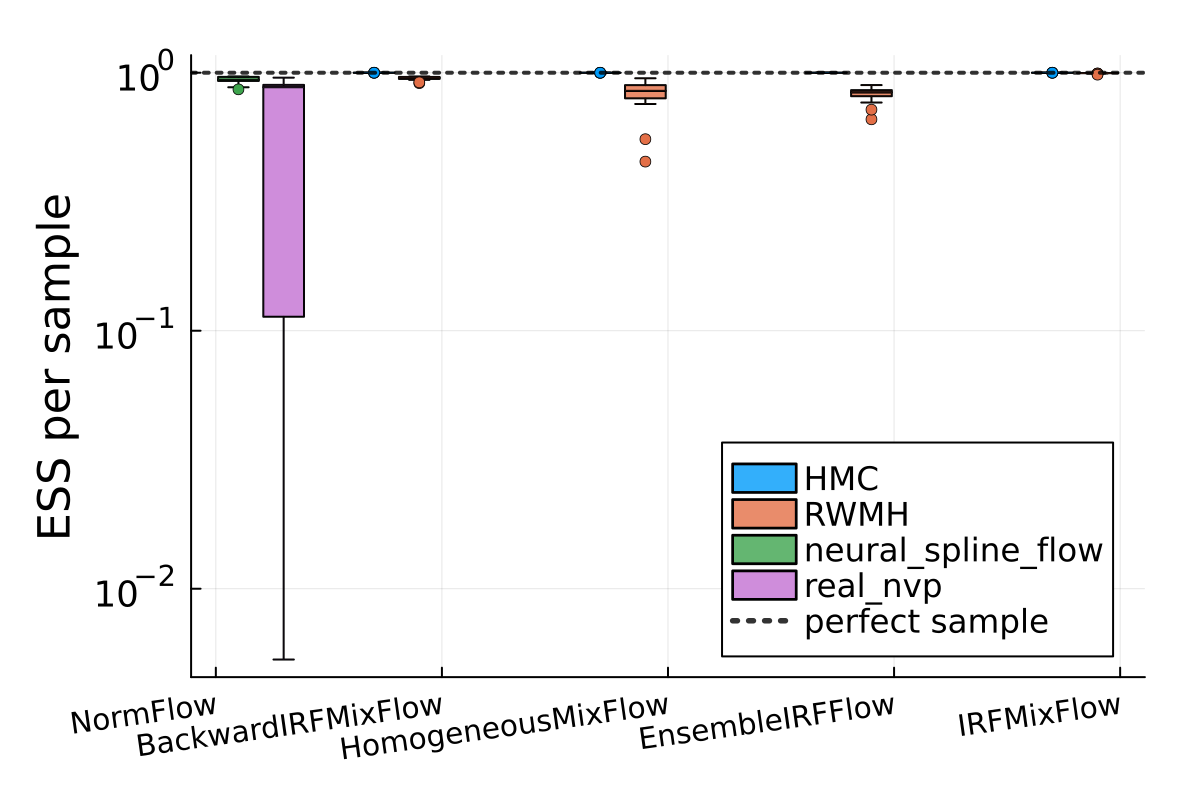}
		\caption{(d) Per-sample importance sampling ESS: (from left to right) funnel, cross, warped Gaussian. }\label{fig:syn_ess_no_banana}
	\end{subfigure}
    \caption{
        Variational approximation quality of IRF Flows versus \texttt{RealNVP} and \texttt{NSF}. 
        Box plots for IRF flows are based on $32$ independent runs, and $10$ runs for the normalizing flows.
    }\label{fig:syn_no_banana}
\end{figure}

\newpage
\subsection{Additional results for real-data experiments} \label{apdx:real_expt}

To approximate the ground truth, we ran an AIS procedure with $4096$ particles with adaptive schedule selection.
The initial temperature schedule was generated via mirror descent
\citep{chopin2024a} with a small step size of $0.005$; the schedule was then refined for
five rounds using the adaptive scheme of \citet{syed2024optimised}, yielding
more than $1000$ annealing steps for each data set. 
All reference values are taken as the median estimates across $10$ independent 
runs of the above procedure.

\begin{figure}[H]
    \begin{subfigure}{\columnwidth}
        \includegraphics[width=0.24\columnwidth, trim=0 100 100 0, clip]{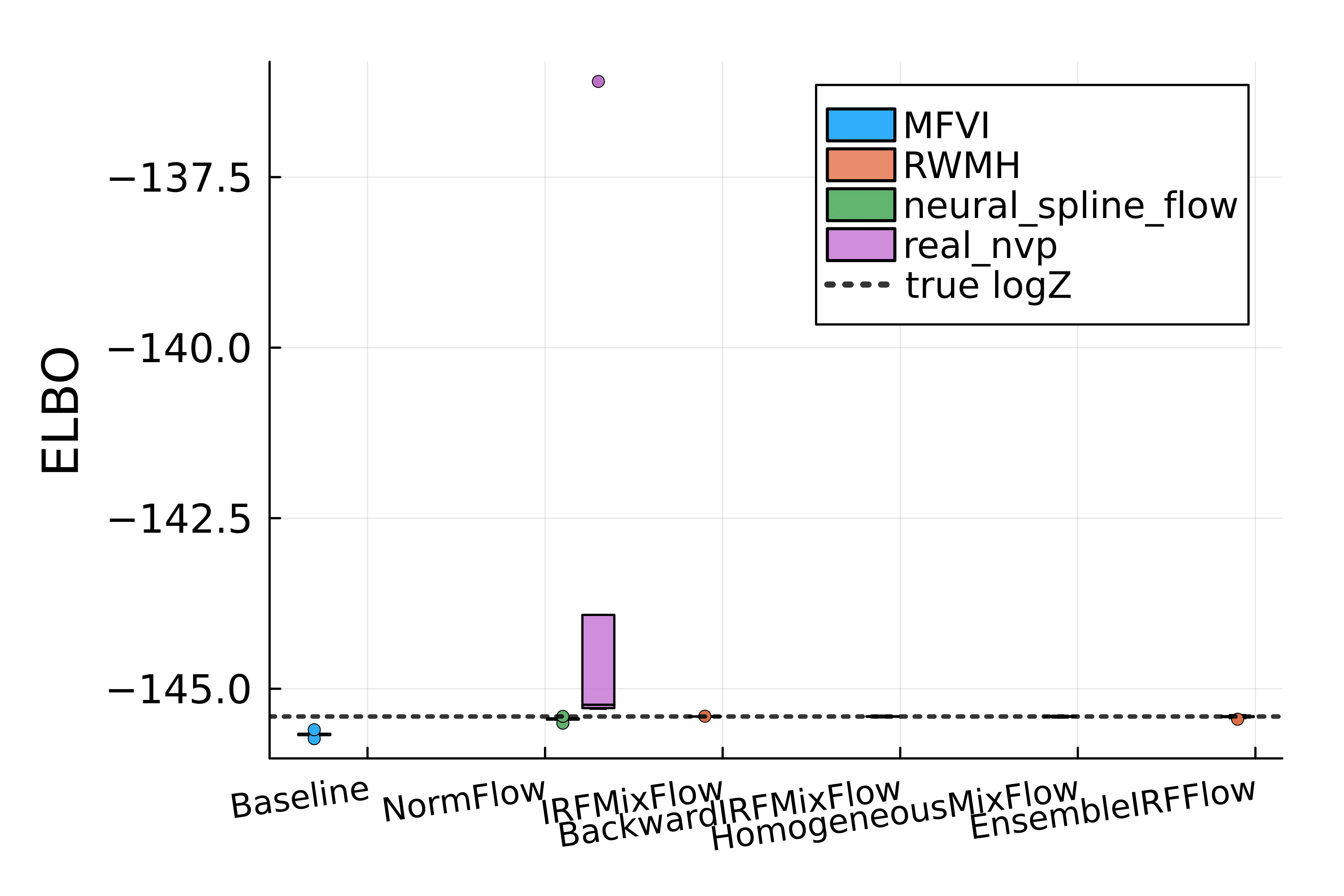}
        \includegraphics[width=0.24\columnwidth, trim=0 100 100 0, clip]{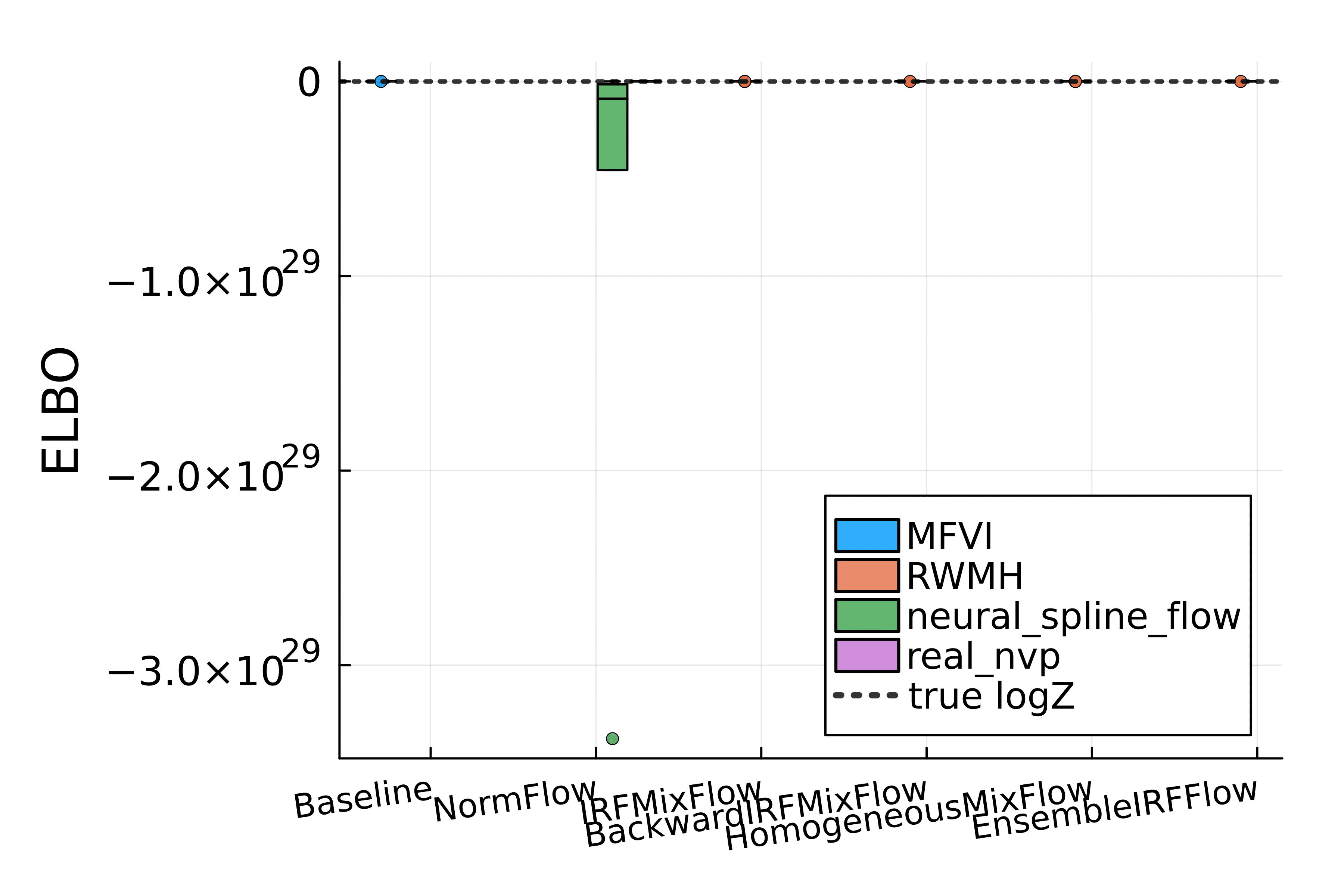}
        \includegraphics[width=0.24\columnwidth, trim=0 100 100 0, clip]{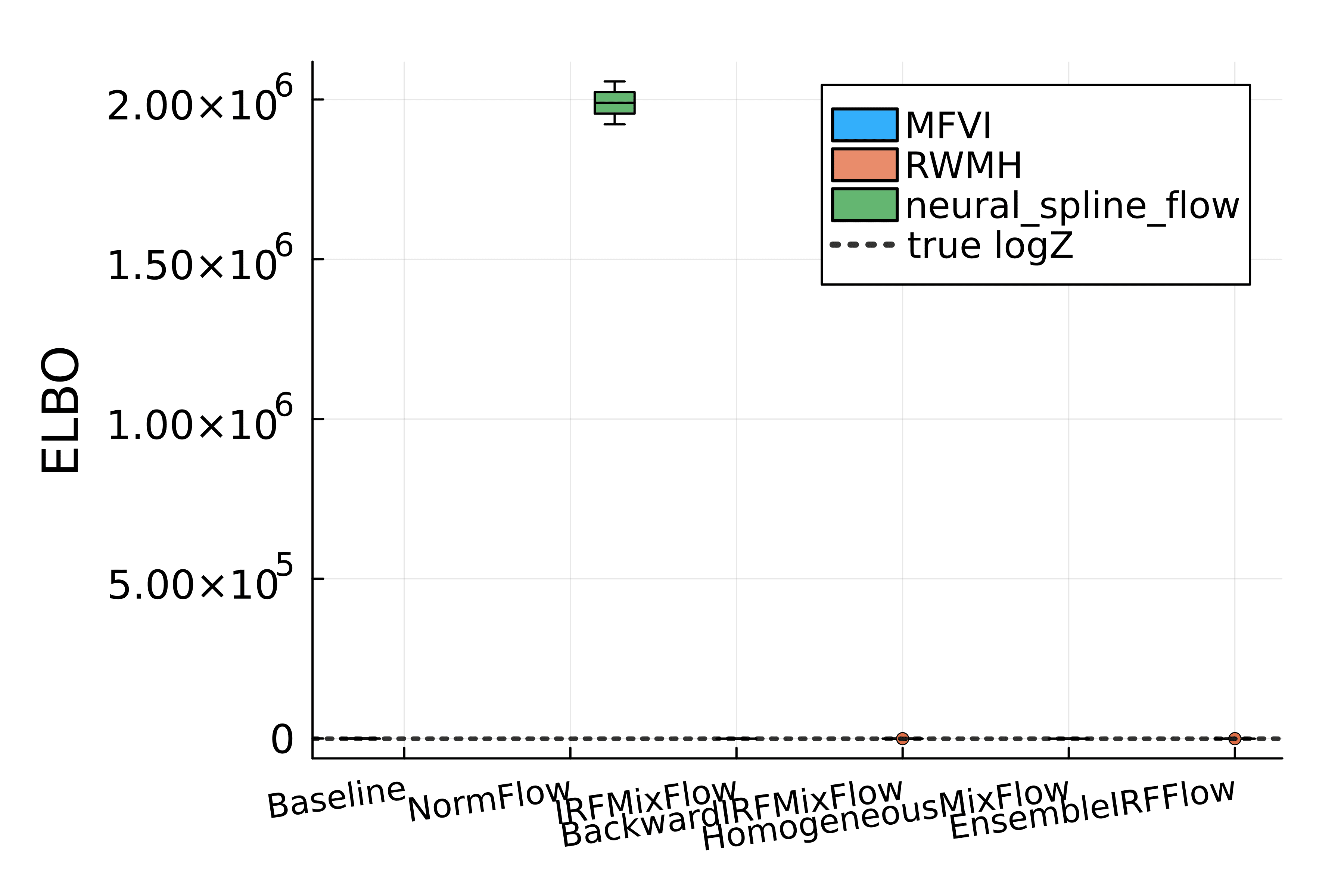}
        \includegraphics[width=0.24\columnwidth, trim=0 100 100 0, clip]{real_vi_LGCP_elbo_raw.png}
		\caption{ELBO in full range.}\label{fig:real_elbo_full_range}
	\end{subfigure}
    \begin{subfigure}{\columnwidth}
        \includegraphics[width=0.24\columnwidth, trim=0 0 100 0, clip]{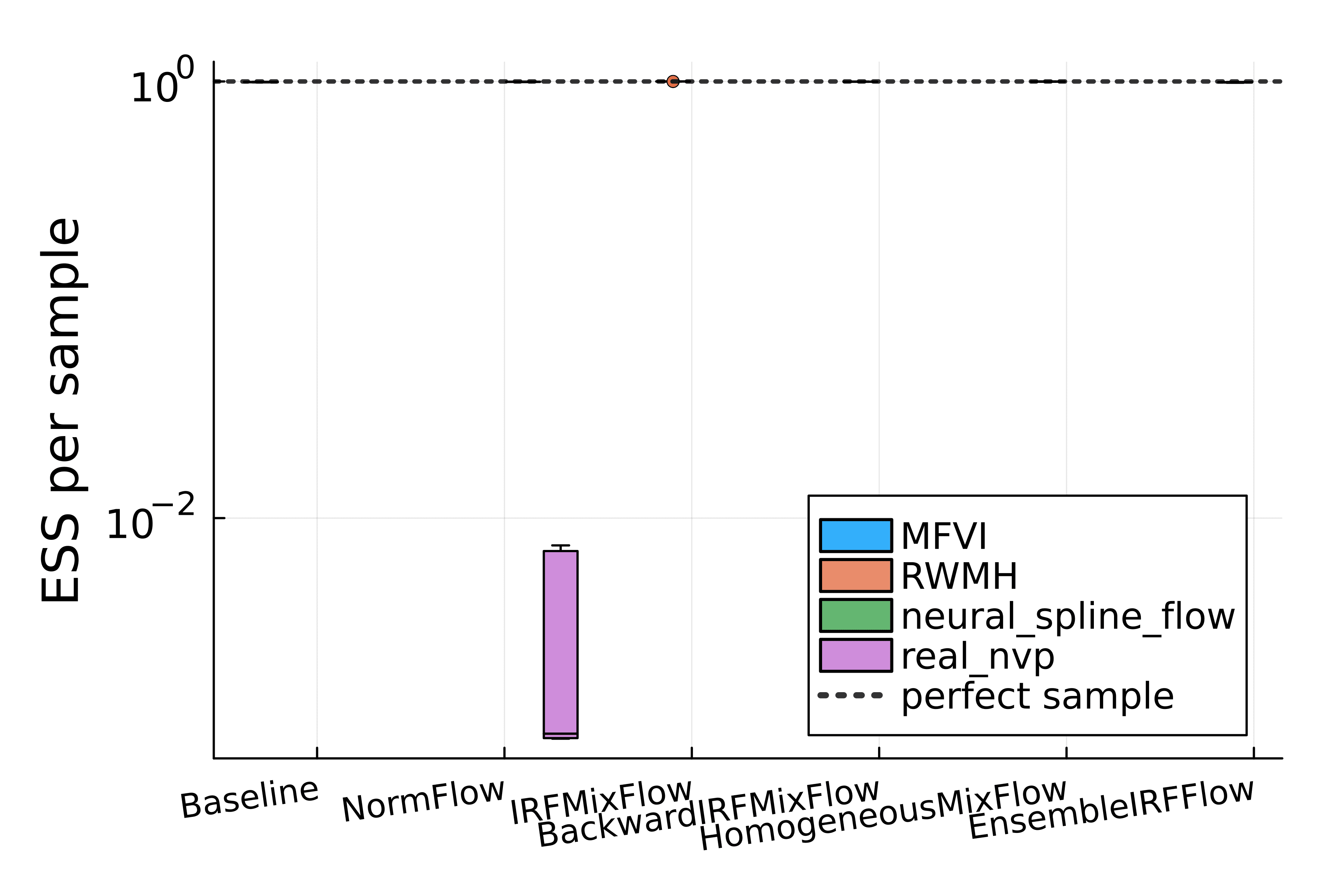}
        \includegraphics[width=0.24\columnwidth, trim=0 0 100 0, clip]{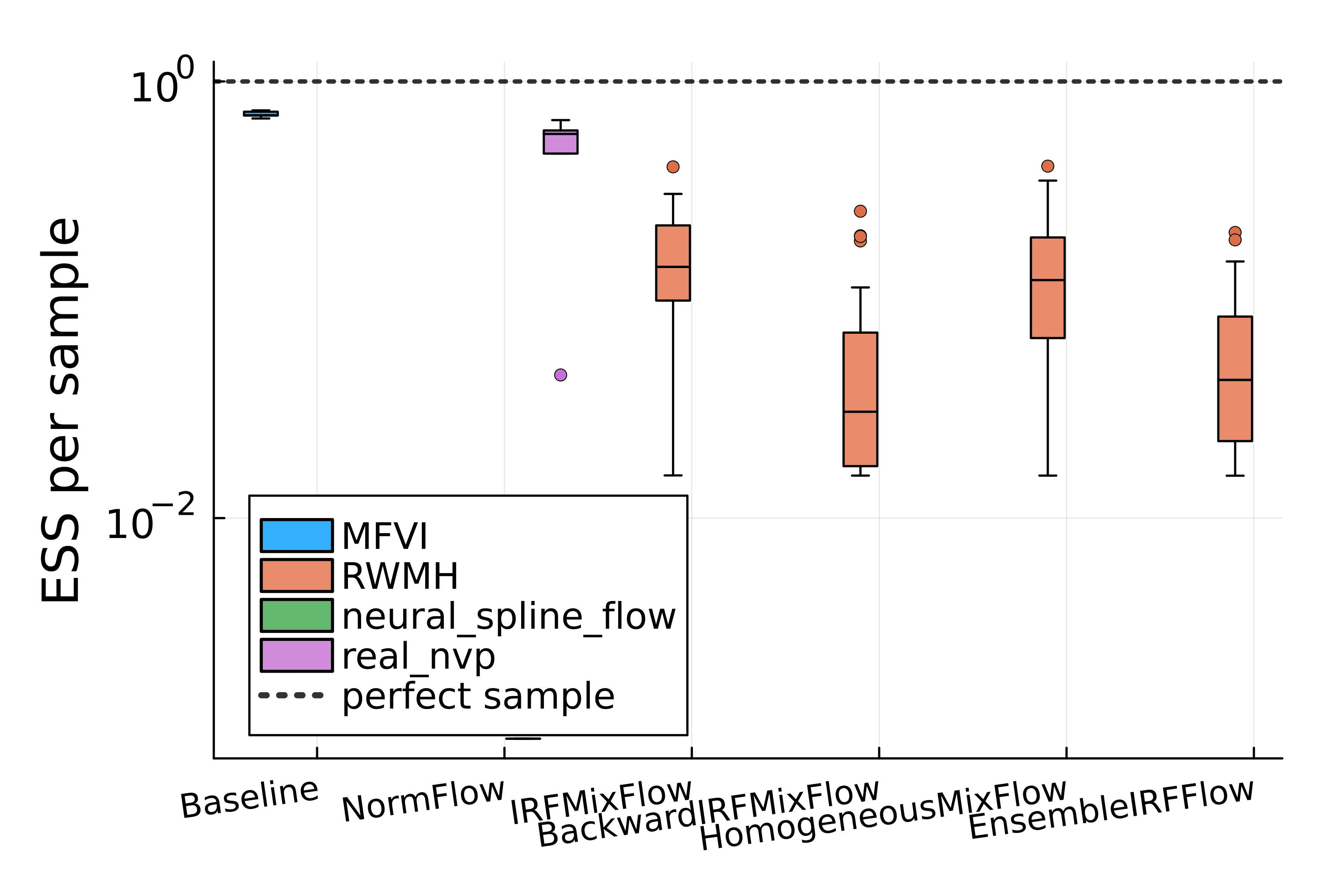}
        \includegraphics[width=0.24\columnwidth, trim=0 0 100 0, clip]{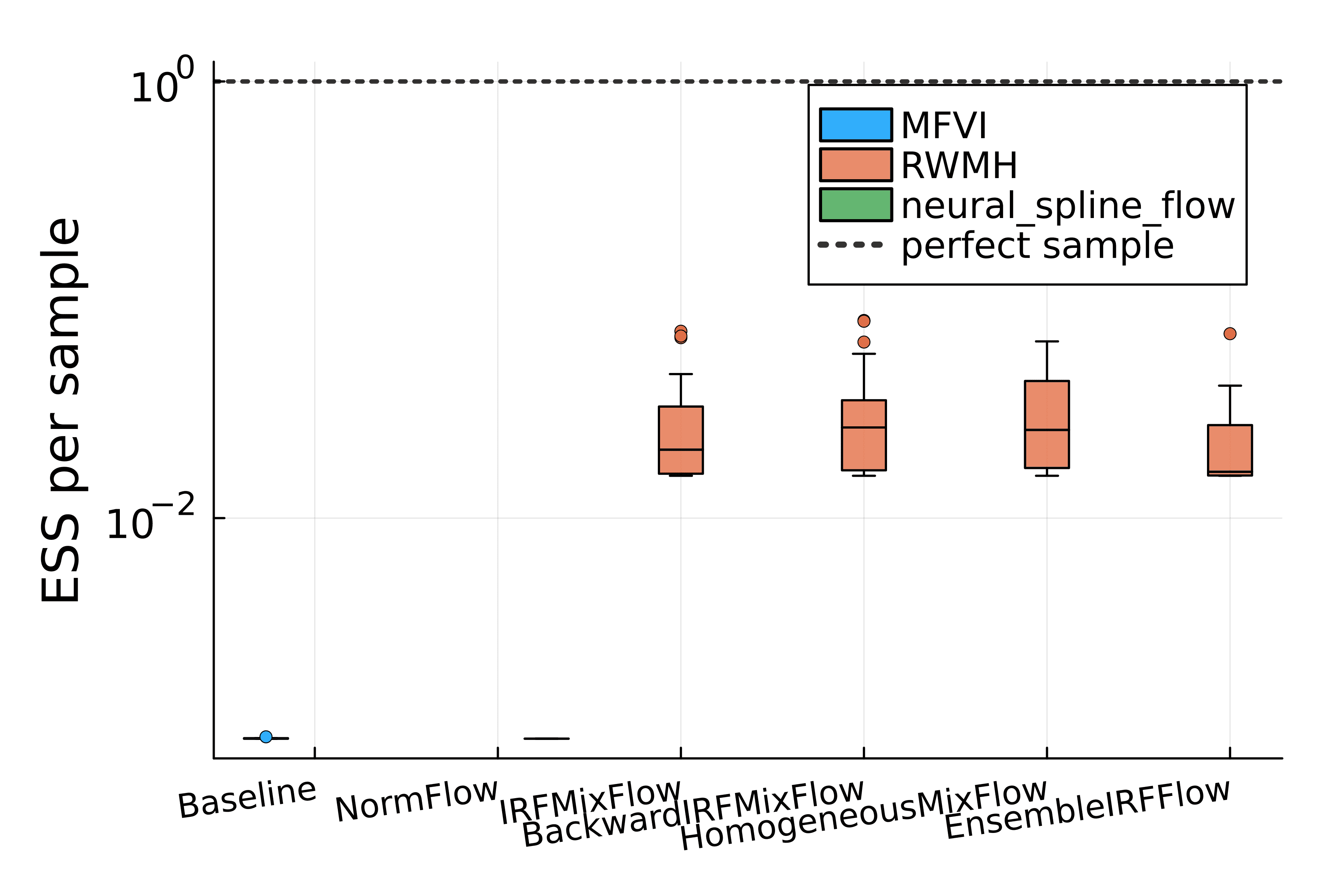}
        \includegraphics[width=0.24\columnwidth, trim=0 0 100 0, clip]{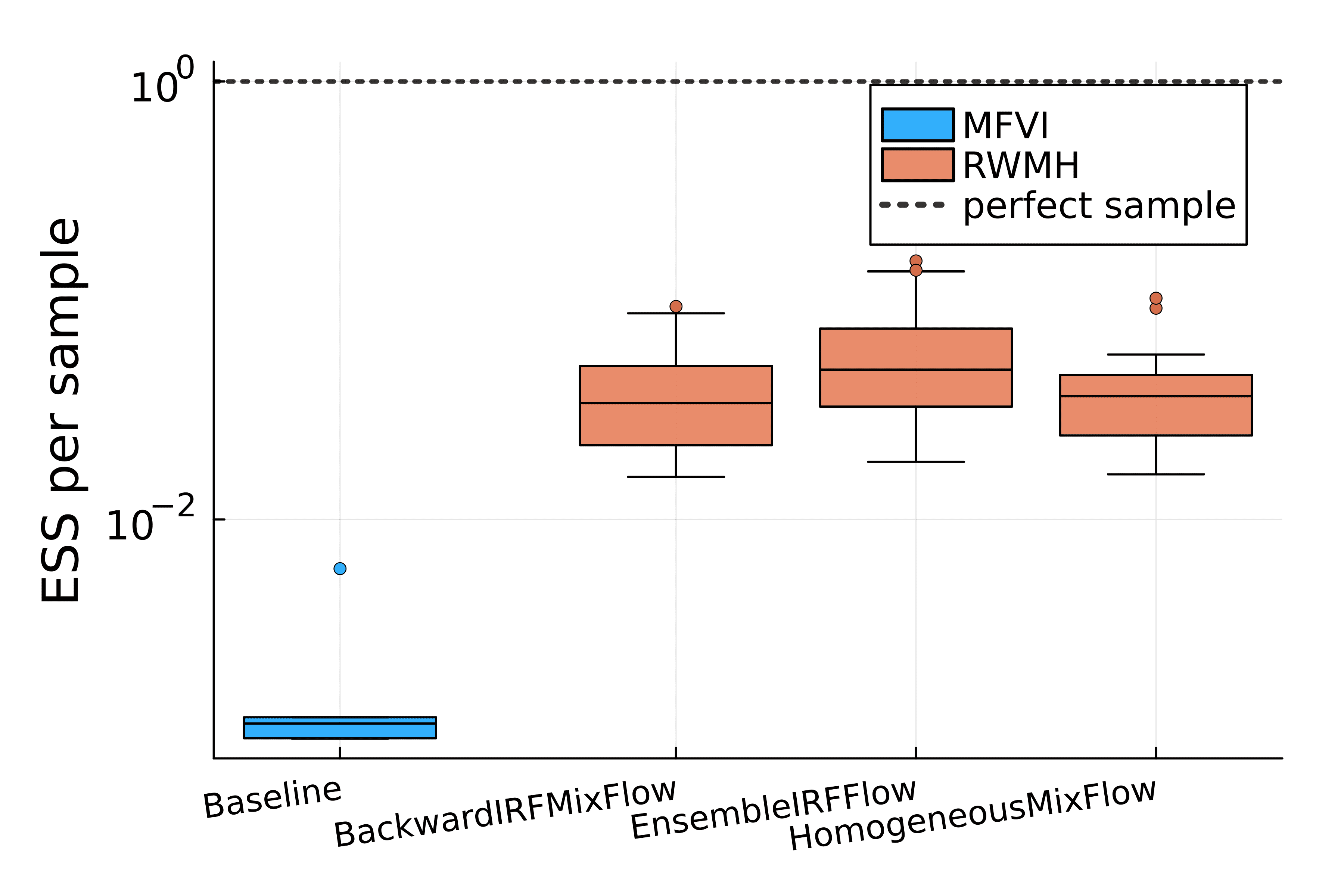}
		\caption{Per-sample importance sampling ESS.}\label{fig:real_ess}
	\end{subfigure}
    \begin{subfigure}{\columnwidth}
        \includegraphics[width=0.24\columnwidth, trim=0 0 0 0, clip]{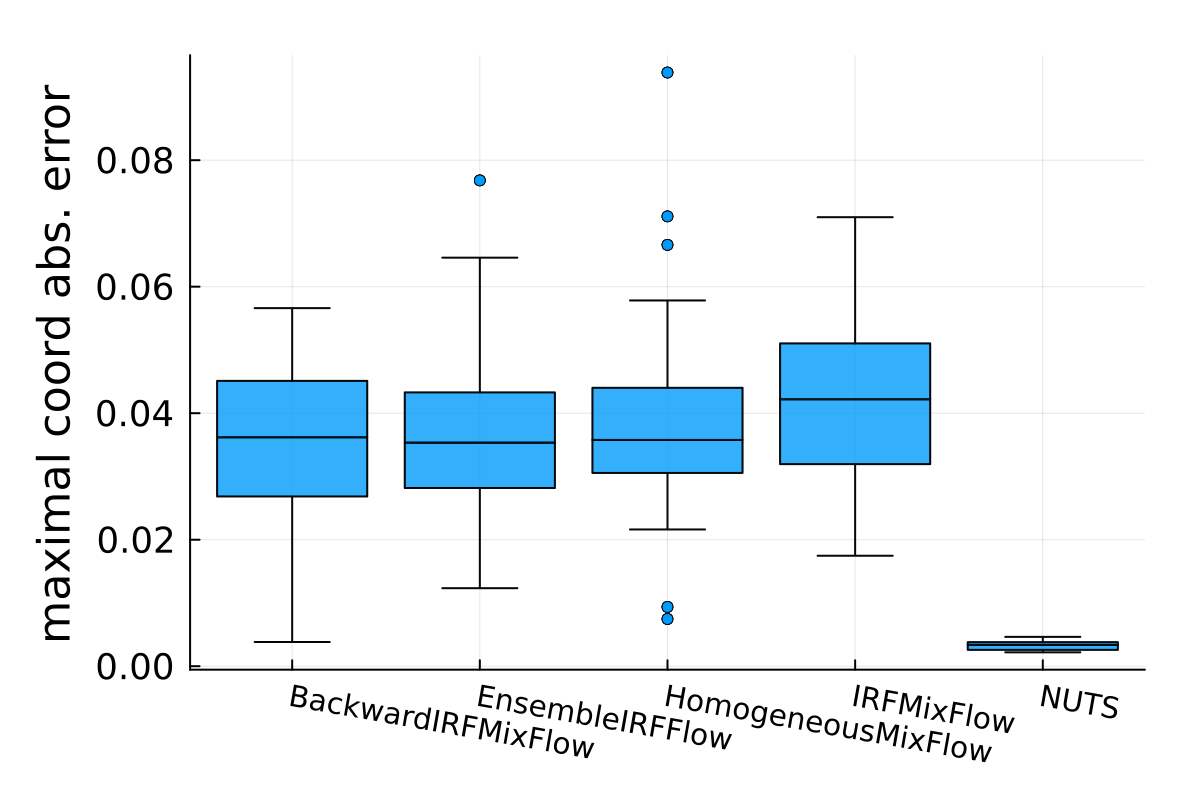}
        \includegraphics[width=0.24\columnwidth, trim=0 0 0 0, clip]{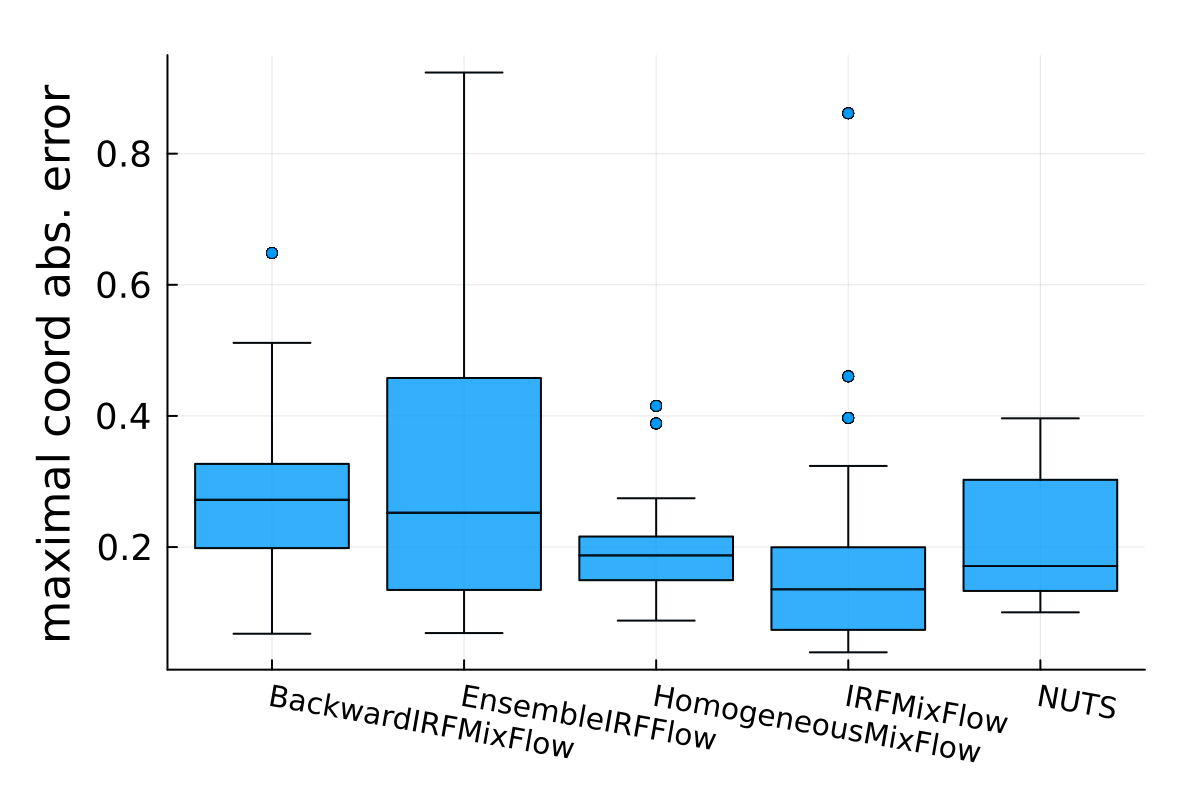}
        \includegraphics[width=0.24\columnwidth, trim=0 0 0 0, clip]{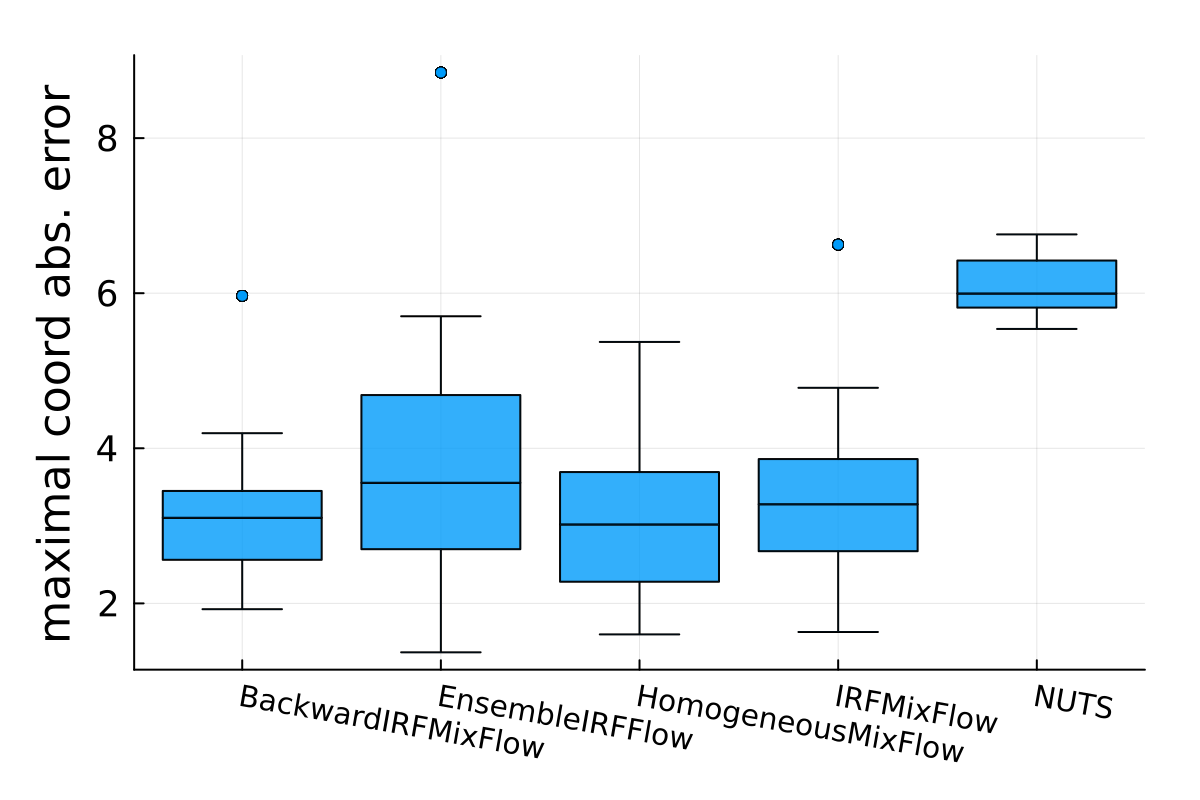}
        \includegraphics[width=0.24\columnwidth, trim=0 0 0 0, clip]{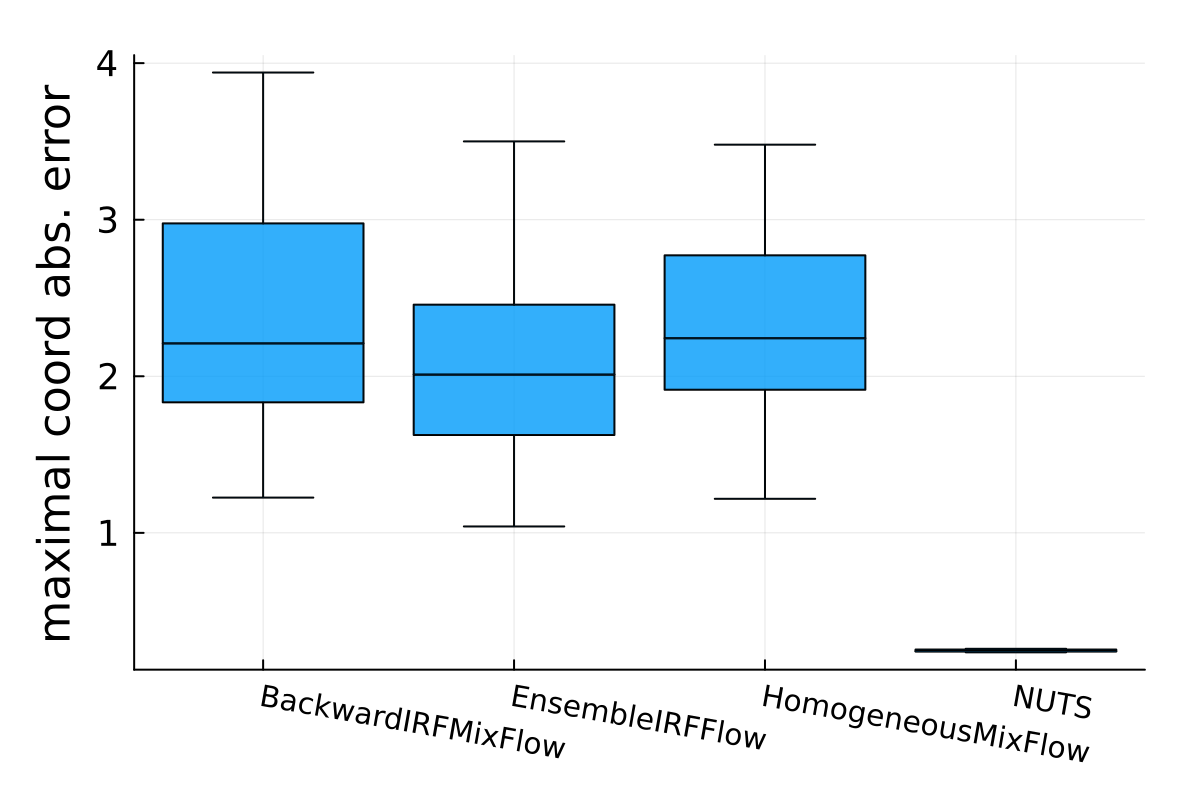}
		\caption{Maximal absolute error of the coordinate mean estimation against NUTS}\label{fig:real_mean}
	\end{subfigure}
\caption{Results on real-data benchmarks (columns, from left to right): 
    \texttt{TReg}($d=4$), \texttt{Brownian}($d = 32$), \texttt{SparseReg} ($d = 83$), and \texttt{LGCP} ($d = 1600$) 
}
\end{figure}

\end{document}